\documentclass[a4paper,11pt]{article}
\pdfoutput=1 

\usepackage{jcappub} 

\usepackage[T1]{fontenc} 

\title{\boldmath \textsc{FlexRT} - A fast and flexible cosmological radiative transfer code for reionization studies I: Code validation}

\author[a]{Christopher Cain\note{Corresponding author.}}
\author[b]{and Anson D'Aloisio}

\affiliation[a]{School of Earth and Space exploration, Arizona State University, Tempe, AZ 85281, USA}
\affiliation[b]{Department of Physics and Astronomy, University of California, Riverside}

\emailAdd{clcain3@asu.edu}
\emailAdd{ansond@ucr.edu}

\abstract{
The wealth of high-quality observational data from the epoch of reionization that will become available in the next decade motivates further development of modeling techniques for their interpretation.  Among the key challenges in modeling reionization are (1) its multi-scale nature, (2) the computational demands of solving the radiative transfer (RT) equation, and (3) the large size of reionization's parameter space.  In this paper, we present and validate a new RT code designed to confront these challenges.  \textsc{FlexRT} (Flexible Radiative Transfer) combines adaptive ray tracing with a highly flexible treatment of the intergalactic ionizing opacity. This gives the user control over how the intergalactic medium (IGM) is modeled, and provides a way to reduce the computational cost of a \textsc{FlexRT} simulation by orders of magnitude while still accounting for small-scale IGM physics. Alternatively, the user may increase the angular and spatial resolution of the algorithm to run a more traditional reionization simulation.   \textsc{FlexRT} has already been used in several contexts, including simulations of the Lyman-$\alpha$ forest of high-$z$ quasars, the redshifted 21cm signal from reionization, as well as in higher resolution reionization simulations in smaller volumes. In this work, we motivate and describe the code, and validate it against a set of standard test problems from the Cosmological Radiative Transfer Comparison Project.  We find that \textsc{FlexRT} is in broad agreement with a number of existing RT codes in all of these tests.  Lastly, we compare \textsc{FlexRT} to an existing adaptive ray tracing code to validate \textsc{FlexRT} in a cosmological reionization simulation.  }

\begin{document}
\maketitle
\flushbottom

\section{Introduction}
\label{sec:intro}

The Epoch of Reionization (EoR) lies at the forefront of current and forthcoming observational efforts.  In its first cycles, JWST has already begun to study galaxies and active galactic nuclei (AGN) throughout the EoR ($6 \lesssim z \lesssim 15$) in unprecedented detail, including their intrinsic ionizing properties \citep{Atek2023,Cameron2023,Saxena2023,Endsley2022b,Whitler2024}.  Ongoing and future cosmic microwave background (CMB) experiments such as the Atacama Cosmology Telescope (ACT;~\cite{Coulton2024}), South Pole Telescope (SPT; \citep{Raghunathan2024}), Simons Observatory \citep{2019JCAP...02..056A}, and CMB-S4~\citep{Jain2023} will use the patchy kinetic Sunyaev Zel'dovich (kSZ) and $\tau_{\rm es}$ effects to place tighter limits on the timing and duration of reionization.  Over the next decade, 21cm experiments such as the Hydrogen Epoch of Reionization Array (HERA; \cite{2017PASP..129d5001D}) and the Square Kilometre Array (SKA; \cite{Koopmans2015}) will continue to improve existing upper limits \citep[e.g.][]{HERA2021b,HERA2022} and possibly detect the EoR signal for the first time.  

Meanwhile, quasar absorption spectra measurements have advanced our understanding of the reionization history substantially, especially for the last phases of reionization.  The tail end of reionization has likely been detected in the statistics of the $5 < z < 6$ Ly$\alpha$ forest, e.g. the mean flux evolution, large-scale $\tau_{\rm eff}$ fluctuations, and dark gap statistics \citep{Kulkarni2019,Keating2020,Nasir2020,Bosman2021, Zhu2021,Zhu2022}.  The rapid evolution in the ionizing photon mean free path observed at $z>5.5$ is also consistent with the tail of reionization~\citep{Becker2021,Cain2021, Zhu2023,Gaikwad2023,Davies2023,Roth2023}. These measurements provide important boundary conditions that must be obeyed by any successful reionization model.  Additionally, the presence of damping wings in $z>6$ quasar spectra help constrain the global neutral fraction at higher redshift ~\citep[e.g.][]{Davies2018,Durovcikova2024}. Quasar absorption spectra will continue to play a prominent role in studying reionization because the number of high-quality quasar spectra will soon grow substantially.  Cosmological surveys by the Vera Rubin Observatory and the Dark Energy Spectroscopic Instrument (DESI) are projected to discover more than 10,000 $z>6$ quasars \citep{2009arXiv0912.0201L, 2016arXiv161100036D}, some fraction of which will be followed up with high-resolution spectrographs.  Most recently, signs of damping wing absorption in the $z < 6$ Ly$\alpha$ forest have strengthened the hypothesis that reionization is ongoing at these redshifts~\cite{Zhu2024,Spina2024}.  

Looking forward, the deepest insights into reionization will come from combining these rich and complimentary data sets. A key problem confronting the field is how to maximize the information contained in them. Together, they are broadly sensitive to not only the distribution of intergalactic hydrogen during reionization, but also the structures and histories of the HI-ionizing radiation background and IGM temperature.  Simulating reionization accurately at this level of detail requires solving, or at least approximating, the radiative transfer (RT) equation, which encodes the physics linking the ionizing sources to the state of the IGM.  But solving the RT equation in this context is a formidable computational challenge for several reasons.  The first is the high dimensionality of the problem.  The RT equation has $7$ dimensions ($3$ spatial, $2$ angular, $1$ frequency, and $1$ time), all of which are important for reionization.  Moreover, reionization is a massively multi-scale problem.  The intrinsic ionizing properties of galaxies are set by astrophysics on parsec scales and below: star formation and feedback processes, AGN activity, as well as the structures of the interstellar and circumgalactic media.  The small-scale clumpiness of the IGM sets its recombination rate and self-shielding properties, which determine the number of photons that the sources must produce to complete reionization~\citep{Haiman2000,Duffy2014,Davies2021b}. The clumping scale may be as low as a kpc.   Finally, simulations in representative volumes ($10$s-$100$s of cMpc on a side) may also contain millions of ionizing sources clustering on $\sim 10$ cMpc scales.   The distribution and sizes of ionized regions depends on the source clustering as well as the small-scale clumpiness of the IGM \citep[e.g.][]{Cain2022b}.  In principle, therefore, reionization simulations require at least $8-9$ orders of magnitude in dynamic range to bridge the gap between galaxy physics (on $\sim$parsec scales) and the large-scale features of reionization.

Because it is impossible to fully resolve every process relevant for reionization, the problem also consists of many dimensions {\it in parameter space}.  Most free parameters typically characterize the highly uncertain ionizing outputs of the sources, e.g. the ionizing escape fraction, which may in general depend on galaxy properties and redshift.  Indeed, constraining the source parameters is a chief goal of the field, since they are often very challenging or otherwise impossible to study directly.  The opacity of the ionized IGM to ionizing photons is also frequently treated as a free parameter, parameterized by either a mean free path (MFP) or a recombination clumping factor~\citep{Pawlik2015,Shukla2016,Mao2019}.  The most recent parameter inference studies include $\sim 5-10$ free parameters, requiring thousands to millions of simulations \citep[e.g.][]{Qin2021,Maity2023}.  

Owing to these difficulties, reionization simulations are divided into two very different approaches, broadly speaking.  One approach avoids the computational cost of explicitly solving the RT equation, using instead numerical recipes that approximate the most important facets of the full RT solution.  These are often referred to as ``semi-numerical'' methods and are designed for probing the vast parameter space of reionization \citep[e.g.][]{Furlanetto2004,Mesinger2007,Grieg2015,Choudhury2018,Trac2021}.  The second approach, which we call ``fully numerical," is to solve the RT equation in detailed cosmological simulations using methods that take advantage of simplifications and/or optimizations afforded by the unique nature of the reionization problem \citep{Mellema2006,Trac2007,Rosdahl2013,Gnedin2014,Ocvirk2016,Rosdahl2018,Kannan2022}.  

Of course, both approaches have their advantages and disadvantages, which largely dictate their uses. Semi-numeric codes can well-approximate the solution of the RT equation in the context of reionization~\citep{Zahn2011}.  Parameter space inferences are impossible without them.   However, because these algorithms do not solve the RT equation, they do not converge to the exact RT solution; improving their accuracy is not simply a matter of increasing the number of resolution elements.  As such, quantifying their modeling errors can be a difficult task.   Additionally, these methods are less suited to modeling the ionizing radiation background and IGM temperature -- both critical elements for modeling quasar absorption spectra (see however \cite{Nasir2020} and \cite{Qin2021,Qin2022} for prescriptions along these lines. )  

On the other hand, the fully numerical approach is well-suited for studying in detail the state and physics of the reionizing IGM. In recent years, this approach has generally moved in the direction of fully coupled radiative hydrodynamics simulations, many of which also simulate the galaxies, including star formation, AGN, and feedback processes \citep[e.g.][]{Gnedin2014,Rosdahl2018,Ocvirk2018, Doussot2019, Kannan2022}.  With this comes the great advantage that the simulations can now leverage the constraints provided by direct observations of the galaxy population.  But the state-of-the-art simulations can cost millions of core hours per run, allowing for only a handful of simulations to be run per year, which precludes parameter space studies.  The limited number of such simulations available also makes it difficult to disentangle the effects of approximations and differing sub-grid modeling choices, which are required in even the most state-of-the-art simulations~\cite[e.g.][]{Vogelsberger2014,Hassan2022}.  It is also worth noting that the moment-based RT methods that are most widely used today truncate the moment hierarchy at the dipole moment of the intensity field by making an ansatz for the quadrupole, and therefore start from an approximate form of the RT equation. Hence, these methods also do not converge to the exact RT solution in any limit of the parameters related to numerical implementation (e.g. number of resolution elements).

We have been developing a new RT code named \textsc{FlexRT} (short for Flexible Radiative Transfer) that attempts to achieve some of the speedup afforded by semi-numeric methods without losing direct contact with the RT equation. The organizing principle of \textsc{FlexRT} is flexibility for the user to tune the level of fidelity up or down as needed for the problem at hand.  Our code solves the RT equation directly, so it converges, at least in principle, to the exact solution in the limit of infinitely many resolution elements.  This allows modeling errors associated with the RT to be straightforwardly quantified.  We also aim for the code to be sufficiently fast (in the regime of fewer resolution elements) and accurate enough for modestly-sized parameter space studies.  To achieve this, \textsc{FlexRT} has two central features: (1) an adaptive ray tracing scheme with numerical parameters that can be adjusted to trade off the speed and accuracy of the RT calculation as desired, and (2) a highly flexible, general prescription for the intergalactic ionizing photon opacity that admits the use of a wide variety of models for un-resolved IGM dynamics. This allows \textsc{FlexRT} to increase the effective dynamic range of the simulation without requiring more RT cells.

The first iteration of \textsc{FlexRT} has already been used in several studies~\citep[e.g.][]{Cain2021,Cain2022b}.  In this first paper of a multi-part series, we will describe \textsc{FlexRT}'s basic framework in detail and present a battery of tests validating its basic functionality and accuracy.  In Paper II, we will present a new and improved version of the sub-grid opacity model first described in Ref.~\cite{Cain2021}, and demonstrate its accuracy in capturing the effects of un-resolved IGM physics down to $\sim$ kpc scales.  Paper III will present performance tests of a further optimized version of \textsc{FlexRT} and demonstrate its use in a simple parameter-space study. 

This work is outlined as follows.  In \S\ref{sec:treatments}, we briefly summarize several of the radiative transfer methods widely used in reionization studies, and describe how \textsc{FlexRT} fits into this context.  \S\ref{sec:adaptive_ray_tracing} describes the ray tracing approach used in \textsc{FlexRT}, while \S\ref{sec:IGMopacity} describes how we solve for the IGM opacity and the ionization state of the gas.  In \S\ref{sec:CRTCP_tests}, we subject \textsc{FlexRT} to the static density RT tests outline in the Cosmological Radiative Transfer Codes Comparison Project~\citep[Paper I, ][]{Iliev2006}.  \S\ref{sec:moving_screen} addresses how \textsc{FlexRT} handles the propagation of cosmological ionization fronts.  In \S\ref{sec:raytracingtests}, we run additional tests that target the ray tracing method used in \textsc{FlexRT}, including a direct test against the \textsc{RadHydro} code of Ref.~\cite{Trac2007}.  We conclude in \S\ref{sec:conc}.  Throughout, we assume the following cosmological parameters: $\Omega_m = 0.305$, $\Omega_{\Lambda} = 1 - \Omega_m$, $\Omega_b = 0.048$, $h = 0.68$, $n_s = 0.9667$ and $\sigma_8 = 0.82$, consistent with results from Ref.~\cite{Planck2018}. All distances are quoted in co-moving units unless otherwise specified. 

\section{The context \& motivation for \textsc{FlexRT}}
\label{sec:treatments}

In this section, we briefly summarize the landscape of existing radiative transfer methods used to simulate reionization. These include moment-based (\S\ref{subsubsec:momentbased}), ray-tracing (\S\ref{subsubsec:raytracingRT}), and Monte-Carlo methods (\S\ref{subsubsec:montecarlo}). This section is not a comprehensive summary; rather, it aims to establish context and motivation for \textsc{FlexRT}. Readers interested in the details of \textsc{FlexRT} may skip this section without loss of continuity.    

\subsection{Radiative transfer implementations}
\label{subsec:RT}


\subsubsection{Moment-Based RT methods}
\label{subsubsec:momentbased}

The high dimensionality of the RT equation makes it one of the most computationally intensive equations to solve numerically, even in relatively simple cases.  A common approach to reducing the dimensionality is to take moments of the RT equation in the angular dimensions, and then truncate the hierarchy at the dipole level with an ansatz for the quadrupole moment, i.e. the Eddington tensor.   The two most prominent ansatzes in use for reionization codes today are the M1 closure \citep{Levermore1984} and the optically thin Eddington tensor~\citep[OTVET;][]{Gnedin2001}.  The equations governing the first two moments -- the radiation energy density and flux -- take on the form of conservation laws, analogous to the Euler equations of fluid dynamics, for which very efficient numerical solvers have been developed.  Such codes are also relatively straightforward to parallelize across many CPUs because of their fluid-like description of the radiation field.  Indeed, the M1 closure has the particularly convenient property that the Eddington tensor becomes a local function of radiation energy density and flux.  Owing to its computational efficiency, the moment approach has become the most widely adopted type of RT algorithm used in almost all of the recent radiation hydrodynamics simulations of reionization, many of which include galaxy formation physics~\cite{Gnedin2014,Ocvirk2016,Ocvirk2018,Rosdahl2018,Kannan2022}.  Indeed, without the moment-based RT implementations, the unprecedented scales and resolutions achieved by these simulations would not be possible.

The accuracy of moment-based codes has been quantified for a suite of idealized test problems in the Cosmological Radiative Transfer Comparison Project \citep{Iliev2006, Iliev2009}, and in other works \citep[e.g.][]{Rosdahl2013, Kannan2019}.  To our knowledge, however, there has been no quantitative study on the inaccuracies introduced by the fluid-like approximation to the radiation field, which arises from the local approximation to the Eddington tensor in the M1 method, in the cosmological reionization problem. There is good motivation for such a study.  Ref.~\cite{Wu2021} considered several idealized RT problems that are relevant for reionization and found that moment-based methods tend to: (1) over-ionize dense, self-shielding absorption systems, which can result in a factor of $\sim 2$ under-prediction of the photoionization rate of intergalactic gas for a given emissivity; and (2) err in the structure of the ionizing radiation background on scales smaller than the mean free path (MFP). For instance, they found that the M1 closure significantly underestimates the fluctuations on these scales.  This effect is visualized nicely in Figure 5 of Ref.~\cite{Gaikwad2023}, who noted that RT simulations run with \textsc{Aton} broadly exhibited a more uniform ionizing background compared to models generated with \textsc{EX-CITE}, which is essentially a ray tracing algorithm. 

In spite of these considerations, moment-RT will remain a premiere method for simulating reionization. One of its greatest advantages is that it allows simulations to have a much larger number of resolution elements compared to other RT methods, which has ushered in a new era of coupling RT, hydrodynamics, and galaxy formation.  

\subsubsection{Ray-tracing}
\label{subsubsec:raytracingRT}

Ray tracing is widely recognized as the most accurate numerical RT method, although the accuracy depends on details of the numerical implementation \citep{Iliev2006, Iliev2009}. It is also the most computationally expensive. These algorithms involve tracing rays around individual sources and calculating the optical depth along each ray, which determines the number of photons absorbed at each location.  Rays are attenuated and eventually disappear at large distances from sources when their optical depths reach $\tau >> 1$. 

Three primary types of ray tracing methods have been applied to simulate reionization: long-characteristics, short-characteristics, and adaptive ray tracing (although not all codes fit neatly into these categories).  The long-characteristics method traces rays between every cell and every source and integrates along the rays to get the associated optical depths.  This is the most expensive method and is mostly intractable for running large reionization simulations for two reasons. First, the number of sources is large (typically $\sim$ millions), and second, radiation travels large distances from the sources after isolated ionized regions overlap, requiring a very large number of rays per source to maintain high spatial resolution.  The short-characteristics method reduces computation time by tracing a limited number of rays from the sources in certain directions and then interpolating the column densities of those directly computed cells to obtain the integrated column density between any given source and cell. \textsc{C$^2$-Ray} is a prominent example of a short-characteristics code that has been used extensively for reionization simulations~\citep{Mellema2006,Hirling2023}.

In adaptive ray tracing, rays split to maintain the level of spatial and angular resolution desired by the user.  For a problem with only one source, for instance, only a small number of rays are cast close to the source, and these split as they move further away (increasing angular resolution) to maintain spatial resolution. An adaptive ray tracing method designed for reionization studies was described in Ref.~\cite{Abel2002}, and implemented in several codes~\citep{McQuinn2007, Trac2007, 2019MNRAS.483.1582H}.  When the radiation from many sources overlaps -- as is the case in reionization, particularly near its end  -- it is often unnecessary to keep track of individual rays from all the sources.  In these instances, rays traveling in approximately the same direction can be merged into a single ray \citep{Trac2007}.  Ray merging is critical for making the reionization problem more computationally tractable with these schemes.      

\subsubsection{Monte-Carlo methods}
\label{subsubsec:montecarlo}

Monte-Carlo methods take a statistical approach to solving the RT equation.  They work by sending ``photon packets'' that travel along random directions drawn from the distribution of angular directions around sources.  This method has the advantage that it preserves the full angular dependence of the RT solution (unlike moment-based methods), and it lends itself naturally to problems that involve scattering~\citep[e.g.][]{Zheng2010, 2018ApJ...863L...6V}.  Indeed, a key feature of the Monte Carlo approach is that, like ray tracing, it converges (in principle) to the exact solution of the RT equation in the limit of an infinite number of photon packets.  Moreover, like moment-based methods, the radiation field at each point in space is defined locally, being determined by the local density of photon packets rather than line integrals from distant sources.  This locality makes Monte-Carlo codes easier to parallelize efficiently than ray tracing methods, since processors working on different rays will be less likely to have to access the same RT cell simultaneously.  However, it also suffers from significant shot noise and poor angular resolution at large distances from isotropic ionizing sources if the number of photon packets is not large enough.  Mitigating these issues requires using a large number of photon packets, which increases computational cost.  An example of a Monte-Carlo code used for reionization studies is CRASH~\citep{Ciardi2001,Maselli2003}.

\subsection{Features of \textsc{FlexRT} \& their motivation}
\label{subsec:FlexRT}

\subsubsection{The RT method of choice for \textsc{FlexRT}}

As mentioned earlier, an organizing principle of \textsc{FlexRT} is flexibility for the user to choose the desired level of accuracy for the problem at hand. Along with this comes the requirement that users should be able to quantify the modeling errors incurred by their choices, e.g. in number of spatial or angular resolution elements.  These requirements dictate that the basic RT engine of \textsc{FlexRT} should be based on the ray tracing and/or Monte Carlo frameworks, as both of these -- at least in principle -- converge to the exact solution to the RT equation in the limit of an infinite number of rays or photon packets.  

The additional need for as much computation speed as possible further narrows the field to either adaptive ray tracing or Monte Carlo RT. Note that adaptive ray tracing allows the user to define the parameters that control how often rays split and merge, making it a naturally flexible method.  Monte-Carlo methods share a similar type of flexibility, and have the additional advantage of being easier to parallelize over many CPUs.  As we will discuss in detail in \S\ref{sec:adaptive_ray_tracing}, the RT method used in \textsc{FlexRT} is based largely on adaptive ray tracing, with some features that are similar to a Monte-Carlo approach. 

\subsubsection{Small-scale IGM physics}
\label{subsubsec:small_scale_physics}

The multi-scale nature of reionization remains the fundamental challenge of simulating it. Some of the most recent reionization simulations run with moment-based RT have begun to capture the small-scale clumpiness and self-shielding of the intergalactic gas in fully coupled hydro-dynamical boxes with $L\gtrsim 100$ Mpc~\citep[e.g.][]{Ocvirk2018,Kannan2022}. This unprecedented dynamic range cannot yet be achieved with adaptive ray tracing or Monte Carlo methods, nor in any such moment-based sim that can be run fast enough for parameter space studies. This fact motivates including the effect of small-scale IGM structure into much faster simulations with coarser resolution in a sub-grid fashion.  We address this issue with a key and novel feature available in \textsc{FlexRT}: its flexible treatment of the ionizing photon opacity. As we describe in \S \ref{sec:IGMopacity}, the opacity in a cell is treated in a highly generalized way that allows the user to specify an opacity model that accounts for the effects of unresolved IGM physics.  This framework can straightforwardly incorporate IGM opacity models based on results from higher resolution radiative hydrodynamics simulations in much smaller volumes.  The model introduced in Ref.~\cite{Cain2021} is one such example.  In Ref.~\cite{Cain2023}, we further demonstrated the flexibility of this approach by adding simple prescriptions for possible effects not accounted for in the original model (e.g. additional opacity from massive star-forming halos, recombination radiation, etc.).  The freedom to specify the opacity of the IGM flexibly not only gives the user a great deal of control over how the effects of the IGM are modeled; it also provides a way to reduce the computational costs of \textsc{FlexRT} by orders of magnitude while still modeling the small-scale physics of the ionizing photon sinks.

To summarize, in this section we have briefly outlined the existing numerical approaches for simulating reionization with RT.  We have also described some of the key features of \textsc{FlexRT} and how they serve our goal of developing a flexible code that is useful for situations in between semi-numeric and massive ``one-off'' RT simulations.   The remainder of this paper will describe in detail the core algorithms of \textsc{FlexRT} and subject them to a battery of validation tests.  

\section{Adaptive Ray Tracing Implementation}
\label{sec:adaptive_ray_tracing}

\subsection{Ray creation and propagation}
\label{subsec:raycasting}

In this section, we describe the adaptive ray tracing algorithm employed in \textsc{FlexRT}.  Our algorithm is similar to that described in Ref.~\cite{Abel2002} and implemented in the RT/hydro code of Ref.~\cite{Trac2007}.  In this method, ray tracing is controlled using HealPix formalism~\citep[the Hierarchical Equal Area isoLatitude Pixelation of a sphere,][]{Gorski1999}.  In each RT cell containing ionizing sources, rays are created at the center of the cell with unit vectors corresponding to the pixels of a HealPix sphere.  The number of rays cast is
\begin{equation}
    \label{eq:Nray}
    N_{\rm ray}^{\rm cast} = 12 \times 4^{l_{\rm hpx}}
\end{equation}
where $l_{\rm hpx} \geq 0$ is the integer ``level'' of the HealPix sphere.  Higher levels correspond to spheres with higher angular resolution.  Each unit vector is assigned a unique integer HealPix pixel number $0 \leq p < N_{\rm ray}^{\rm cast}$ (using the NESTED numbering scheme).  

In traditional time-independent ray tracing schemes, rays would be traced away from the source until they either terminate in a neutral region or all their photons are absorbed by the IGM~\cite{Sokasian2001,Mellema2006,Gaikwad2023}.  In \textsc{FlexRT}, ``rays'' are treated instead like photon packets that travel a distance $\Delta x_{\rm RT}$, the RT cell size, during each time step.  Specifically, a ray starting a time step at position $(x_0,y_0,z_0)$ travels to $(x_1,y_1,z_1) = (x_0,y_0,z_0) + \hat{v} \Delta x_{\rm RT}$ by the end of the time step, where $\hat{v}$ is the unit vector pointing along the original direction of the ray.  Photons are absorbed into each cell intersected over this distance following the RT equation (see next section).  Each ray retains its original unit vector assigned at casting until it is either split into child rays or merged with other rays (see next sub-sections).  This approach, inspired by Monte-Carlo codes, naturally lends itself to the finite speed of light and ``localizes'' the rays, making \textsc{FlexRT} easier to parallelize (see \S\ref{subsubsec:montecarlo}).  As we will see, this approach makes our procedure for merging of rays particularly straightforward.  

To avoid numerical artifacts introduced by the HealPix structure, during each time step we randomly rotate the HealPix spheres used to cast rays around each source by a randomly chosen set of direction angles, $\alpha$, $\beta$, and $\gamma$.  Each ray retains knowledge of these angles, so that the original HealPix unit vectors associated with it can be recovered when needed.  We find that this procedure eliminates discernible structural artifacts even for simulations with just one or very few ionizing sources.  Indeed, such artifacts can cause steady-state a-spherical structures in the radiation field around isotropic sources, and so removing them is necessary.  An unavoidable side effect of this procedure is that the flux incident on a given cell will ``flicker'' from one time step to the next as the HealPix directions rotate, averaging to the correct value only over several time steps.  This introduces artificial dependence on the ratio of the RT time step and the recombination timescale\footnote{Note that {\it all} Monte Carlo RT (e.g. CRASH, Ref.~\cite{Ciardi2001}) codes are subject to this effect at some level, since they randomize the directions of photon packets emitted from sources.  }.  
 Fortunately, as we will see, in most relevant applications, either the RT time step is much smaller than the recombination time scale or the local radiation field receives contributions from many sources in a way that averages out this flickering effect.  We will comment on this point when it becomes relevant in subsequent sections.

\subsection{Ray splitting}
\label{subsec:ray_splitting}

Rays can split into ``child rays'' to maintain a desired level of angular resolution around isolated point sources.  A ray of level $l_{\rm hpx}$ is split when it has traveled from its source a distance 
\begin{equation}
    \label{eq:r_split}
    r_{\rm split}(l_{\rm hpx}) = \Delta x_{\rm RT} 2^{l_{\rm hpx}} \left(\frac{3}{\pi N_{\rm ray}^{\min}}\right)^{1/2}
\end{equation}
where $N_{\rm ray}^{\min}$ is (approximately) the minimum number of rays contained in each cell surrounding the source.  This is a free parameter that is set by the user, which can be adjusted to control for the angular resolution of the radiation field and the speed of the computation.  Each ``parent'' ray is split into $4$ ``child'' rays, which have their HealPix parameters updated to
\begin{equation}
    \label{eq:lhpx}
    l_{\rm hpx} \rightarrow l_{\rm hpx} + 1 \hspace{1cm}
    p \rightarrow 4p + i, i \in [0,4)
\end{equation}
In the NESTED pixel numbering scheme, these pixels specify four unit vectors that uniformly sub-sample the solid angle corresponding to ($l_{\rm hpx}, p$).  Figure 2 of Ref.~\cite{Gorski1999} and Figures 1 and 2 of Ref.~\cite{Abel2002} visualize the sub-division of the HealPix sphere and the splitting procedure used here.  Child rays are assigned these new directions and re-positioned to occupy the centers of the new pixels to which they belong.  They also retain the same random rotation angles ($\alpha$, $\beta$, and $\gamma$) as their parent rays.  

\subsection{Ray merging}
\label{subsec:ray_merging}

A crucial time (and memory)-saving feature of \textsc{FlexRT} is that rays traveling in similar directions can merge into a single ray.  Ref.~\cite{Trac2007} demonstrated that ray merging can dramatically speed up ray tracing reionization simulations without substantially sacrificing accuracy.  Crucially, as we will see, merging makes it possible to make the scaling of the computation independent of the number of sources (see \S\ref{subsubsec:raytracingRT}).  

Our implementation of the ray merging is similar to the implementation in the \textsc{RadHydro} code of Ref.~\cite{Trac2007} (see \S\ref{sec:raytracingtests}).  We first determine which rays are eligible to be combined with other rays.  All rays are rank-ordered by the number of photons they contain, and the $N_{\rm ex} N_{\rm RT}$ rays with the most photons are considered {\it ineligible} for merging.  Here, $N_{\rm RT}$ is the number of RT cells and $N_{\rm ex}$, the number of rays per cell ``exempt'' from merging, is a free parameter chosen by the user.  As we will see later, excluding the ``brightest'' rays from the merging procedure helps reduce associated shot noise.  All other rays are first grouped by the RT cell they occupy.  In each cell, we throw down a randomly oriented HealPix sphere of level $l_{\rm merge}$ (also a user-set parameter) that specifies $12 \times 4^{l_{\rm merge}}$ independent directions.  Finally, rays in the same cell that share the same pixel on that sphere are merged into a single ray traveling in the direction specified by its pixel.  Photons are summed over all merged rays, and all other properties (including the ray's new position) are photon-weighted averages of those of its constituent rays.  These ensure that after rays are merged at the end of each time step, the maximum number of rays per cell in the box is
\begin{equation}
    \label{eq:Nray_per_cell_max}
    N_{\rm ray/cell}^{\max} = N_{\rm ex} + 12 \times 4^{l_{\rm merge}} \equiv N_{\rm ex} + N_{\rm dir}
\end{equation}
where $N_{\rm dir} \equiv 12 \times 4^{l_{\rm merge}}$ is the number of independent directions used for merging.   

\begin{figure}
    \centering
    \includegraphics[scale=0.9]{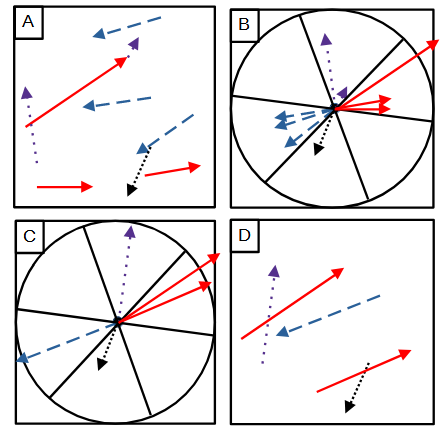}
    \caption{2D representation of how ray merging works in \textsc{FlexRT}.  Panel A shows a cell containing several rays.  The lengths of the arrows represent the number of photons in each ray, and rays with same color/line style point in similar directions. 
The location of the bases of the arrows denotes the $(x,y)$ positions of the rays.  In panel B, we throw down a randomly rotated ``pizza'' (analogous to a HealPix sphere in 3D) and group rays by the ``slice'' that they occupy, placing their bases in the center of the panel to aid comparison of their directions.  We merge rays occupying the same slice in panel C, excluding the longest (red) ray, since it has enough photons to be exempt.  Merged rays point along the bisector of the slice they occupy (i.e. the HealPix pixel direction).  In panel D, we show that the positions of merged rays are the photon-weighted average of their constituents.  }
    \label{fig:merging_vis}
\end{figure}

Figure~\ref{fig:merging_vis} shows a 2D representation of the steps in our merging scheme.  Panel A shows an RT cell containing a set of rays.  The lengths of the rays represent how many photons they contain, and rays with the same colors/line styles point in similar directions.  The locations of the bases of the arrows denote the $(x,y)$ positions of the rays.  In panel B, we throw down a randomly rotated ``pizza'' (analogous to a HealPix sphere in 3D) and assign each ray to the slice it occupies.  Note that we have consolidated all rays to the center of the panel (ignoring for now their spatial positions) to make it easy to compare their directions.  In panel C, rays within the same slice are merged.  Note that two of the slices are empty and one of them has only a single ray, so no merging occurs in those slices.  In the slice with the red rays, the longest ray had enough photons to be exempt from merging, and the other two were combined.  Panel D shows again $(x,y)$ positions of the merged rays, which are the photon-weighted averaged positions of the constituent rays (see again panel A). Note that rays that were not merged with any others (long solid red and dotted black) have not changed their positions.   

\begin{figure}
    \centering
    \includegraphics[scale=0.35]{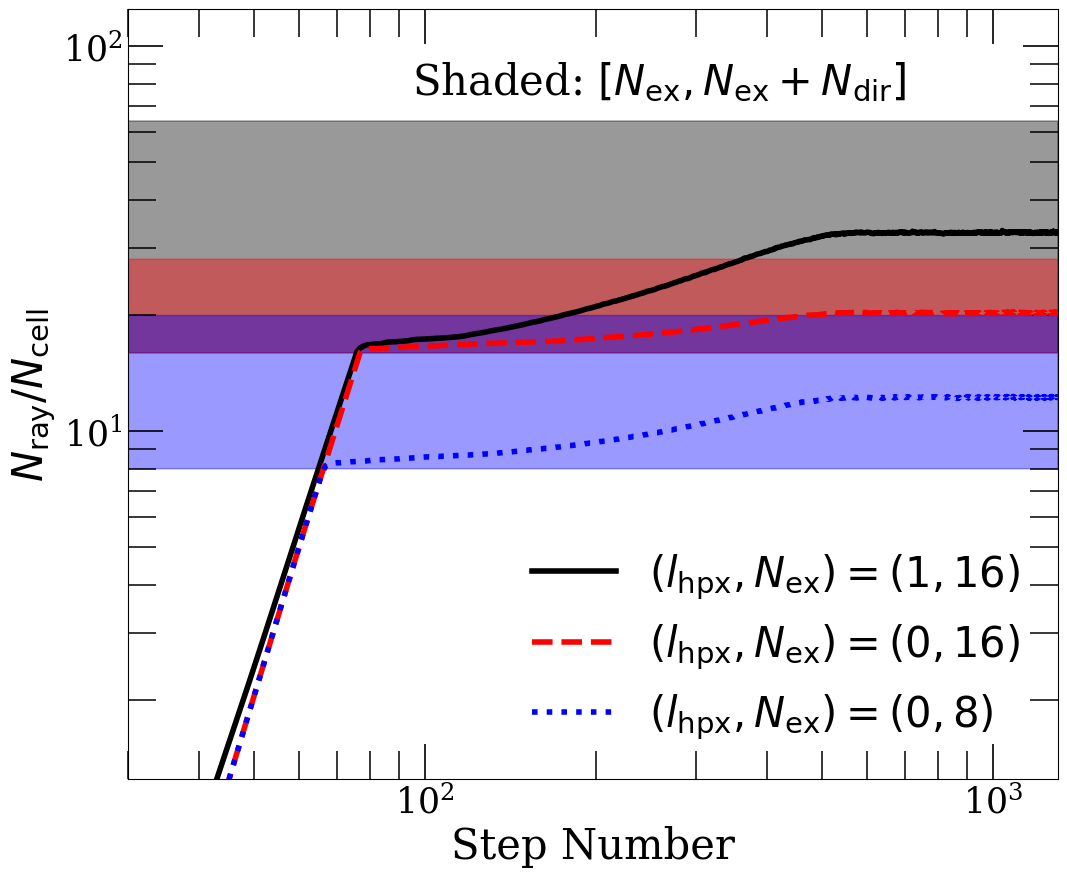}
    \caption{Examples of how merging limits the number of rays in \textsc{FlexRT} simulations.  The curves show the number of rays per cell vs. number of RT steps for a simulation with $N_{\rm cell} = 200^3$, and several values of $l_{\rm hpx}$ and $N_{\rm ex}$ (see legend).  The shaded region the range between $N_{\rm ex}$ and $N_{\rm ex} + N_{\rm dir}$, where $N_{\rm dir} = 12 \times 4^{l_{\rm hpx}}$ is the number of independent directions allowed in the HealPix merging scheme.  The number of rays per cell grows like a power law with the number of step until reaching $N_{\rm ex}$, after which it grows much more slowly.  The number of rays (after the merging step completes) can never exceed $N_{\rm ray/cell}^{\max} = N_{\rm ex} + N_{\rm dir}$, and may be somewhat less than that, since most cells do not contain rays pointing in every allowed direction.  In all cases, $N_{\rm ray}/N_{\rm cell}$ plateaus slightly below $N_{\rm ray/cell}^{\max}$ once all cells in the volume are ionized.  }
    \label{fig:ray_count_merging_example}
\end{figure}

Figure~\ref{fig:ray_count_merging_example} demonstrates how merging limits the number of rays in a \textsc{FlexRT} simulation.  We show the number of rays per cell vs. number of RT steps for several version of the same $N_{\rm RT} = 200^3$ \textsc{FlexRT} simulation with cosmological sources (see \S\ref{subsec:hy_cosmo} for details), with several combinations of merging parameters (see legend).  The shaded regions denote the interval $[N_{\rm ex},N_{\rm ex} + N_{\rm dir}]$.  For $N_{\rm ray}/N_{\rm RT} < N_{\rm ex}$, all rays are exempt from merging, and the number of rays grows with time as a steep power law.  Upon reaching $N_{\rm ex}$, $N_{\rm ray}/N_{\rm cell}$ stalls as merging begins restricting the number of rays in ionized cells.  The ray count increases gradually thereafter, plateauing slightly below $N_{\rm ray/cell}^{\max}$ (Eq.~\ref{eq:Nray_per_cell_max}) once the entire volume ionizes.  By adjusting $l_{\rm hpx}$ and $N_{\rm ex}$, the user can control the number of rays in the simulation (determining its computational cost), and the angular/spatial resolution of the radiation field.  

\section{Treatments of the IGM Opacity}
\label{sec:IGMopacity}

As stated in \S\ref{sec:intro} and \ref{subsubsec:small_scale_physics}, \textsc{FlexRT}'s other key feature is its flexible treatment of the opacity of the ionized IGM.  In fact, \textsc{FlexRT} has two ``modes'' for doing this, between which the user can choose.  The first, which we will refer to as the ``Standard Mode'' and will describe in \S\ref{subsec:standardRT}, solves directly for the chemistry of the gas and its opacity using a method very similar to that employed in the C2-Ray code of Ref.~\citep{Mellema2006}.  The other approach, described in \S\ref{subsec:subgrid_mode}, uses a highly general treatment for the IGM opacity that relies on flexible external input.  We refer to this as ``Generalized Opacity Mode''.  

\subsection{Direct solution of the chemistry equations (``Standard Mode'')}
\label{subsec:standardRT}

In this mode, we assume that all small-scale fluctuations in the HI number density ($n_{\rm HI}$) are fully resolved by our simulation.  That is, $n_{\rm HI}$ does not fluctuate below the scale of the RT cell, $\Delta x_{\rm RT}$.  Under this assumption, we can write an initial guess for the HI photo-ionization rate $\Gamma_{\rm HI}$ at the beginning of an RT time step of length $\Delta t = \Delta x_{\rm RT}/c$ as: 
\begin{equation}
    \label{eq:gamma_initial_standard}
    \Gamma_{\rm HI}^{i,0} =\sum_{j=1}^{N_{\rm rays}} \sum_{\nu=1}^{N_{\rm freq}} \frac{N_{\gamma}^{ij\nu}(1 - \exp[-n_{\rm HI}^{i,0} \sigma_{\rm HI}^{\nu} \Delta s_{ij}]/\Delta t)}{V_{\rm cell}^{i} n_{\rm HI}^{i,0}}
\end{equation}
Here, $\Delta s_{ij}$ the path length of ray $j$ through $i$ during $\Delta t$.  $N_{\gamma}^{ij\nu}$ is the number of photons initially in frequency bin $\nu$ of ray $j$, and the sum runs over all rays that intersect cell $i$ during $\Delta t$.  The initial HI number density in cell $i$ is $n_{\rm HI}^{i,0}$, $V_{\rm cell}^{i}$ is the cell volume, and $\sigma_{\rm HI}^{\nu}$ is the ionization cross-section of HI in frequency bin $\nu$.  The numerator is the number of ionizations in cell $i$ and the denominator is the initial {\it number} of HI atoms.  

The ionization balance equation for HI is 
\begin{equation}
    \label{eq:ion_balance_eqn}
    \frac{dn_{\rm HI}}{dt} = -(\Gamma_{\rm HI}n_{\rm HI} + C_{\rm HI}(T)n_e) + \alpha(T) n_e n_{\rm HII}
\end{equation}
where $T$ is temperature, $C_{\rm HI}(T)$ is the collisional ionization coefficient of HI, $\alpha(T)$ is the recombination coefficient of HII, $n_e$ is the electron number density and $n_{\rm HII}$ is the HII number density.  Updating Eq.~\ref{eq:ion_balance_eqn} using Eq.~\ref{eq:gamma_initial_standard} would ignore the fact that $n_{\rm HI}$ and $\Gamma_{\rm HI}$ may evolve during $\Delta t$.  Following Ref.~\cite{Mellema2006}, we account for this by solving for the time-dependence of the ionized fraction $x_i$ during $\Delta t$ assuming that the recombination rate remains constant (Eq. 12-14 of Ref.~\cite{Mellema2006}), 
\begin{equation}
    \label{eq:xi_of_t}
    x_i(t) = x_{\rm eq} + (x_0 - x_{\rm eq})e^{-t/t_i}
\end{equation}
where $x_0$ is the ionized fraction at the beginning of the time step, and 
\begin{gather}
    \label{eq:xeq}
    x_{\rm eq} = \frac{\Gamma_{\rm HI} +  C_{\rm HI}(T)n_e}{\Gamma_{\rm HI} +  C_{\rm HI}(T)n_e + \alpha(T) n_e}\\
    \label{eq:ti}
    t_i = \frac{1}{\Gamma_{\rm HI} + C_{\rm HI}(T)n_e + \alpha(T) n_e}
\end{gather}
Then the time-averaged ionized fraction during the time step is
\begin{equation}
    \label{eq:avg_xi}
    \langle x_i \rangle = x_{\rm eq} + (x_0 - x_{\rm eq})(1 - e^{-\Delta t/t_i}) \frac{t_i}{\Delta t}
\end{equation}
From this we can calculate the time-averaged average HI number density, $\langle n_{\rm HI} \rangle$, and re-calculate $\Gamma_{\rm HI}$: 
\begin{equation}
    \label{eq:gamma_updated_standard}
    \Gamma_{\rm HI}^{i} = \sum_{\nu=1}^{N_{\rm freq}}\sum_{j=1}^{N_{\rm rays}} \frac{N_{\gamma}^{ij\nu}(1 - \exp[-\langle n_{\rm HI}^{i} \rangle \sigma_{\rm HI}^{\nu} \Delta s_{ij}]/\Delta t)}{V_{\rm cell}^{i} \langle n_{\rm HI}^{i} \rangle}
\end{equation}
We iterate Eq.~\ref{eq:xi_of_t}-\ref{eq:gamma_updated_standard} several times to achieve convergence.  Ref. \cite{Mellema2006} pointed out that this procedure gives the exact solution for $n_{\rm HI}$, no matter how large $\Delta t$ is, if recombinations are neglected.  Generally, $n_{\rm HI}$ evolves most quickly inside ionization fronts, which are generally very thin ($10$s of ckpc~\cite{Wilson2024a}) and move very fast ($10^3-10^4$ km/s), such that the integrated number of recombinations taking place {\it inside} I-fronts is small compared to the number in ionized gas elsewhere.  Thus, in the situations where the iteration procedure described above is most important, it is generally a good approximation to neglect recombinations.  Note that the ``inside'' of the I-front refers to gas within the thin boundary between highly ionized ($x_{\rm HI} \sim 0$) and fully neutral ($x_{\rm HI} \sim 1$) gas.  

However, one problem with Eq.~\ref{eq:gamma_updated_standard} is that if $\Delta x_{\rm RT}$ is much larger than the I-front width $\Delta x_{\rm IF}$, the I-front will be ``smeared out'' uniformly over the cell containing it.  This is because $n_{\rm HI}$ is assumed to have a uniform value everywhere in the cell.  A more accurate treatment would divide the cell into a highly ionized and fully neutral component, and track the fraction of the cell in each state as the I-front crosses it.  As we will see in the next sub-section, this is precisely the approach used in the Generalized Opacity Mode of \textsc{FlexRT}.  In Standard Mode, the error resulting from this smearing effect is that the recombination rate in partially ionized cells is too low by approximately a factor of $\langle x_i \rangle$ whenever $\Delta x_{\rm RT} >> \Delta x_{\rm IF}$.  As such, Standard Mode can under-estimate the integrated recombination rate if $\Delta x_{\rm RT}$ is large enough - we will return to this point in \S\ref{sec:moving_screen}.  


The temperature in each cell is given by the solution of Ref.~\citep{Hui1997},
\begin{equation}
    \label{eq:temp_evolution_standard}
    \frac{dT}{dt} = \frac{2}{3 k_{\rm B} n_{\rm tot}}(\mathcal{H} - \Lambda) - 2H(z)T - \frac{T}{n_{\rm tot}}\frac{d n_{\rm tot}}{dt} + \frac{2T}{3\Delta} \frac{d\Delta}{dt}
\end{equation}
In the first term on the RHS, $\mathcal{H}$ and $\Lambda$, are the net radiative heating and cooling rates, respectively, $n_{\rm tot}$ is the total number density of gas particles (neutrals, ions, and electrons), and $k_{\rm B}$ is Boltzmann's constant.  We include heating by photo-ionizations as well as cooling by recombinations, collisional ionizations and excitations, free-free interactions, and Compton scattering off the CMB.  The second term, wherein $H(z)$ is the Hubble parameter, accounts for adiabatic cooling by cosmic expansion.  The third term accounts for changes in temperature from an increase or decrease in the number of particles in the gas.   The last term accounts for heating/cooling by compression/expansion, and is proportional to the change in the density in units of the mean, $\Delta$.  
 
\subsection{General treatment (``Generalized Opacity Mode'')}
\label{subsec:subgrid_mode}

\textsc{FlexRT}'s second mode is built around relaxing the assumption that small-scale fluctuations in $n_{\rm HI}$ are fully resolved.  Relaxing this assumption allows for each cell to have some un-resolved spatial distribution of $n_{\rm HI}$ inside it that would not be accounted for in Standard Mode.  In this case, the equations in the previous section become ill-defined, since they rely on assuming a single value for $n_{\rm HI}$ in each cell.  A commonly-employed method to correct for the effect of un-resolved density fluctuations is to boost the recombination rate (and thus, the residual $n_{\rm HI}$ in photo-ionization equilibrium) in each cell by a sub-grid clumping factor~\citep[e.g.][]{Mao2019}.  However, this is not the most general approach, since it assumes that the ionizing opacity is determined uniquely by the recombination rate (e.g. Eq. 17 of Ref.~\cite{Emberson2013}).  This is not true in the presence of dense, self-shielding structures that harbor neutral gas, wherein the assumption of photo-ionization equilibrium may not hold.  Specifically, neutral gas embedded in dense clumps may be in the process of photo-evaporation, or low-density ionized gas may be accreting onto denser structures and rapidly recombining.  In these scenarios, the ionization rate can be larger or smaller than the recombination rate, respectively.  

A more general approach is to forgo trying to model $n_{\rm HI}$ entirely, and instead focus on the effective frequency-dependent absorption coefficient $\kappa_{\nu}$ associated with each cell.  The opacity to ionizing photons across a path length $\Delta s$ through the cell is given by
\begin{equation}
    \label{eq:tau_cell}
    \tau_{\nu}\equiv \kappa_{\nu} \Delta s \equiv \Delta x_{\rm RT}/\lambda_{\nu}
\end{equation}
where we have defined $\lambda_{\nu} \equiv 1/\kappa_{\nu}$ to be the effective mean free path (note that we will use both $\kappa_{\nu}$ and $\lambda_{\nu}$ in what follows).  Here, $\kappa_{\nu}$ is defined to be the mean ``effective'' absorption coefficient in the cell.  This means that $\kappa_{\nu}$ is defined such that on average, Eq.~\ref{eq:tau_cell} returns the correct opacity for any length $\Delta s$ along any direction.  In \S\ref{subsec:standardRT}, where we assumed a uniform $n_{\rm HI}$ in each cell, $\kappa_{\nu}= n_{\rm HI} \sigma^{\nu}_{\rm HI}$ and $\tau_{\nu} = n_{\rm HI} \sigma^{\nu}_{\rm HI} \Delta s$.  Another example is the sub-grid model described in Refs. \cite{Cain2021,Cain2022b}.  There, $\lambda_{\nu}$ is calculated from a suite of high-resolution, small-volume simulations using a definition that satisfies these conditions (see Appendix C of Ref.~\cite{Cain2022b} for a derivation).  

We begin by making an initial guess for $\Gamma_{\rm HI}$ that does not require knowledge of either $n_{\rm HI}$ or $\kappa_{\nu}$.  We can get this by taking the limit of Eq.~\ref{eq:gamma_initial_standard} for $\tau_{\nu}  << 1$, which simplifies to
\begin{equation}
    \label{eq:gamma_initial_subgrid}
    \Gamma_{\rm HI}^{i,0} = \sum_{\nu=1}^{N_{\nu}}\sum_{j=1}^{N_{\rm rays}} \frac{N_{\gamma}^{ij\nu} \sigma_{\rm HI}^{\nu} \Delta s_{ij}}{V_{\rm cell}^{i} \Delta t} = \sum_{\nu=1}^{N_{\nu}} F_{\gamma}^{i,\nu} \sigma_{\rm HI}^{\nu}
\end{equation}
where $F_{\gamma}^{i,\nu}$ is the photon number flux in frequency bin $\nu$ incident on cell $i$\footnote{To get the second equality, we have used the fact that $\sum_{j=1}^{N_{\rm rays}} \frac{N_{\gamma}^{ij\nu} \Delta s_{ij}}{V_{\rm cell}^{i} \Delta t}$ is the total incident ionizing photon number flux in frequency bin $\nu$.}.  Eq.~\ref{eq:gamma_initial_subgrid} is crucial because, as we will see, {\it any} reasonable procedure for determining $\lambda_{\nu}$ requires first having at least a guess for $\Gamma_{\rm HI}$.  

The next step is to determine $\kappa_{\nu}$.  A highly general functional form for $\kappa_{\nu}$ at some redshift $z$ is given by
\begin{equation}
    \label{eq:lambda_nu_general}
    \kappa_{\nu}(z) = G_{\kappa}(z, t, \Delta(z,t), \Gamma_{\rm HI}(z,t), x_i(z,t), T(z,t), \nu)
\end{equation}
Here, $\Delta$, $\Gamma_{\rm HI}$, $x_i$, and $T$ are the density in units of the mean, photo-ionization rate, ionized fraction, and temperature of a given cell, respectively.  We use $z$ to refer to the current time, while $t$ indicates the history of a quantity prior to $z$.  The statement here is that $\kappa_{\nu}$ can, in general, be some function not only of the current physical conditions in a cell, but also of the prior history of those conditions.  One example, encapsulated in $x_i(z,t)$, is the dependence of $\kappa_\nu$ on the prior reionization history.  Indeed, several studies (e.g. Refs.~\cite{Park2016,DAloisio2020,Chan2023}) have found that the opacity of the IGM in some region depends not only on the current redshift, but also on when that region was reionized.   

Before continuing, it is worth emphasizing that the approach described in this section is so versatile because Eq.~\ref{eq:lambda_nu_general} can have practically any form imaginable, and the equations given below will still hold. Ref.~\cite{Cain2021} provides but one example. There, $G_{\kappa}$ takes the form of a sub-grid model for the ionizing opacity calibrated to the results of high-resolution radiative hydrodynamics simulations.  In Ref.~\cite{Cain2023}, we explored adding other contributions to $\kappa_\nu$, highlighting the flexibility of our framework (see \S3.6 and Appendix D of that work).      

Our final task is to provide a general expression for $\Gamma_{\rm HI}$ given knowledge of $\kappa_{\nu}$.  Indeed, if $\Delta x_{\rm RT}$ were assumed to always be $<< \lambda_{\nu}$, Eq.~\ref{eq:gamma_initial_subgrid} would be sufficient.  However, as already mentioned, a key feature of \textsc{FlexRT} is that it should admit the use of large (up to several Mpc) RT cells without substantial loss of accuracy.  As such, we must allow for scenarios in which $\lambda_{\nu} \lesssim \Delta x_{\rm RT}$.  We can do this as follows.  First, we assume (without loss of generality, as we will see) that all rays $j$ during the time step $\Delta t$ are incident on one face of cell $i$ traveling in the direction normal to that face.  We can write the contribution of ray $j$ to $\Gamma_{\rm HI}$ at a distance $x_j$ into the cell as
\begin{equation}
    \label{eq:gamma_x_subgrid}
    \Gamma_{\rm HI}^{ij}(x_j) = \sum_{\nu=1}^{N_{\nu}}  F_{\gamma}^{ij\nu}\exp(-x_j/\lambda_{\nu}) \sigma_{\rm HI}^{\nu} = 
\end{equation}
\begin{equation*}
    \sum_{\nu=1}^{N_{\nu}} \frac{N_{\gamma}^{ij\nu}}{A_{\rm cell}^{i} \Delta t}\exp(-x_j/\lambda_{\nu}) \sigma_{\rm HI}^{\nu}
\end{equation*}
where $A_{\rm cell} \equiv \Delta x_{\rm RT}^2$ and we have averaged the incident ionizing flux across the transverse directions.  The first expression follows from $\lambda_{\nu}$ being constant across the cell by definition.\footnote{The astute reader may wonder if this assumption contradicts our earlier observation that $n_{\rm HI}$ is not constant within cells.  It does not because the isotropically-averaged opacity over some in-homogeneous region of the IGM still captures the average effect of the density fluctuations.  The assumption here is that all information about the un-resolved in-homogeneity in $n_{\rm HI}$ is encoded in $\lambda_{\nu}$ (via whatever function is used for Eq.~\ref{eq:lambda_nu_general}).  }  Next, we average $\Gamma_{\rm HI}^{ij}$ in the incident direction across the length of the cell, $\Delta x_{\rm RT}$, and sum over all rays.  The path length traversed by the ray is $\Delta s_{ij}$, which is always $\leq \Delta x_{\rm RT}$ (as enforced by our choice of time step, $\Delta t = \Delta x_{\rm RT}/c$).  This gives


\begin{equation}
    \label{eq:gamma_avg_subgrid}
    \langle \Gamma_{\rm HI} \rangle^{i} = \sum_{j=1}^{N_{\rm rays}} \sum_{\nu=1}^{N_{\nu}}  \frac{1}{\Delta x_{\rm RT}} \int_{0}^{\Delta s_{ij}}dx_j \frac{N_{\gamma}^{ij\nu}}{A_{\rm cell}^{i} \Delta t}\exp(-x_j/\lambda_{\nu}) \sigma_{\rm HI}^{\nu}
\end{equation}
 \begin{equation*}
    = \sum_{j=1}^{N_{\rm rays}}\sum_{\nu=1}^{N_{\rm freq}} \frac{N_{\gamma}^{ij\nu}[1 - \exp(-\Delta s_{ij}/\lambda_{\nu}^{i})] \sigma_{\rm HI}^{\nu} \lambda_{\nu}^{i} }{V_{\rm cell}^{i} \Delta t}
\end{equation*}
In the second equality, we have evaluated the integral and used the fact that $A_{\rm cell} \Delta x_{\rm RT} = V_{\rm cell}$.  Importantly, we see that Eq.~\ref{eq:gamma_avg_subgrid} simplifies to Eq.~\ref{eq:gamma_updated_standard} in the limit that $\lambda_{\nu} = 1/[\sigma_{\rm HI}^{\nu} n_{\rm HI}]$, which is the case in Standard mode.  This demonstrates that Eq.~\ref{eq:gamma_avg_subgrid} does not depend on the plane-parallel geometry assumed in its derivation.  Eq.~\ref{eq:lambda_nu_general} and~\ref{eq:gamma_avg_subgrid} are updated iteratively until a self-consistent solution for $\lambda$ and $\Gamma_{\rm HI}$ is reached.    

We note that the components of Eq.~\ref{eq:gamma_avg_subgrid} have straightforward interpretations.  First of all, $N_{\gamma}^{ij\nu}[1 - \exp(-\Delta s_{ij}/\lambda_{\nu}^{i})]/\Delta t$ is simply the number of photons absorbed into cell $i$ from ray $j$ per unit time, and the sum of this over rays is the total ionization rate.  Since $\Gamma_{\rm HI}$ is defined to be the number of ionizations per unit time divided by the number of neutral atoms, then the remaining factor of $V_{\rm cell}/[\sigma_{\rm HI}^{\nu} \lambda_{\nu}]$ can be interpreted as the number of neutral atoms subject to ionization by radiation in the cell (excluding any self-shielded systems, where $\Gamma_{\rm HI}$ would locally be $0$ and no atoms can be ionized).  Indeed, this is trivially the case if $n_{\rm HI}$ is assumed uniform, since $\lambda_{\nu} = 1/\kappa_{\nu} = 1/n_{\rm HI}\sigma^{\nu}_{\rm HI}$.  We showed in Appendix B of Ref.~\cite{Cain2022b} that for the sub-grid opacity model used in that work, $\langle \lambda \rangle_{\nu} = 1/n_{\rm HI}^{\Gamma}\langle \sigma_{\rm HI}\rangle_{\nu}$ to a good approximation, where $n_{\rm HI}^{\Gamma}$ is the $\Gamma_{\rm HI}$-weighted average neutral hydrogen density and $\langle ... \rangle_{\nu}$ denotes an appropriately defined average over frequency.  For the specific definition of $\lambda_{\nu}$ adopted in Ref.~\cite{Cain2023}, Eq.~\ref{eq:gamma_avg_subgrid} is equivalent to their Eq. 7, although Eq.~\ref{eq:gamma_avg_subgrid} is more general.  

Eq.~\ref{eq:gamma_initial_subgrid}-\ref{eq:gamma_avg_subgrid} apply to highly ionized regions.  To track the growth of these regions, we assume that the boundary between highly ionized and fully neutral gas (that is, the I-front) is infinitely thin - this is the so-called ``moving screen'' approximation\footnote{Note that a similar assumption is used in the \textsc{RadHydro} code.  }.  This is equivalent to assuming that $\Delta x_{\rm IF} << \Delta x_{\rm RT}$.  Indeed, it is the opposite assumption to that made in the \S\ref{subsec:standardRT}, namely that $n_{\rm HI}$ is uniform within partially ionized cells.  Given our earlier discussion of the intended application of \textsc{FlexRT}'s Generalized Opacity formalism, it is reasonable to expect that this condition should hold in most relevant contexts (that is, when RT cells are $\sim$ a few Mpc).  In subsequent sections, we will highlight this assumption whenever it becomes relevant (as it will in several instances).  We will also explictly demonstrate the regime in which this approximation is valid in \S\ref{sec:moving_screen}.  The evolution of the ionized fraction  $x_{ion}$ in a cell is (in the non-relativistic limit)
\begin{equation}
    \label{eq:dxidt}
    \frac{d x_{\rm ion}}{dt} = \frac{v_{\rm IF}}{\Delta x_{\rm RT}} = \frac{F_{\gamma}}{(1 + \chi) n_{\rm H} \Delta x_{\rm RT}}
\end{equation}
where $v_{\rm IF}$ is the I-front speed, $F_{\gamma}$ is the ionizing flux incident on the neutral component of the cell, $n_{\rm H}$ is the hydrogen number density, and the factor of $1 + \chi$ accounts for single ionization of helium along with hydrogen\footnote{We note that $x_i$ in Eq.~\ref{eq:dxidt} is defined as the fraction of a given cell that is in a highly ionized (vs. fully neutral) state, which is not quite the same as the ionized fraction of the cell.  This is because highly ionized gas still retains a small fraction of neutral gas that is not included in the definition of this quantity.  }.  For partially ionized cells, we evaluate $\Gamma_{\rm HI}$ (Eq.~\ref{eq:gamma_avg_subgrid}) for the ionized component by making the substitutions
\begin{equation}
    \Delta s_{ij} \rightarrow x_{\rm ion}  \Delta s_{ij} \hspace{1cm}
    V_{\rm cell}^{i} \rightarrow x_{\rm ion}V_{\rm cell}^{i}
\end{equation}
Note that photons contribute to $F_{\gamma}$ in Eq.~\ref{eq:dxidt} only after they have traveled a distance $x_{\rm ion} \Delta s_{ij}$ through a cell.  This condition assumes that rays enter partially ionized cells on the ``ionized side'' of the cell, which is not always the case.  However, this geometry is physically reasonable since rays will generally move in the same direction as the I-front.  

The peak gas temperature immediately behind an I-front, $T_{\rm reion}$, is set by the heating and radiative cooling inside of the I-front \cite{DAloisio2019, Zeng2021}.  Generalized Opacity Mode is designed for large-volume simulations of reionization that do not resolve the small distance scales and short timescales necessary to capture this physics. Hence, in this mode, $T_{\rm reion}$ is a user-specified parameter, for which our default model is that of Ref. \cite{DAloisio2019}. They demonstrated that $T_{\rm reion}$ depends mainly on the velocity of the I-front and the spectral index of the flux driving it - with the former being the dominant effect as long as the spectrum is not very soft or the I-front very fast.  For all simulations presented in this paper, $T_{\rm reion}$ is obtained from the fitting function of Ref.~\cite{DAloisio2019}, which assumes $\alpha = 1.5$.\footnote{Their fitting function requires I-front speeds as input, which we compute using the flux-based method also described in Ref.~\cite{DAloisio2019}.} We note, however, that any model for $T_{\rm reion}$ can be implemented into the code. For example, one may choose to implement the more general model provided by the interpolation tables of Ref. \cite{DAloisio2019}, which include the dependence on $\alpha$.  We track the subsequent thermal evolution of the gas in Generalized Opacity mode using Eq.~\ref{eq:temp_evolution_standard}.  However, since we do not know $n_{\rm HI}$, we cannot directly compute the photo-heating rate or the collisional cooling rates (ionization and excitation).  To estimate the photo-heating rate, we assume that the number of ionizations in ionized gas is balanced by the number of recombinations (that is, photo-ionization equilibrium).  Assuming the ionizing radiation has a power law spectrum with index $\alpha$ and a cutoff at $4$ Ryd, the heating rate due to HI ionizations is given by
\begin{equation}
    \label{eq:analytical_heating_HI}
    \mathcal{H}_{\rm HI} = E_{\rm HI} \left(\frac{\alpha + \beta}{\alpha + \beta - 1}\frac{1 - 4^{1 - \beta - \alpha}}{1 - 4^{-\beta - \alpha}} - 1\right)\alpha_{\rm A}(T) n_e n_{\rm HII}
\end{equation}
where $\beta = 2.75$ is the approximate power-law scaling of $\sigma_{\rm HI}$ with $\nu$ and $E_{\rm HI} = 1$ Ryd is the ionization energy of HI.  We include an analogous expression to account for heating due to HeI ionizations.  The collisional cooling rates can be estimated using a similar approximation. 

\section{Cosmological RT Comparison Project (2006) Tests}
\label{sec:CRTCP_tests}

In this section, we will subject \textsc{FlexRT} to the four static density tests introduced in the Cosmological Radiative Transfer Comparison Project\footnote{\url{https://astronomy.sussex.ac.uk/~iti20/RT_comparison_project/index.html}} (CRTCP)~\citep[][henceforth I06]{Iliev2006}.  We will perform each test in both Standard and Generalized Opacity mode, and compare them to each other and results from the CRTCP.  

\subsection{Iso-thermal Stromgren Sphere (Test \#1)}
\label{subsec:stromgren_test}

The first test in I06 is perhaps the simplest cosmologically relevant RT problem - the growth of an HII region around an isotropic point source in a homogeneous, isothermal region.  This is one of the only RT problems with a closed-form analytic solution.  Assuming the source produces $\dot{N}_{\gamma}$ ionizing photons per unit time, and that the boundary between ionized and neutral gas is infinitely thin, the radius and velocity of the I-front are given by (Eq. 5-7 of I06),
\begin{equation}
    \label{eq:r_of_t}
    r(t) = r_S(1 - \exp[-t/t_{\rm rec}])^{1/3}
\end{equation}
\begin{equation}
    \label{eq:v_of_t}
    v(t) = \frac{r_S}{3 t_{\rm rec}}\frac{\exp[-t/t_{\rm rec}]}{(1 - \exp[-t/t_{\rm rec}])^{2/3}}
\end{equation}
where the Stromgren radius is given by
\begin{equation}
    \label{eq:rS}
    r_S \equiv \left(\frac{3 \dot{N}_{\gamma}}{4 \pi \alpha_{\rm B}(T) n_{\rm H}}\right)^{1/3}
\end{equation}
is the $t \rightarrow \infty$ limit of $r(t)$. Here, $\alpha_{\rm B}$ is the case B recombination rate of the gas, $T$ is the temperature, and $n_{\rm H}$ is the hydrogen number density.  The recombination timescale is
\begin{equation}
    \label{eq:trec}
    t_{\rm rec} \equiv \frac{1}{\alpha_{\rm B}(T) n_{\rm H}}
\end{equation}

Following I06, we assume $n_{\rm H} = 10^{-3}$ cm$^{-3}$, $T = 10^4$ K, and $\dot{N}_{\gamma}$ = $5 \times 10^{48}$ photons/s, and a photon energy of $13.6$eV (the ionization energy of hydrogen).  Our computation volume is a box with $L = 13.2$ pkpc with the source at the center\footnote{As noted by I06, $r_{\rm S} = 5.4$ pkpc for this test, motivating this box size.  }.  We use $N = 100^3$ RT cells and a reduced speed of light of $\tilde{c} = 0.01$.  The reduced speed of light is appropriate here because the I-front speed, $v_{\rm IF}$, is much less than $\tilde{c}$ except at the beginning of the test.  

As discussed previously, in Generalized Opacity mode we do not (a-priori) specify a prescription for $\lambda_{\nu}$ (Eq.~\ref{eq:lambda_nu_general}).  We must therefore do so here before proceeding with this and subsequent tests.   Assuming no unresolved density fluctuations (which is trivially the case here), we can write for each cell (see \S\ref{subsec:standardRT}),
\begin{equation}
    \label{eq:subgrid_lambda}
     \lambda_{\nu} = \frac{1}{\sigma_{\rm HI}^\nu n_{\rm HI}}
\end{equation}
For highly ionized gas, we can assume photo-ionization equilibrium, 
\begin{equation}
    \label{eq:equilibrium_condition}
    \Gamma_{\rm HI} n_{\rm HI} = \alpha(T) n_e n_{\rm HII}
\end{equation}
which can be solved for $n_{\rm HI}$ to give
\begin{equation}
    \label{eq:subgrid_nHI}
    n_{\rm HI} = \frac{\alpha(T) n_e n_{\rm HII}}{\Gamma_{\rm HI}} \approx \frac{\alpha(T) n_{\rm H}^2}{\Gamma_{\rm HI}}
\end{equation}
The last equality is motivated by our moving-screen I-front prescription (Eq.~\ref{eq:dxidt}), which assumes that $n_{\rm HI} << n_{\rm H}$ behind the I-front.  We will use Eq.~\ref{eq:subgrid_lambda} and~\ref{eq:subgrid_nHI} to solve for $\lambda_{\nu}$ and estimate $n_{\rm HI}$ in Generalized Opacity mode in all the tests that follow in this section.  Note that this prescription is just one very special case of Eq.~\ref{eq:lambda_nu_general}.  

\begin{figure}
    \centering
    \includegraphics[scale=0.27]{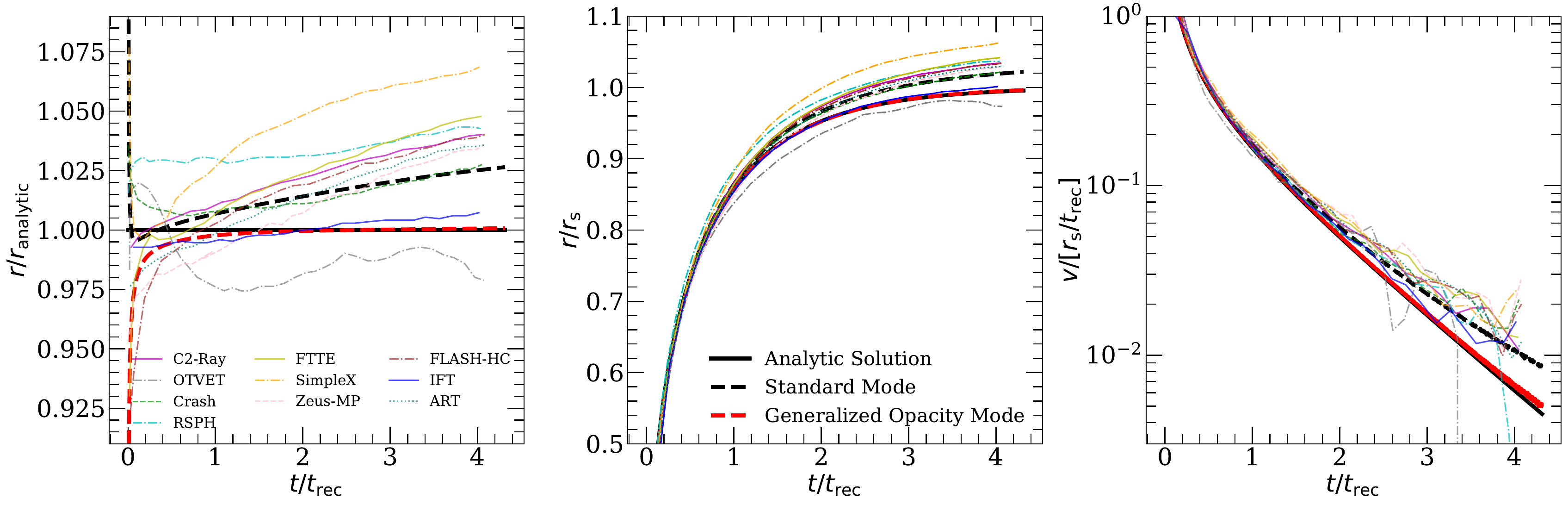}
    \caption{Growth of an isothermal, spherically symmetric HII region around a single source.  The left and middle panels show the ratio of the simulated $r(t)$ with the analytic solution (Eq.~\ref{eq:r_of_t}) and the Stromgren radius (Eq.~\ref{eq:rS}), respectively, vs. time in units of $t_{\rm rec}$ (Eq.~\ref{eq:trec}).  The right panel shows the I-front velocity in units of $r_{\rm S}/t_{\rm rec}$.  The black and red dashed curves show results for Standard and Generalized Opacity mode, respectively, while the solid black line denotes the analytic solution.  The other curves denote results from codes presented in I06 \cite{Mellema2006,Gnedin2001,Ciardi2001,Susa2000,Nakamoto2001,Razoumov2005,Ritzerveld2003,Whalen2006,Rijkhorst2006,Alvarez2006}.  Both modes of \textsc{FlexRT} agree well with the analytic prediction.  Generalized Opacity mode yields sub-percent agreement since, like the analytic model, it assumes an infinitely sharp I-front.  Like most CRTCP codes, Standard mode predicts a slightly larger HII radius than the analytic prediction, owing to the finite width of the I-front.}  
    \label{fig:stromgren_rs}
\end{figure}

Figure~\ref{fig:stromgren_rs} shows the evolution of the I-front position and velocity for Test \#1.  The left panel shows the ratio of the isotropically-averaged I-front radius with the analytic prediction (Eq.~\ref{eq:r_of_t}), while the middle and right panels show the evolution of $r(t)$ and $v(t)$.  Following I06, we express distance and time in units of $r_s$ and $t_{\rm rec}$, respectively.  The black and red dashed curves show results for Standard and Generalized Opacity mode, respectively, while the black solid curve shows the analytic result.  The other thin curves are all results from CRTCP codes.  In Standard mode, the HII region expands slightly faster than the analytic solution and reaches a slightly larger volume.  Indeed, most of the CRTCP codes do this, reaching final radii a few percent larger than $r_s$.  This is in part because the I-front is not infinitely sharp\footnote{And perhaps also due to the an-isotropic nature of the Cartesian mesh. 
 }, as the analytic solution assumes.  Generalized Opacity mode agrees with the analytic prediction within $<< 1\%$ at $t > 2 t_{\rm rec}$.  This is because it treats the I-front as infinitely sharp, matching the assumption of the analytic solution.  We also note that Generalized Opacity mode is very close to the result of from I-Front Tracker (IFT), which makes the same assumption.  

\begin{figure}
    \centering
    \includegraphics[scale=0.325]{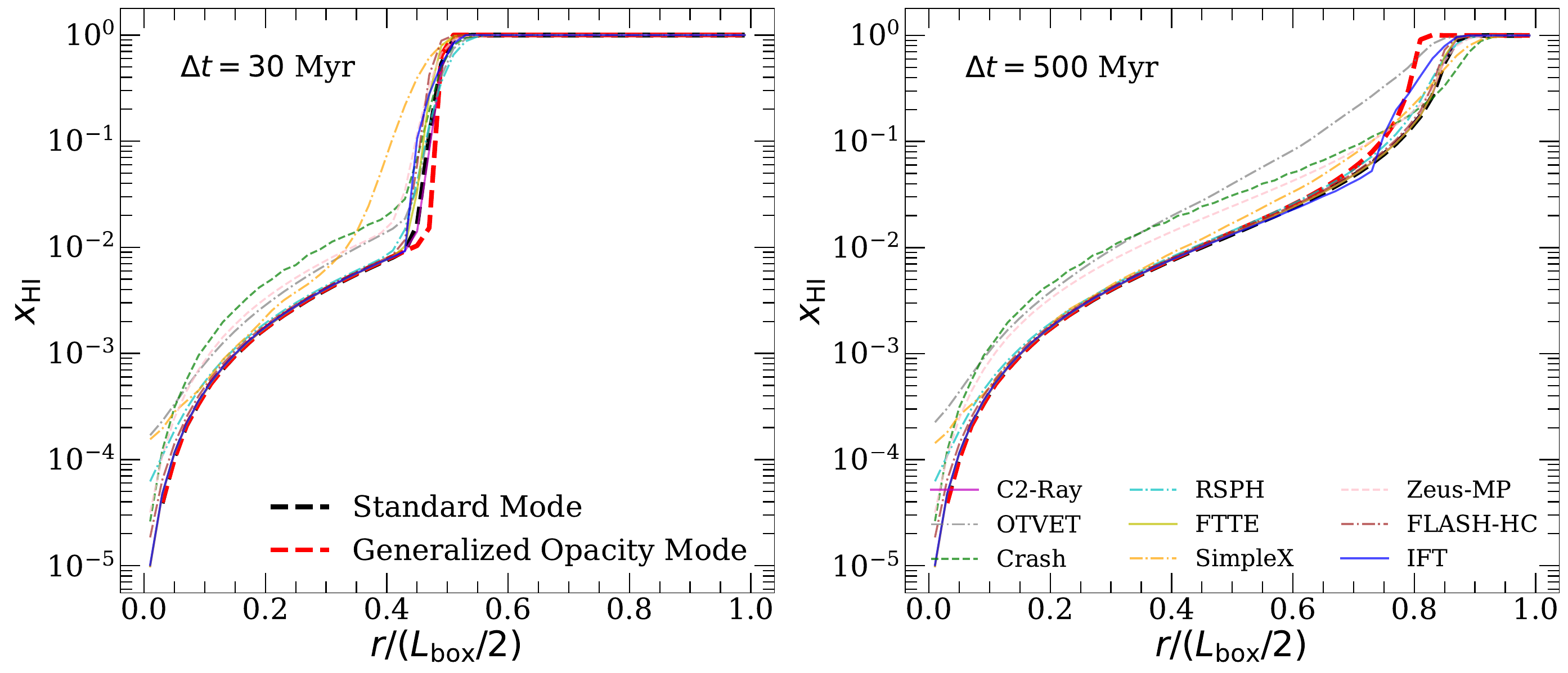}
    \caption{Spherically averaged HI fraction profile for Test \#1 at $\Delta t = 30$ Myr (left) and $500$ Myr (right).  The two modes agree closely for gas behind the I-front.  The gas transitions to neutral more abruptly at $\Delta t = 500$ Myr in Generalized Opacity mode, owing to the assumption of sharp I-fronts.  The profiles are also in good agreement with CRTCP codes.  Standard mode closely matches C2-Ray, as expected since the ionization solvers in those codes are very similar.}  
    \label{fig:stromgren_profile}
\end{figure}

Figure~\ref{fig:stromgren_profile} shows the angle-averaged HI fraction ($x_{\rm HI}$) profile as a function of distance from the source.  The left and right panels show results at $\Delta t = 30$ and $500$ Myr, respectively.  Note that in Generalized Opacity mode, the HI fraction is not calculated explicitly, and so we must invert Eq.~\ref{eq:subgrid_lambda} to get it.  The two modes agree well with each other at $t = 30$ Myr, when the I-front is sharpest.  At $t = 500$ Myr, when the I-front has nearly stopped, a larger difference emerges.  The transition from highly ionized to neutral is wider in Standard mode and extends to a larger distance from the source.  Behind the I-front, the profile of residual HI is in good agreement.  Standard mode is in tight agreement with several CRTCP modes, particularly C2-Ray.  This makes sense because standard mode and C2-Ray have very similar ionization solvers.  Once again, Generalized Opacity mode agrees best with IFT.  

\begin{figure*}
    \centering
    \includegraphics[scale=0.177]{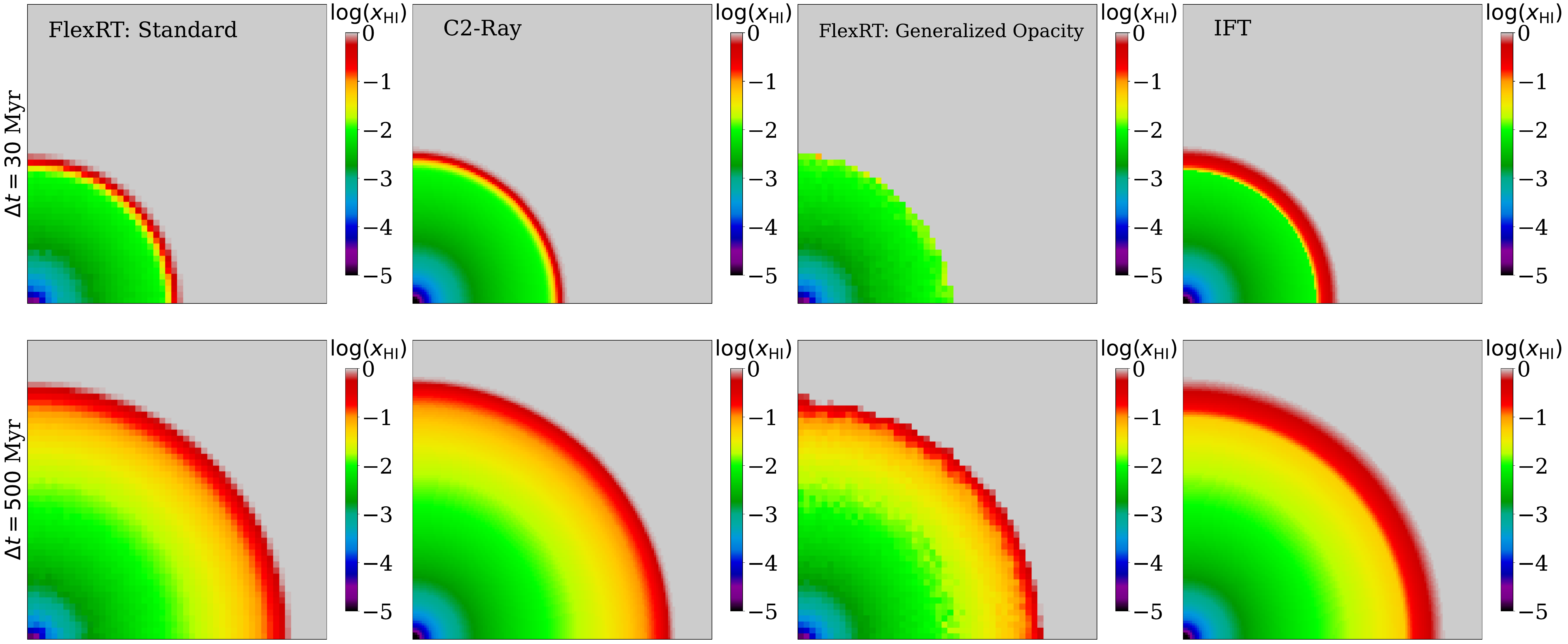}
    \caption{Visualization of Test \#1 - expansion of an isothermal, spherically symmetric HII region around a point source.  We show quarter-slices of $\log(x_{\rm HI})$ through the center of the HII region at $\Delta t = 30$ Myr (top row) and $500$ Myr (bottom row).  From left to right, the columns show results for Standard mode, C2-Ray (from I06), Generalized Opacity mode, and IFT (from I06).  Standard and C2-Ray are in excellent visual agreement.  In Generalized Opacity mode, the boundary between highly ionized and neutral gas is much sharper at $\Delta t = 30$ Myr, owing to its sharp I-front treatment.  In IFT, there is a similarly sharp jump from highly ionized gas (with $x_{\rm HI} <  0.01$) to significantly neutral gas (with $x_{\rm HI}  > 0.1$).  However, IFT has an extended region with $x_{\rm HI} < 1$ in front of this jump, while in Generalized Opacity mode $x_{\rm HI}$ goes straight to unity.  Generalized Opacity mode also displays a noisier $x_{\rm HI}$ field, since $x_{\rm HI}$ is estimated in each cell based on its instantaneous value of $\Gamma_{\rm HI}$, and is more sensitive to ray-tracing related artifacts in the radiation field than the time-integrated $x_{\rm HI}$ in Standard mode.  }
    \label{fig:stromgren_vis}
\end{figure*}

Figure~\ref{fig:stromgren_vis} shows slices of the HI fraction through the center of the box at $t = 30$ and $500$ Myr (top and bottom rows, respectively)\footnote{ We only show one corner of the HII region, since the CRTCP tests do not simulate the entire sphere.}.  In the 1st and 2nd column, we compare Standard mode to C2-Ray, since these use similar methods to solve for the ionization state of the gas.  The 3rd and 4th columns compare Generalized Opacity mode to IFT, since these both assume sharp boundaries between ionized and neutral gas.   We find excellent visual agreement between \textsc{FlexRT} and C2-Ray at both times.  Generalized Opacity mode displays a much sharper boundary between ionized and neutral gas, especially at $\Delta t = 30$ Myr, reflecting its sharp I-front treatment.  In IFT, there is a similarly sharp jump between highly ionized ($x_{\rm HI} < 0.01$) gas and highly neutral gas, although $x_{\rm HI}$ does not jump immediately to unity as it does in Generalized Opacity mode.  There is also some noise noticeable in Generalized Opacity mode, particularly at $\Delta t = 500$ Myr.  This is a manifestation of the ``flickering'' effect arising from rotating the HealPix unit vectors associated with the source on each time step.  The reason these are visible in Generalized Opacity Mode and not Standard Mode is because, in the former, $\lambda$ is calculated using the instantaneous value of $\Gamma_{\rm HI}$ (Eq.~\ref{eq:subgrid_lambda}-Eq.~\ref{eq:subgrid_nHI}), such that the neutral fraction effectively reacts instantaneously to changes in incident flux.  Note that this is not the case in the sub-grid model described in Refs.~\cite{Cain2021,Cain2022b}, which includes some dependence of $\lambda$ on the prior history of $\Gamma_{\rm HI}$.  In Standard Mode, $n_{\rm HI}$ is the result of time integration (Eq.~\ref{eq:avg_xi}), and is thus much less sensitive to this effect. Fortunately, the tight agreement of Generalized Opacity Mode with the analytic expectation in Figure~\ref{fig:stromgren_rs} suggests that this does not significantly affect the recombination rate in this test.


This test validates the most basic functions of the code - the ray transport and splitting (\S\ref{subsec:ray_splitting}) algorithms, and the two opacity solvers (\S\ref{sec:IGMopacity}).  Since this is a test with only one source, we do not use the merging algorithm (\S\ref{subsec:ray_merging}), which is expected to work well only in the limit of many sources (tests of the merging algorithm are presented in \S \ref{sec:raytracingtests}.)  It also highlights some of the key similarities and differences between the Standard and Generalized Opacity modes, including how they handle I-fronts and how sensitive they are to ray-tracing artifacts in the radiation field.  

\subsection{Stromgren sphere with $T$ evolution (Test \#2)}
\label{subsec:test2}

\begin{figure}
    \centering
    \includegraphics[scale=0.263]{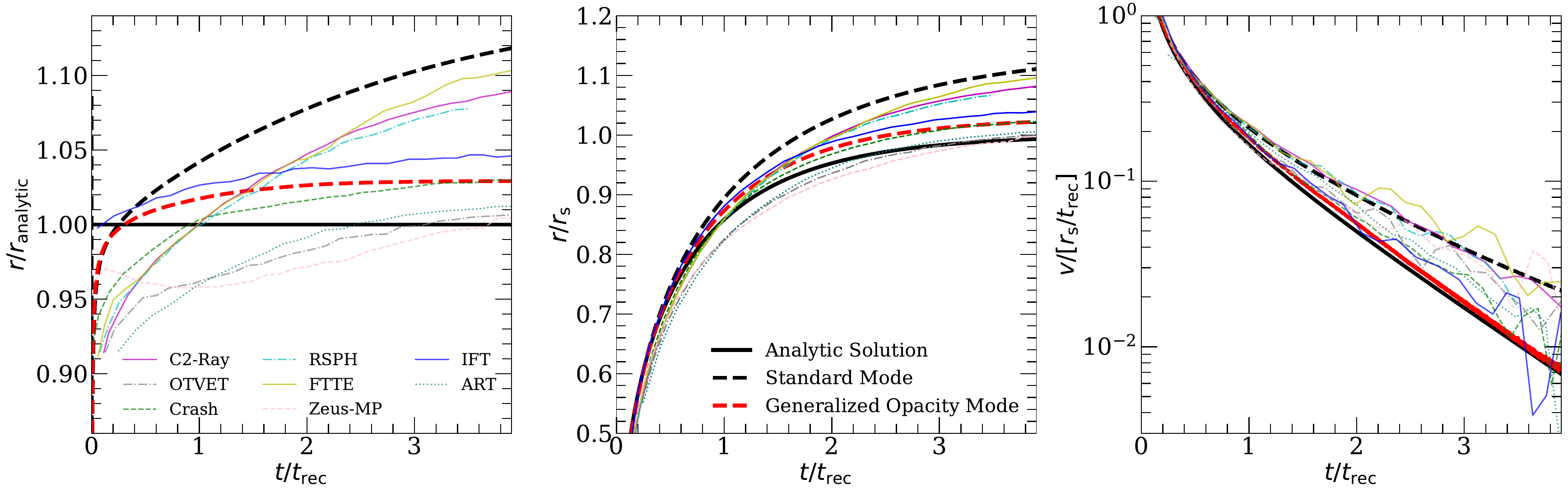}
    \caption{Same as Figure~\ref{fig:stromgren_rs}, but for Test \#2 including temperature evolution.  For consistency with I06, we continue to show the constant-temperature ($T=10^4$ K) analytic solution from Figure~\ref{fig:stromgren_rs} as a reference, even though it does not describe the evolving temperature case accurately. The HII region grows larger than the analytic expectation in both modes of \textsc{FlexRT}, since most of the gas reaches temperatures larger than $10^4$K (and thus has a lower recombination rate).  As in Test\#1, Generalized Opacity mode produces a smaller HII region. Both modes are in broad agreement with CRTCP results.  Standard mode produces a slightly larger Stromgren sphere than the other codes, possibly due to differences in the temperature structure near the I-front and/or differences in how the I-front position is defined (see text).  Generalized Opacity mode is close to IFT, but with a slightly smaller HII region.  }
    \label{fig:stromgren_rs_plot_T}
\end{figure}

Our next test has the same setup as \#1, but with evolving gas temperature.  Following I06, we set the initial gas temperature to $100$ K and use a blackbody spectrum with temperature $T_{\rm bb} = 10^5$K for the source.  For this and subsequent tests, we have included collisional cooling processes in Generalized Opacity mode by using Eq.~\ref{eq:subgrid_nHI} to get $n_{\rm HI}$, then calculating the rates using cooling coefficients from Ref.~\cite{Hui1997}.  Figure~\ref{fig:stromgren_rs_plot_T} shows the growth of the HII region in the same format as Figure~\ref{fig:stromgren_rs}.  Note that for consistency with I06, we continue to show the constant-temperature analytic solution from Figure~\ref{fig:stromgren_rs} as a reference, even though it does not describe the evolving temperature case accurately.  The I-front grows slightly faster than it did in Test \#1, finishing with a radius $\approx 12\%$ ($4\%$) larger than the analytic prediction for Standard (Generalized Opacity) mode.   This is because the temperature in the ionized region is higher than $10^4$K, which reduces the recombination rate and increases $r_S$ (Eq.~\ref{eq:rS}) relative to Test \#1.  

Both modes of \textsc{FlexRT} agree broadly with the CRTCP codes.  One notable difference is that Standard mode produces a slightly larger Stromgren sphere than all the CRTCP codes.  This could be in part due to differences in the amount of pre-heating ahead of the I-front, which changes the recombination rate near the edge of the ionized region.  This effect is determined by the frequency resolution of the ionizing spectrum and the assumed physics of pre-heating, which differ considerably between the CRTCP codes (as pointed out by I06).  Another factor could be that our estimate of $r$ is based on the total ionized fraction in the box, assuming a sharp boundary between ionized and neutral gas. 
 Since the I-front in standard mode is asymmetric in the radial direction, this definition may slightly over-estimate the true I-front location.  Generalized Opacity mode agrees reasonably well with IFT, but has a slightly smaller Stromgren sphere.  This may owe to the gas being slightly cooler in \textsc{FlexRT}.   

\begin{figure}
    \centering
    \includegraphics[scale=0.325]{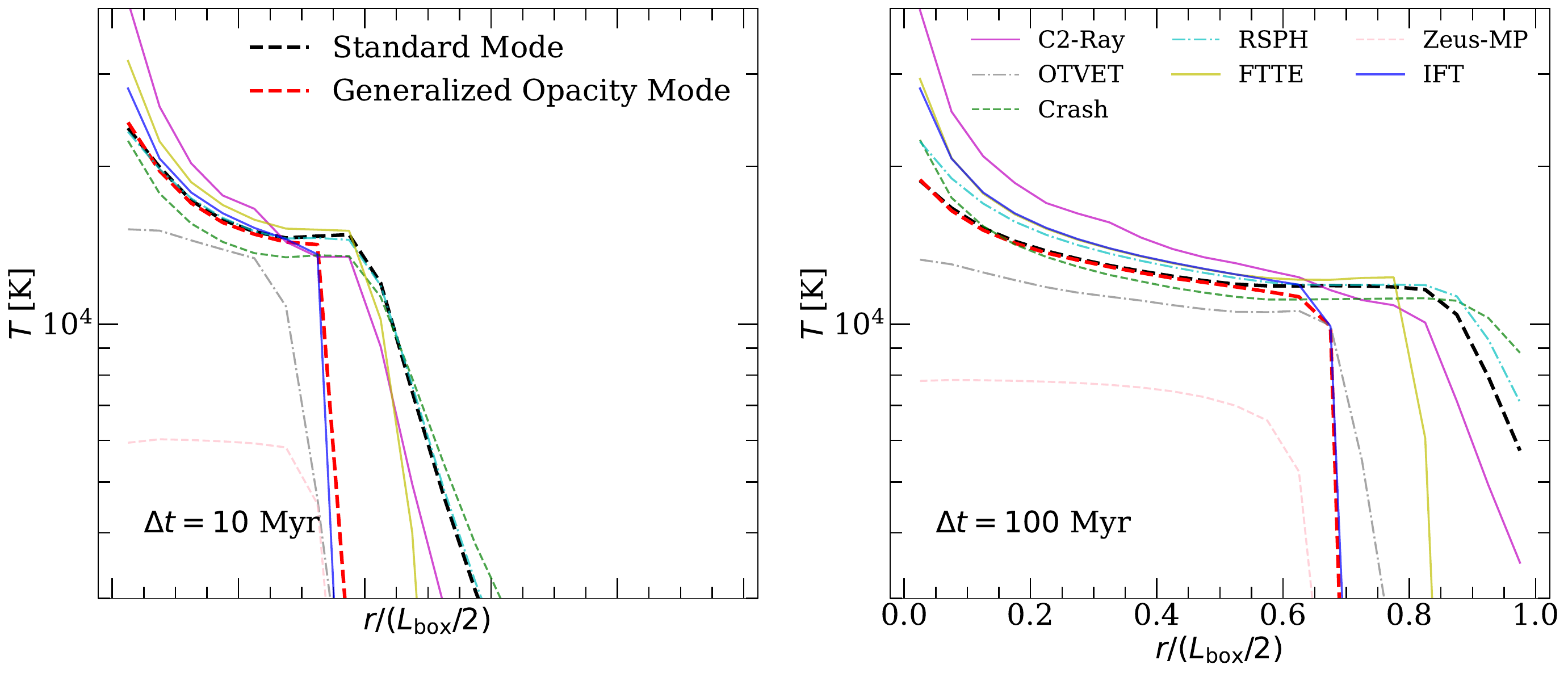}
    \caption{Spherically averaged $T$ profile for Test \#2 at $\Delta t = 10$ Myr (left) and $100$ Myr (right).  Behind the I-front, the \textsc{FlexRT} modes agree well with each other and the CRTCP results.  In Generalized Opacity mode, the temperature drops abruptly ahead of the I-front, since gas there cannot receive any radiation.  In standard mode, the hardest ionizing photons pre-heat the neutral gas ahead of the I-front, raising its temperature to $10^3-10^4$K.  Behind the I-front, the temperature in \textsc{FlexRT} is colder than C2-Ray and varies less with distance from the source. Standard mode displays similar levels of pre-heating as RSPH and CRASH.  Generalized Opacity mode agrees most well with IFT, as in Figure~\ref{fig:stromgren_profile}.  }
    \label{fig:profile_plot_T}
\end{figure}

Figure~\ref{fig:profile_plot_T} shows the spherically averaged temperature profile for $\Delta t = 10$ Myr (left) and $100$ Myr (right).  At both times, the two modes agree closely with each other behind the I-front.  In Generalized Opacity mode, $T$ drops abruptly at the I-front position, since the sharp I-front approximation prevents pre-heating of neutral gas.  In Standard mode, there is significant pre-heating ahead of the I-front.  By $\Delta t = 100$ Myr, the gas is pre-heated to $\sim 10^4$ K to almost the edge of the box.  Again, the differences between the modes arise from their different I-front treatments.  The temperature profiles behind the I-front in both modes are in broad agreement with the CRTCP codes.  Both agree most closely with RSPH and IFT at $\Delta t = 10$ Myr, and at $\Delta t = 100$ Myr, we find the best agreement with RSPH and CRASH, although the $T$ profile has a slightly different shape than both.  The gas in FlexRT is colder close to the source than in C2-Ray, but drops off less quickly approaching the I-front.  Ahead of the I-front, Standard mode displays similar levels of pre-heating as RSPH and CRASH.  Generalized Opacity mode agrees well with IFT near the I-front.  

\begin{figure*}
    \centering
    \includegraphics[scale=0.17]{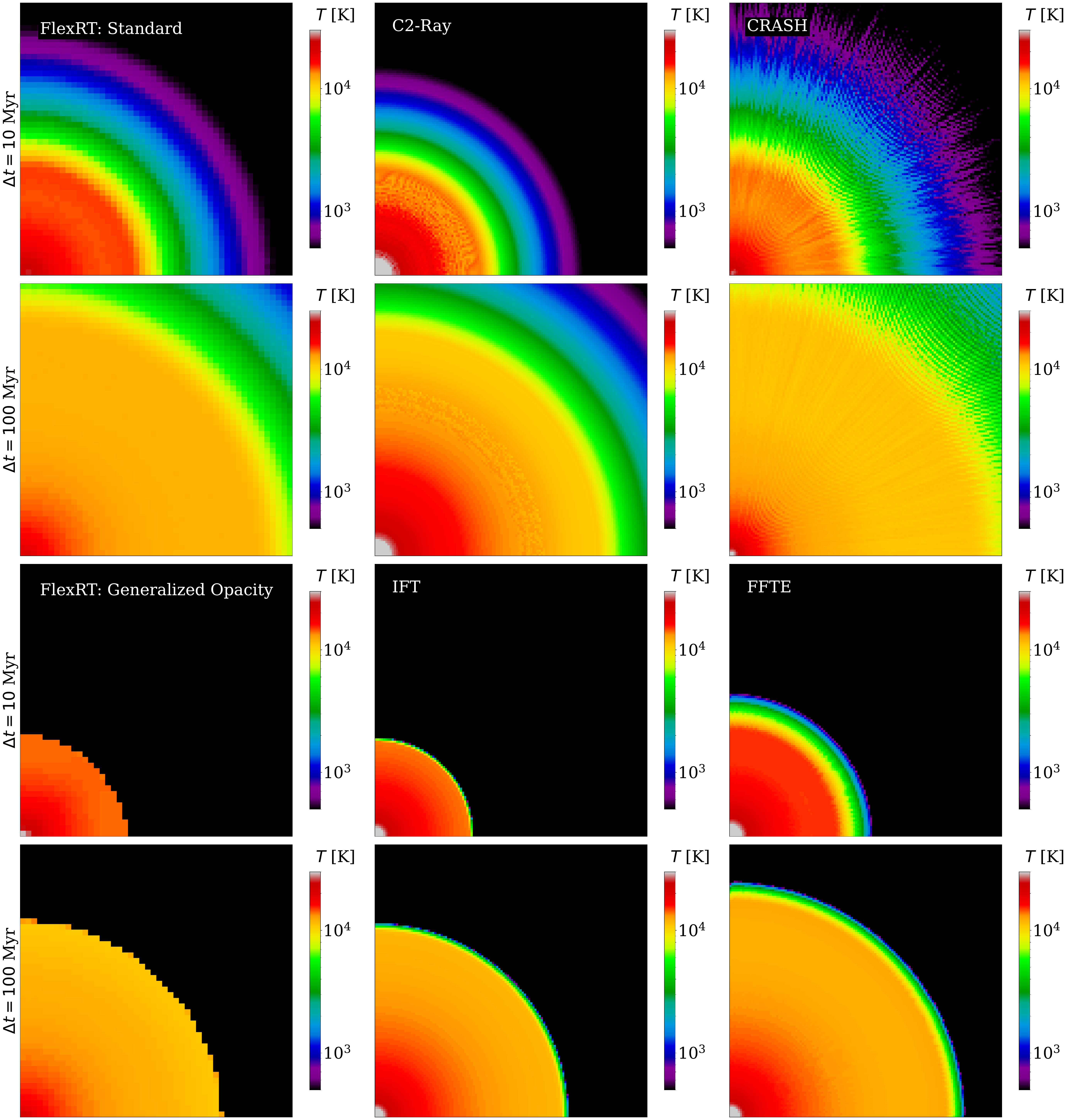}
    \caption{Slices through the gas temperature for Test\#2.  The 1st and 2nd rows compare Standard mode, C2-Ray, and CRASH (from left to right) at $\Delta t = 10$ and $100$ Myr, respectively.  The 3rd and 4th rows show the same comparison for Generalized Opacity mode, IFT, and FFTE.  The $T$ maps are qualitatively similar in Standard mode, C2-Ray, and CRASH, with \textsc{FlexRT} displaying more (less) pre-heating ahead of the I-front than C2-Ray (CRASH).  This likely owes to differences in the frequency resolution of the ionizing spectrum.  Generalized Opacity mode displays no pre-heating due its sharp I-front model, in good agreement with IFT and FFTE.  The temperature structure inside the I-front is qualitatively similar in these three codes.  The gas is cooler at $\Delta t = 100$ Myr in \textsc{FlexRT} than the other codes, suggesting it includes more equilibrium cooling. }
    \label{fig:vis_plot_T}
\end{figure*}

In Figure~\ref{fig:vis_plot_T}, we show slices through the temperature around the source using the same format as Figure~\ref{fig:stromgren_vis}, at $\Delta t = 10$ and $100$ Myr.  The top two rows compare Standard mode, C2-Ray, and CRASH.  The bottom two rows make the same comparison for Generalized Opacity mode, IFT, and FFTE.  Standard mode pre-heats the gas ahead of the I-front more than C2-Ray, but less than CRASH, likely because the codes all resolve the high-energy end of the spectrum differently.  The temperature in C2-Ray is hotter near the source than in \textsc{FlexRT} and CRASH.  This could be due to the lack of collisional ionization cooling in C2-Ray (at the time of writing of I06, see their Table 2, 6th column).  The gas near the I-front (edge of the red/orange region) is hotter in \textsc{FlexRT} than in CRASH or C2-Ray at $\Delta t = 10$ Myr, suggesting differences in the post-I-front gas temperature.  In the bottom two rows, we see that Generalized Opacity mode compares favorably to both IFT and FFTE, since neither of these allow significant pre-heating of the gas ahead of the I-front.  The temperature inside the I-front is similar in all three codes at $10$ Myr, but is slightly cooler in \textsc{FlexRT} at $100$ Myr than the other codes.  This suggests that \textsc{FlexRT} may have higher equilibrium cooling rates.  

Test \#2 demonstrates that the two modes of \textsc{FlexRT} are in good agreement with each other and with the CRTCP codes in terms of thermal physics.  As in \S\ref{subsec:stromgren_test}, the differences between the two modes of \textsc{FlexRT} and the CRTCP codes arise mainly from their different I-front treatments, differences in the frequency resolution of the ionizing spectrum, and perhaps differences in the detailed implementation of the heating/cooling physics.  

\subsection{I-front trapping by a dense clump (Test \#3)}

\begin{figure*}
    \centering
    \includegraphics[scale=0.215]{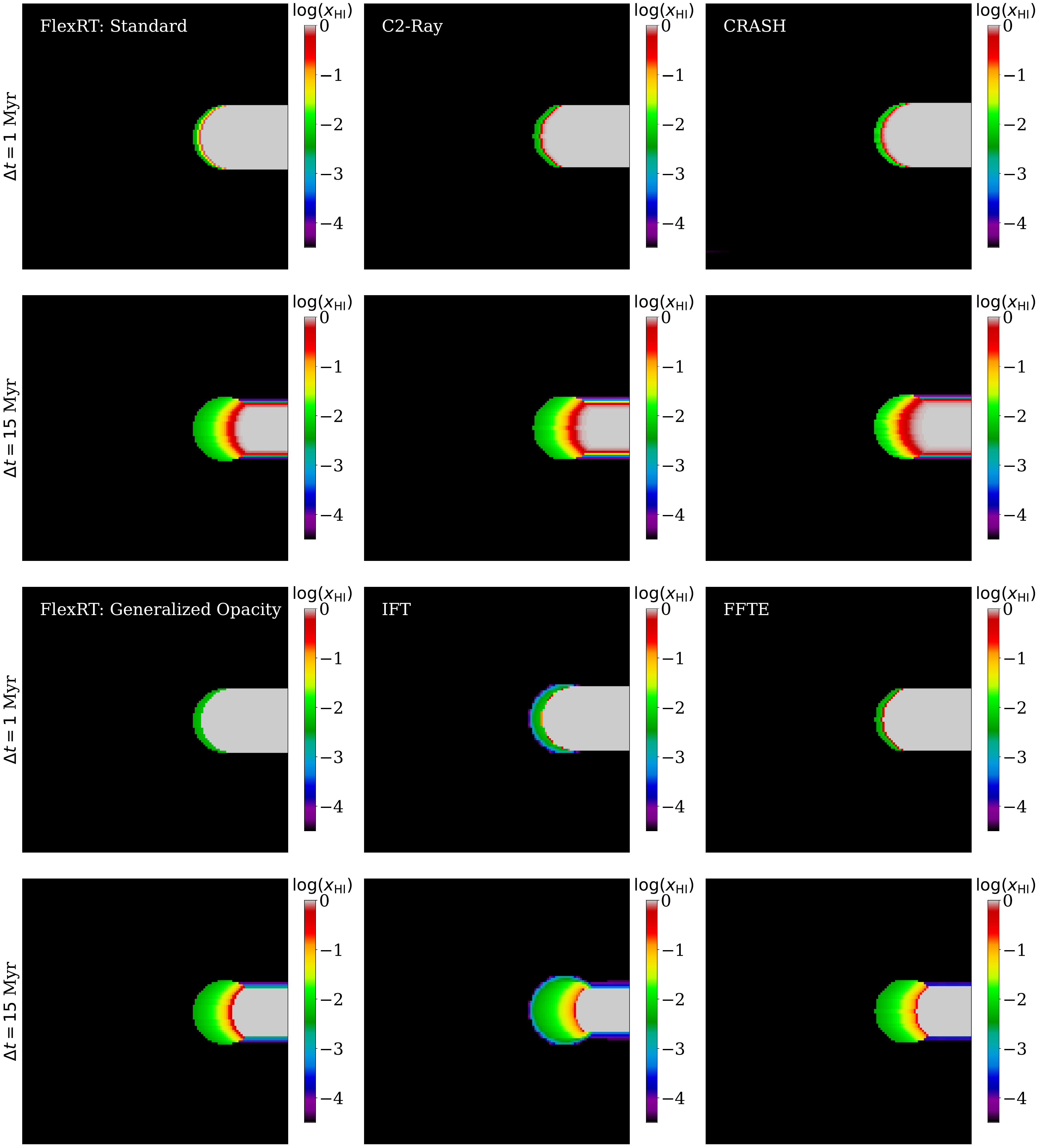}
    \caption{Visualization of $x_{\rm HI}$ for Test \#3 - shadowing of a plane-parallel cosmological I-front by a dense clump (see description in text).  The layout of the plot is the same as that in Figure~\ref{fig:vis_plot_T}, with Standard mode compared to C2-Ray and CRASH in the top rows, and Generalized Opacity mode compared to IFT and FFTE in the bottom rows.  All codes display fairly similar results at $\Delta t = 1$ Myr - a small amount of gas at the edge of the clump is ionized, with the rest remaining shielded.  Again, the boundary between ionized and neutral gas is noticeably sharper in Generalized Opacity mode than in Standard mode, as it is in IFT and FFTE.  At $\Delta t = 15$ Myr, the I-front has penetrated about halfway into the clump and stalled.  Standard mode and C2-Ray agree well, while in CRASH the I-front doesn't get as far into the clump.  Generalized Opacity mode and FFTE are in similarly good agreement, while the front penetrates furthest into the clump in IFT.  In these three cases, the boundary between ionized and neutral gas remains sharp throughout.  }
    \label{fig:shadow_HI_vis}
\end{figure*}

Test \#3 models the trapping of a cosmological I-front by a dense gas clump embedded in a low-density medium.  This test will demonstrate how \textsc{FlexRT} handles self-shielding and I-front trapping.   We use the same setup and test parameters described in \S3.4 of I06, which include a $10^5$K blackbody spectrum and a clump that is $200\times$ denser than the gas around it.  Figure~\ref{fig:shadow_HI_vis} visualizes the HI fraction at $\Delta t = 1$ Myr and $15$ Myr.  The layout of the plot is the same as that of Figure~\ref{fig:vis_plot_T}, with Standard mode compared to C2-Ray and CRASH in the top two rows, and Generalized Opacity mode compared to IFT and FFTE in the bottom two.  At $\Delta t = 1$ Myr, the codes in the top row are all similar - there is a thin boundary separating ionized gas on the left edge of the clump from the shadowed interior.  At $\Delta t = 15$ Myr, the I-front has penetrated about halfway through the clump and stalled, leaving much of the gas behind the clump still shadowed.  The $x_{\rm HI}$ maps are similar in Standard mode and C2-Ray, while in CRASH the I-front doesn't go as far into the clump.  I06 attributed this to differences in the spectral energy distribution between CRASH and C2-Ray.  

\begin{figure*}
    \centering
    \includegraphics[scale=0.22]{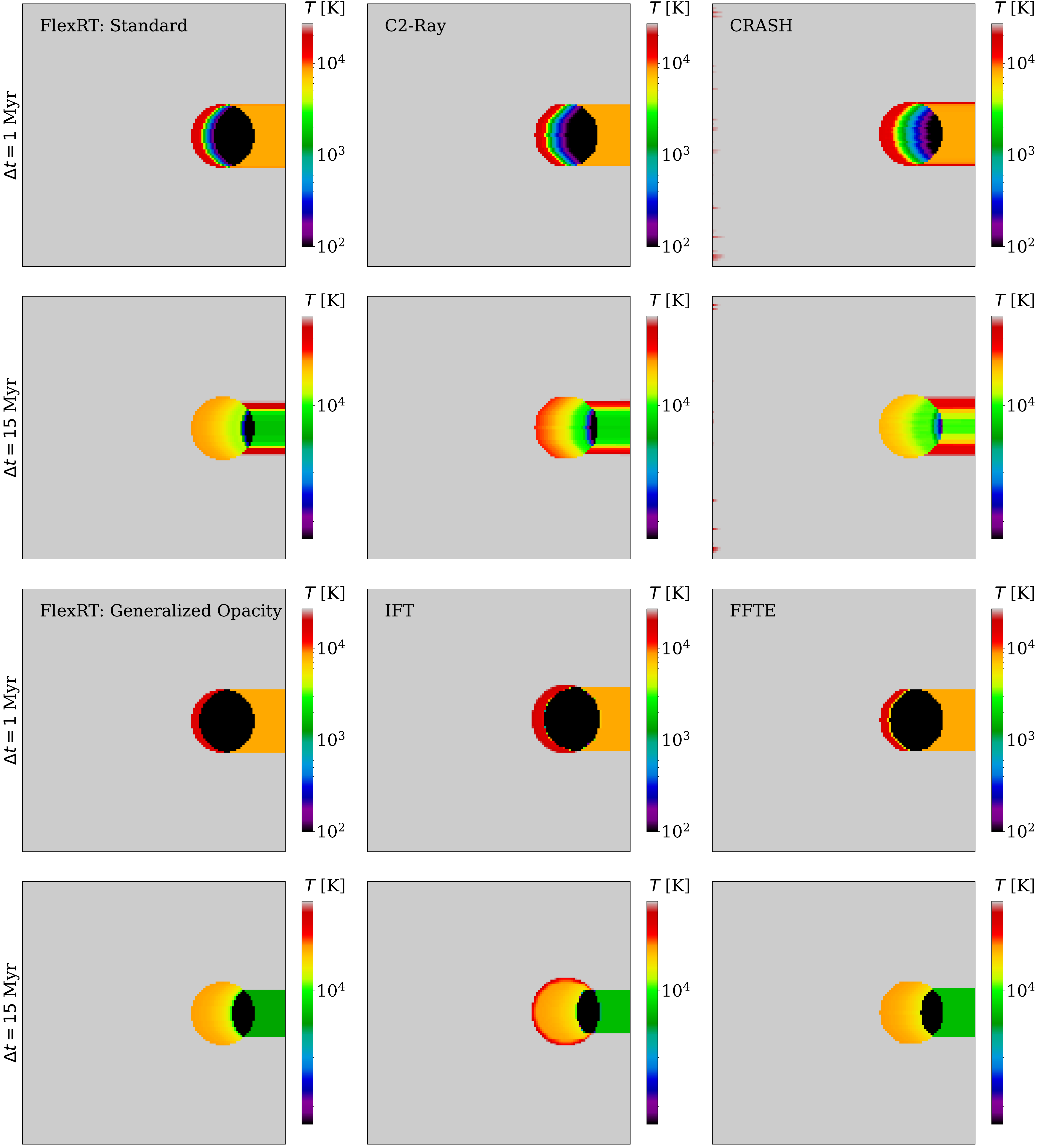}
    \caption{Same as Figure~\ref{fig:shadow_HI_vis}, but for gas temperature.  The top two rows are qualitatively similar, with Standard mode showing slightly less pre-heating than C2-Ray or CRASH.  At $\Delta t = 15$ Myr, the temperature inside the clump is similar in Standard mode and CRASH, but is slightly hotter near the left edge of the clump in C2-Ray.  The temperature to the right of the clump is similar in all three codes, with the edges heated by hard photons and the gas in between remaining near its initial temperature.  As in Figure~\ref{fig:vis_plot_T}, we see good agreement in the temperature of Generalized Opacity mode, IFT, and FFTE in the bottom rows.  The gas at the left edge of the clump is hot at $\Delta t = 1$ Myr, with no pre-heating behind it.  At $\Delta t = 15$ Myr, Generalized Opacity mode and FFT display very similar results, while IFT there is a hot ring around the edge of the clump.  The temperature behind the clump is similar in all three codes. }
    \label{fig:shadow_temp_vis}
\end{figure*}

The Generalized Opacity mode compares favorably to the CRTCP codes in the bottom rows.  At both $\Delta t = 1$ and $15$ Myr, the boundary between ionized and neutral gas is sharper than in the top rows, especially in Generalized Opacity mode.  In IFT, the I-front penetrates slightly further into the clump than in Generalized Opacity mode or FFTE, and displays a ring of highly ionized gas round the edge of the clump.  Generalized Opacity mode and FFTE are very similar in this comparison.  Note that once again, in the ionized part of the clump, Standard and Generalized Opacity modes display similar results.  

\begin{figure*}
    \centering
    \includegraphics[scale=0.218]{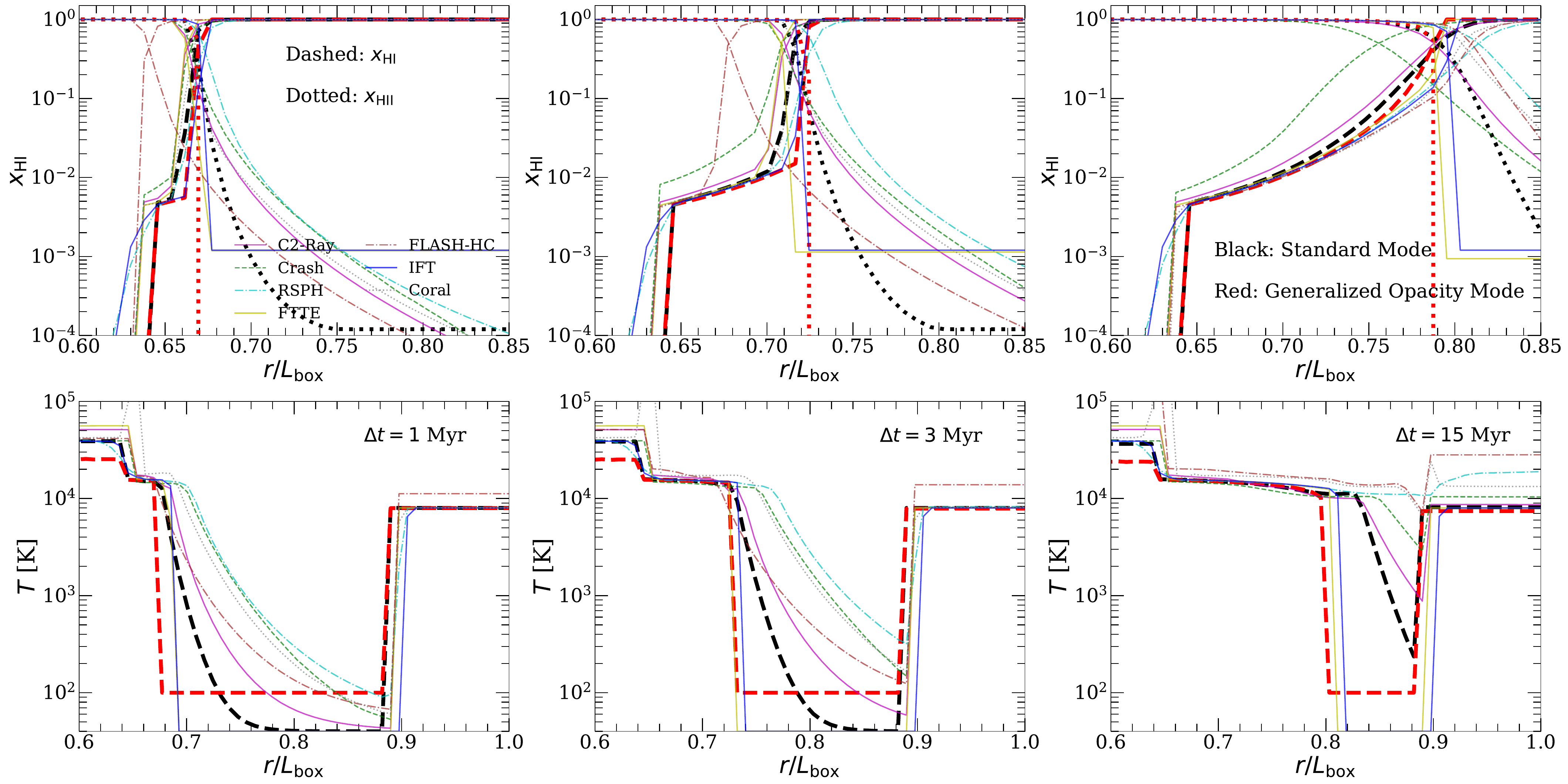}
    \caption{Profiles of $x_{\rm HI}$ (top) and $T$ (bottom) along the symmetry axis of the clump.  In the top row, the dashed curves denote $x_{\rm HI}$ and the dotted curves $x_{\rm HII}$. The $x_{\rm HI}$ profiles behind the I-front inside the clump are in good agreement in Standard and Generalized Opacity mode, as are the I-front positions.  In Generalized Opacity mode, there is no ionization behind the front owing to the sharp I-front treatment.  The temperature profiles behind the I-front are also similar.  In Standard mode, there is significant pre-heating ahead of the I-front, with the whole clump being heated by $\Delta t = 15$ Myr, while Generalized Opacity mode lacks pre-heating entirely.  Both codes broadly agree with the CRTCP results.  The temperature profiles ahead of the I-front inside the clump differ significantly between the CRTCP codes.  The amount of pre-heating in Standard mode is most similar to C2-Ray, while Generalized Opacity mode has no pre-heating and thus best matches FFTE and IFT.  }
    \label{fig:shadowing_profiles}
\end{figure*}

Figure~\ref{fig:shadow_temp_vis} shows the gas temperature in the same format as Figure~\ref{fig:shadow_HI_vis}.  At $\Delta t = 1$ Myr, Standard mode and C2-Ray agree well - the gas temperature is $\approx 2 \times 10^{4}$ K at the left edge of the clump and smoothly decreases to the initial temperature of $40$ K near the clump's center.  This smooth decline arises due to pre-heating by the hardest ionizing photons.  CRASH displays a similar temperature structure, but with the pre-heating penetrating deeper into the clump due to the inclusion of harder ionizing photons in CRASH.  At $\Delta t = 15$ Myr, most of the clump has been heated to $T > 10^4$ K, with only a small amount of cold gas remaining at the right edge of the clump.  The temperature in the ionized part of the clump is similar in FlexRT and CRASH, while the left edge of the clump is hotter in C2-Ray.  This may owe to the lack of collisional ionization cooling in C2-Ray (see \S\ref{subsec:test2}), which is important at these densities and temperatures.  The temperature structure behind the clump is similar in all three codes, with $T < 10^4$ K gas behind the shadowed part of the clump and hot gas outside.  
We see similar results in the bottom rows, with some notable differences.  As in Figure~\ref{fig:vis_plot_T}, Generalized Opacity mode, IFT, and FFTE display very sharp boundaries between hot and cold gas, due to their lack of pre-heating.  All three codes are very similar at $\Delta t = 1$ Myr.  At $\Delta t = 15$ Myr, Generalized Opacity mode and FFTE are also very similar.  IFT displays a ring of very hot gas around the edge of the clump\footnote{I06 notes a similar effect for the CORAL code (not displayed here) that arose from inaccuracies in the temperature solver in cells bordering regions with high and low opacity.  We speculate that a similar effect may exist in IFT.  }, but is otherwise similar to Generalized Opacity mode and FFTE.  Again, we note that the temperature of ionized gas is very similar in Standard and Generalized Opacity modes of \textsc{FlexRT}.  

 \begin{figure*}
     \centering
     \includegraphics[scale=0.23]{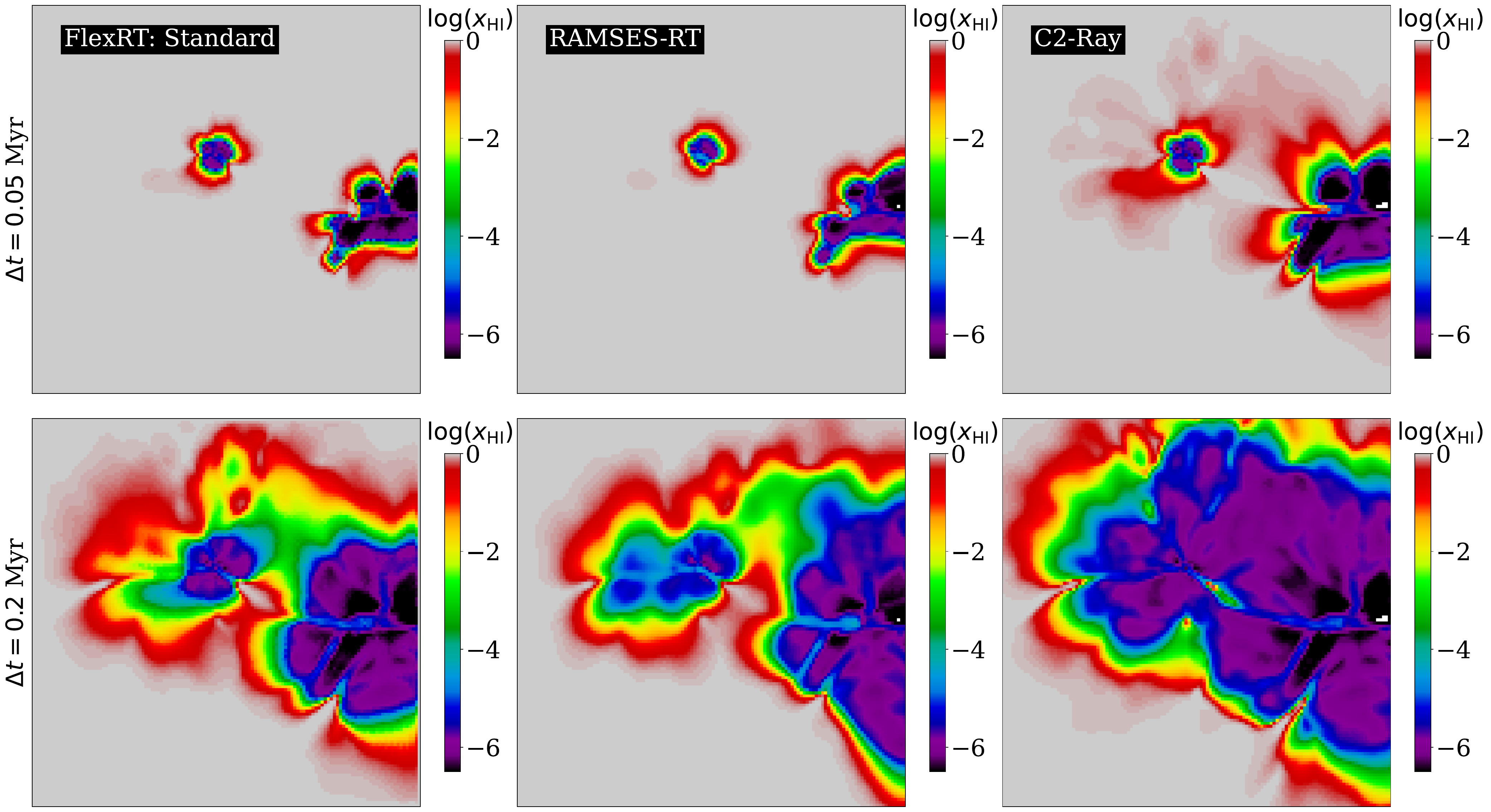}
     \caption{Visualization of $x_{\rm HI}$ for Test \#4 - multiple sources in a cosmological density field.  We show $\Delta t = 0.05$ Myr and $0.2$ Myr in the top and bottom rows (respectively).  From left to right, the columns show results for Standard mode, RAMSES-RT, and C2-Ray.  The shapes and sizes of the ionized regions in \textsc{FlexRT} and RAMSES-RT are similar at both times, with some differences in the ionization structure inside them that may arise from differences in the ionization solvers and/or temperature treatments.  In C2-Ray, the ionized regions are much larger at both times than in the other two codes, due to its use of the infinite speed of light approximation.  However, the ionization state of the gas in the interior of the ionized regions is similar in \textsc{FlexRT} and C2-Ray, reflecting the similarity of their ionization solvers.   }
     \label{fig:cosmo_HI_vis}
\end{figure*}

Figure~\ref{fig:shadowing_profiles} shows the HI (top) and temperature (bottom) profiles cut through the center of the clump along the y direction at $\Delta t = 1$, $3$, and $15$ Myr (left to right).  We compare against all the codes in I06 (see their Figures 27 and 28).  The I-front location and the $x_{\rm HI}$/temperature profiles behind it in \textsc{FlexRT}'s two modes agree well with each other.  The HI fraction and temperature profiles in \textsc{FlexRT} broadly agree with the CRTCP results, but there are some notable differences.  Behind the I-front, \textsc{FlexRT} and the CRTCP codes all agree reasonably well.  The CRTCP codes differ with each other most in the temperature profile, and associated levels of pre-ionization, ahead of the I-front.  Standard mode is most similar to C2-Ray, but differs in that the hardest ionizing photons do not penetrate as far into the clump, which results in less pre-heating of the neutral gas.  Generalized Opacity mode is closest to FFTE and IFT, since neither of these codes capture pre-heating. The codes also differ in the temperature of the low-density gas shadowed by the clump. \textsc{FlexRT} agrees with C2-Ray and IFT, finding that this gas has not yet been heated significantly by $\Delta t = 15$ Myr.  This comparison demonstrates that \textsc{FlexRT} accurately handles RT and shielding effects in very over-dense regions.

\subsection{Test with Cosmological Sources (Test \#4)}

Our final test in this section is CRTCP Test \#4 - a cosmological density field with multiple sources.  The setup for this test is described in \S3.5 of I06.  The box is $0.5$ $h^{-1}$Mpc on a side, and ionizing radiation is emitted from the $16$ most massive halos in the box with a constant volume ionizing emissivity of $\dot{N}_{\gamma} = 8.27 \times 10^{53}$ ph/s/cMpc$^{3}$.  The redshift is fixed to $z = 9$ during the test.  In Figure~\ref{fig:cosmo_HI_vis}, we show slices through the HI fraction at $\Delta t = 0.05$ Myr (top row) and $0.2$ Myr (bottom row).  From left to right, we show Standard mode, results for the same test using the RAMSES-RT code~\citep{Rosdahl2013}, and C2-Ray.  Note that we do not show Generalized Opacity mode here because I06 does not include a realization of Test \#4 using IFT, to which Generalized Opacity mode is most directly comparable.  We first note that the sizes and shapes of the ionized regions are similar in \textsc{FlexRT} and RAMSES-RT, while the ionization state of the ionized gas inside differs slightly.  This may owe to the different ionization solvers in the two codes.  C2-Ray, on the other hand, displays significantly larger ionized regions than the other codes, due to its use of the infinite speed of light approximation (as pointed out by Ref.~\cite{Rosdahl2013}).  The ionization structure deep inside the ionized regions is similar in FlexRT and C2-Ray due to the similarity of their ionization solvers.  

\begin{figure*}
     \centering
     \includegraphics[scale=0.232]{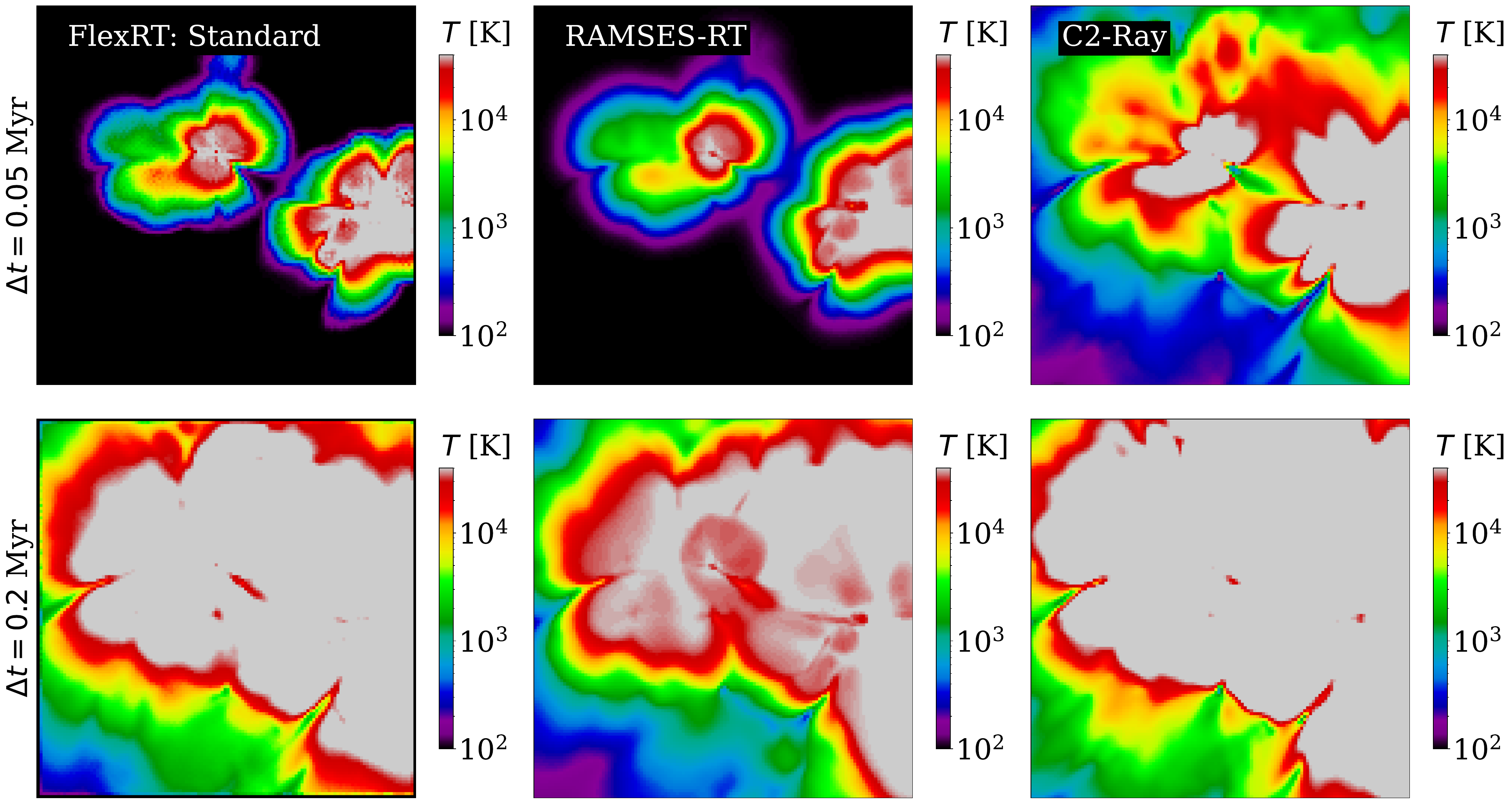}
     \caption{Visualization of temperature for Test \#4 with the same layout as Figure~\ref{fig:cosmo_HI_vis}.  The thermal structure in Standard mode is in good qualitative agreement with RAMSES-RT and C2-Ray.  The gas inside the highly ionized  regions is hot ($T > 3 \times 10^4$K), and the neutral gas outside is pre-heated by the hardest ionizing photons.  The ionized gas in RAMSES-RT is slightly cooler than in \textsc{FlexRT} and C2-Ray, particularly near the over-dense filaments.  The gas near the edges of the box is hotter in C2-Ray since the ionized regions are larger.  }
     \label{fig:cosmo_temp_vis}
 \end{figure*}

Figure~\ref{fig:cosmo_temp_vis} shows maps of the gas temperature in Test \#4 with the same layout as Figure~\ref{fig:cosmo_HI_vis} .  The codes show qualitatively similar results - the gas in ionized regions is very hot ($T > 3 \times 10^4$ K), with layers of progressively colder gas outside the ionized regions.  The gradual fall-off in temperature owes, again, to pre-heating by hard photons in the neutral gas.  This pre-heating is more significant in C2-Ray than in \textsc{FlexRT} or RAMSES-RT because of the infinite speed of light.  The gas within the ionized regions is slightly cooler in RAMSES-RT than in \textsc{FlexRT}, possibly due to differences in the binning of ionizing spectrum and/or the implementation of heating and cooling physics.

\begin{figure*}
     \includegraphics[scale=0.245]{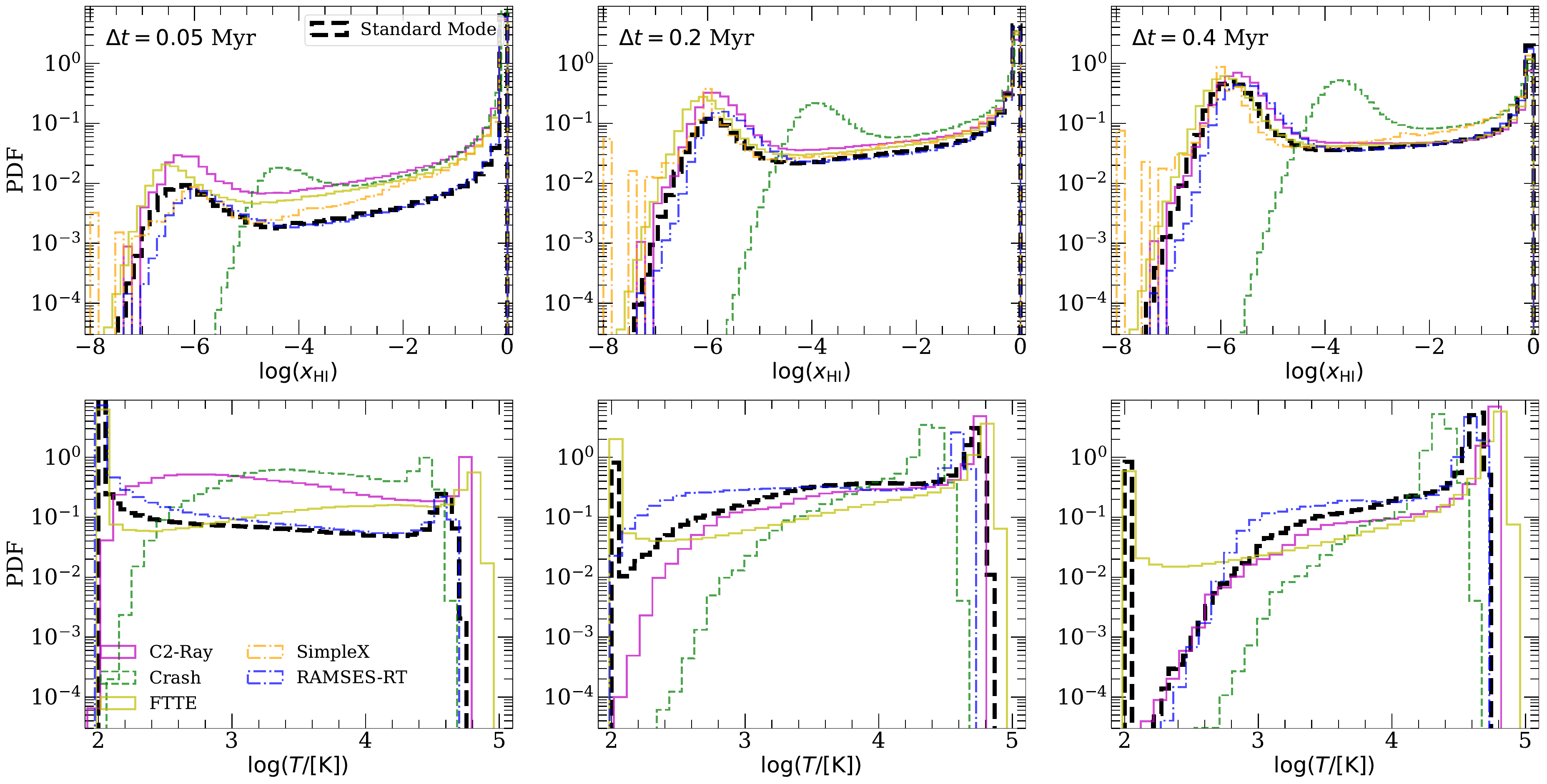}
     \caption{Distribution of $x_{\rm HI}$ (top) and $T$ (bottom) in Test \#4 at $\Delta t = 0.05$, $0.2$, and $0.4$ Myr (left to right).  The shape of the $x_{\rm HI}$ distribution in standard mode agrees well with all the CRTCP codes except CRASH, which has a peak at higher $x_{\rm HI}$ than the other codes.  We also find good agreement with RAMSES-RT, especially at $\Delta t = 0.05$ Myr when the finite speed of light is most important. \textsc{FlexRT} has more cold cells at $\Delta t = 0.05$ and $0.2$ Myr than the CRTCP codes due to its finite speed of light (and correspondingly smaller ionized regions), but again we see good agreement with RAMSES-RT.  At $\Delta t = 0.4$ Myr, the high-T end of the distribution best matches RAMSES-RT, while the low-T, preheated tail best matches that code and C2-Ray.  }
     \label{fig:cosmo_PDFs}
 \end{figure*}

Figure~\ref{fig:cosmo_PDFs} shows the distributions of $x_{\rm HI}$ (top) and $T$ (bottom) for $\Delta t = 0.05$, $0.2$, and $0.4$ Myr (left to right) in Standard mode, RAMSES-RT, and several CRTCP codes.  The distribution of $x_{\rm HI}$ in Standard mode agrees well with those of C2-Ray, FFTE, SimpleX, and RAMSES-RT at all times.  At $\Delta t = 0.05$ Myr, the normalization is somewhat lower than the CRTCP codes because fewer cells have been ionized in \textsc{FlexRT}, due to the finite speed of light.  The agreement with RAMSES-RT, which also uses the finite speed of light, is much better.  At $\Delta t = 0.2$ Myr, the $T$ distribution is similar to that of C2-Ray and RAMSES-RT, reflecting the visual agreement in Figure~\ref{fig:cosmo_temp_vis}.  At $\Delta t = 0.4$ Myr, the location of the high-$T$ peak agrees best with RAMSES-RT, and the shape of the distribution at lower temperatures is intermediate between that of RAMSES-RT and C2-Ray.  This suggests that the equilibrium heating/cooling in \textsc{FlexRT} is closest to RAMSES-RT, and the amount of pre-heating (set by the shape and binning of the ionizing spectrum) agrees well with both codes.  

The tests in this section demonstrate that \textsc{FlexRT} is in broad agreement with previously tested and validated reionization RT codes in a variety of scenarios.  They also highlight some of the important differences between the two modes of \textsc{FlexRT}.  These tests provide key validation for the basic functions of \textsc{FlexRT} and show that it solves the RT equation accurately in a variety of relevant contexts.  


\section{Validity of the Moving-Screen I-front method}
\label{sec:moving_screen}

In the previous section, we noted that the moving-screen I-front approximation employed in Generalized Opacity mode resulted in significant differences with Standard mode in CRTCP tests \#1-3.  In this section, we address the regime in which this approximation is valid, i.e. the regime in which Standard and Generalized Opacity modes yield very similar results.  

\begin{figure}
    \centering
    \includegraphics[scale=0.27]{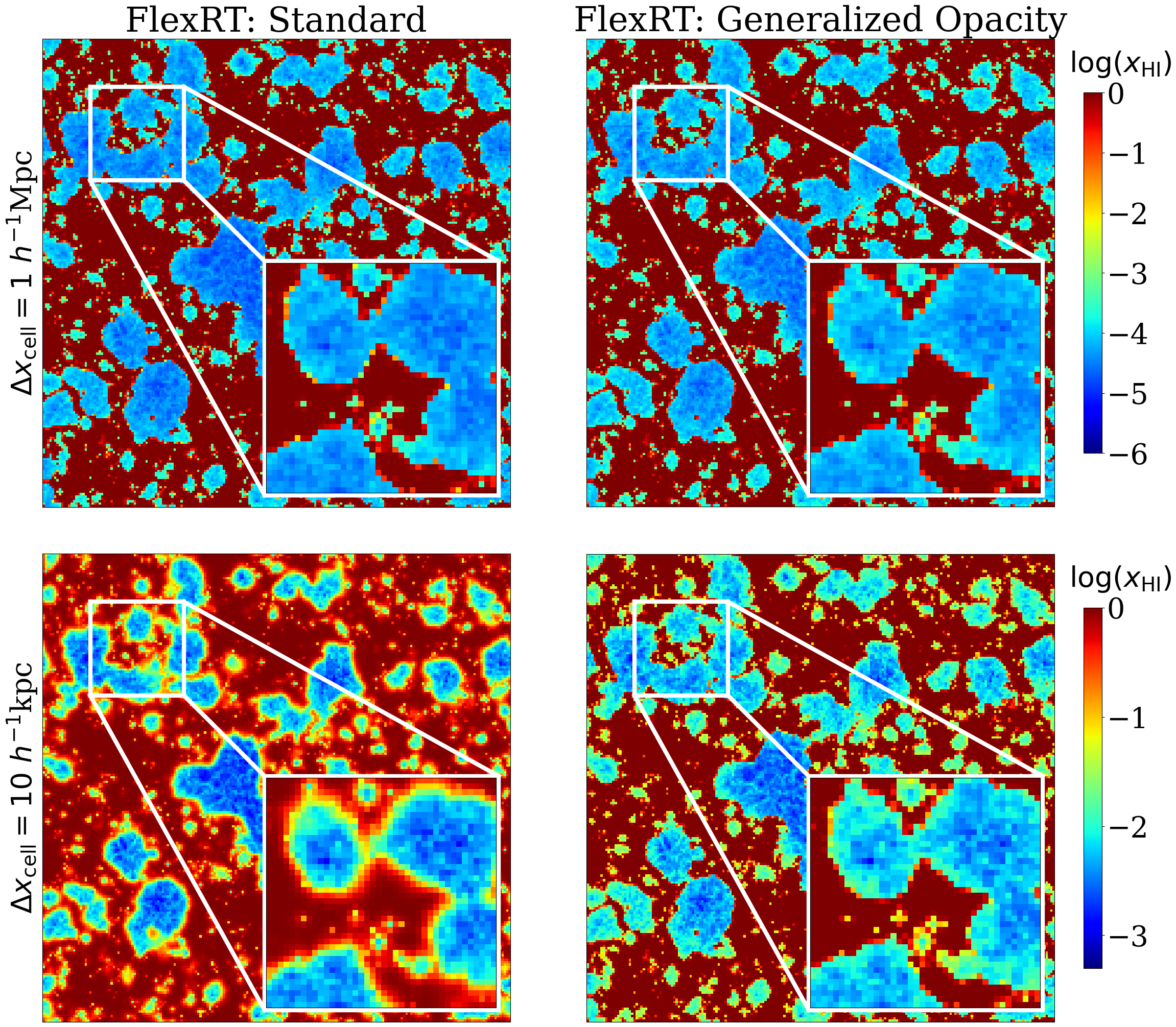}
    \caption{Demonstration of the regime in which the moving-screen I-front approximation used in Generalized Opacity mode is valid.  In the top row, we show maps of $x_{\rm HI}$ from a cosmological simulation in a $200$ $h^{-1}$Mpc with $\Delta x_{\rm RT} = 1$ $h^{-1}$Mpc in both modes.  The insets show zoom-ins to highlight the structure of I-fronts.  The boundary between neutral (red) and highly ionized (blue/green) cells is one cell thick, indicating that the I-fronts are un-resolved. 
 In this limit, Standard and Generalized Opacity modes produce the same result.  In the bottom  row, we scale the box volume and ionizing emissivity down by a factor of $100^3$, so that $\Delta x_{\rm RT} = 10$ $h^{-1}$kpc.  In this case, the I-fronts are partially resolved in Standard mode (bottom left), differing from the result of the moving-screen approximation seen in the bottom right.  This comparison shows that Generalized Opacity mode should be applied in cases where I-fronts are un-resolved.  This turns out to be its intended regime of applicability.  Conversely, the comparison of the case with large $\Delta x_{\rm cell}$ shows that the I-front ``smearing'' effect discussed in \S\ref{subsec:standardRT} does not affect the total recombination rate significantly, at least not for this test.  }
    \label{fig:moving_screen_vis}
\end{figure}

The moving-screen I-front approximation assumes that the I-front is entirely unresolved - namely, the typical width of an I-front, $\Delta x_{\rm IF}$, is much smaller than $\Delta x_{\rm RT}$.  To demonstrate this, we have run a cosmological simulation in a $200$ $h^{-1}$Mpc box with a cosmological distribution of halos acting as ionizing sources.  The cells in the box have $\Delta x_{\rm RT} = 1$ $h^{-1}$Mpc, considerably larger than the widths of typical cosmological I-fronts ($\Delta x_{\rm IF} \sim 10-100$ $h^{-1}$kpc).  The details of this simulation are described in the next section, so we will omit them here - save to emphasize that the moving-screen approximation is expected to be valid in this case because of the large RT cells.  The top-left and top-right panels of Figure~\ref{fig:moving_screen_vis} show maps of $x_{\rm HI}$ (in log scale) for this test in Standard and Generalized Opacity modes, respectively.  The inset shows a zoom-in to highlight the structure of the I-fronts.  In both cases, the boundary between highly ionized (blue) and neutral (red) cells is only one cell thick, showing that the I-front is un-resolved.  The two modes agree very well in this limit. 

In the bottom panels, we show a contrasting example in which the moving-screen approximation fails.  To do this, we artificially scale the volume of our $200$ $h^{-1}$Mpc box down by a factor of $100^3$ so that it becomes a $2$ $h^{-1}$Mpc box.  We also scale the ionizing emissivity of sources down by the same factor so that reionization progresses at the same rate. We then rerun the simulation in the two modes.  These runs have $\Delta x_{\rm RT} = 10$ $h^{-1}$kpc, such that I-fronts are partially resolved.  We see that in the bottom-left panel, the boundary between highly ionized and fully neutral gas is several cells thick, whereas it remains one cell thick on the bottom-right.  This comparison illustrates the regime in which the moving-screen approximation employed in Generalized Opacity mode is valid: $\Delta x_{\rm RT} >> \Delta x_{\rm IF}$.  We emphasize that Generalized Opacity mode is designed for large-volume simulations of reionization (see e.g. \cite{Cain2021, Cain2022b}).  Standard mode should be used for problems in which the I-fronts are resolved and their detailed structure is important.    

Conversely, our test shows that in the case with $\Delta x_{\rm RT} = 1$ $h^{-1}$Mpc, the growth of ionized regions is largely unaffected by the error in the recombination rate inside I-fronts incurred in Standard Mode (see discussion in S\ref{subsec:standardRT}).  This suggests that, at least in this test, the recombinations taking place in partially ionized cells do not contribute significantly to the total ionizing budget in the simulation.  However, we caution that this result may not hold in the case where the recombination rate is being boosted at the sub-grid level to account for missing small-scale structure.  In general, a test like the one shown in this section - comparing the moving-screen method to the Standard Mode procedure - can be used to assess the importance of such recombinations and determine if the treatment in Standard Mode is sufficiently accurate for a given situation.  

\section{Testing the Merging Scheme}
\label{sec:raytracingtests}

In \S\ref{sec:CRTCP_tests}, we demonstrated good qualitative and quantitative agreement between \textsc{FlexRT} and the CRTCP codes.  Those tests validate the basic functionality of \textsc{FlexRT}, including ray transport and splitting, absorption, gas chemistry, and thermal evolution.  However, none of them employ one of the key speed-up mechanisms in \textsc{FlexRT}: ray merging.  Indeed, we turned off merging in all of those tests to enable a maximally clean comparison to the CRTCP results.  In this section, we describe two more tests designed to validate the merging scheme in FlexRT.  In \S\ref{subsec:CRTCP4_again}, we will re-visit the CRTCP cosmological test, this time with merging included.   In \S\ref{subsec:hy_cosmo}, we run a cosmological test on a much larger scale and directly compare to results from the \textsc{RadHydro} code of Ref.~\cite{Trac2007}, which uses a similar ray merging approach.  

\subsection{CRTCP Test \#4 (again)}
\label{subsec:CRTCP4_again}

We return here to CRTCP Test \#4, but this time vary the parameters that control merging, which are described in \S\ref{subsec:ray_merging}.  We consider a case where rays adaptively split but do not merge (``No Merging''), and two scenarios with merging, both of which track $12$ unique directions for merged rays (that is, $l_{\rm merge} = 0$).  We set the number of rays per cell exempt from merging, $N_{\rm ex}$, to $16$ for one run and $4$ for the other.  Both runs have $N_{\rm ray}^{\min} = 2$ (Eq.~\ref{eq:r_split}).  In the case with $N_{\rm ex} = 16$, the box can reach $\approx$ half ionized before rays begin merging.  Since rays are rank-ordered for by their photon counts to determine if they will be exempt from merging, only rays near the edges of ionized regions will be merged in this case.  For $N_{\rm ex} = 4$, rays begin merging when the box is only $1/8$th ionized, and by the end of the simulation most rays are eligible for merging.  Importantly, the $N_{\rm ex} = 4$ ran $\approx 7$ times faster and requires a factor of $\approx 2$ less memory than the No Merging case, reflecting the significant computational savings afforded by merging.  

\begin{figure}
    \centering
    \includegraphics[scale=0.3]{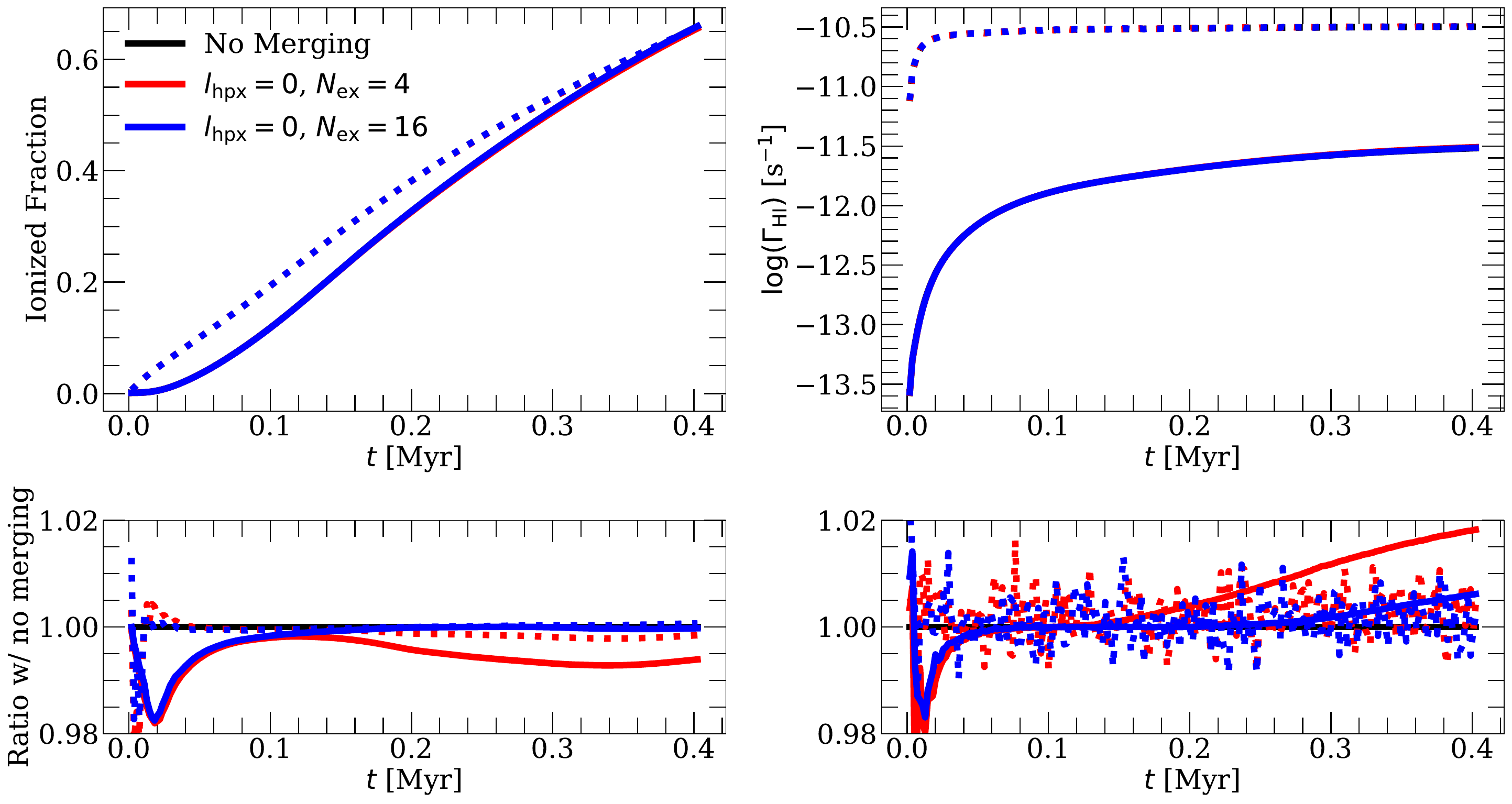}
    \caption{Mean ionized fraction and $\Gamma_{\rm HI}$ in versions of Test \#4 with different merging parameters.  Solid (dotted) curves denote volume (mass)-weighted averages.  The black curve has no merging, while for the red curve we allow for aggressive ray merging that speeds up the calculation by a factor of $\approx 7$.  The blue curve is an intermediate case.  The bottom panels show the ratios of each quantity with the no-merging run. We find $1-2\%$ level agreement at all times in these quantities for very different merging scenarios. }
    \label{fig:merging_global_test}
\end{figure}

Figure~\ref{fig:merging_global_test} shows the global ionized fraction vs. time (top left) and mean photo-ionization rate $\Gamma_{\rm HI}$ (top right) for these three runs.  The solid and dashed curves denote volume and mass-weighted averages.  The bottom panels show the ratio of the $N_{\rm ex} = 4$ and $16$ cases with the No Merging case.  We see that the curves mostly overlap in the top panels.  The bottom panels show that the differences between the averaged quantities are at most $1-2\%$ in all cases, demonstrating that merging has a small effect on globally averaged quantities\footnote{The $\sim 2\%$ differences at the beginning of the sims may arise from different choices for random rotations of the rays in the ray casting scheme - see \S\ref{subsec:raycasting}.  }.

\begin{figure*}
    \centering
    \includegraphics[scale=0.39]{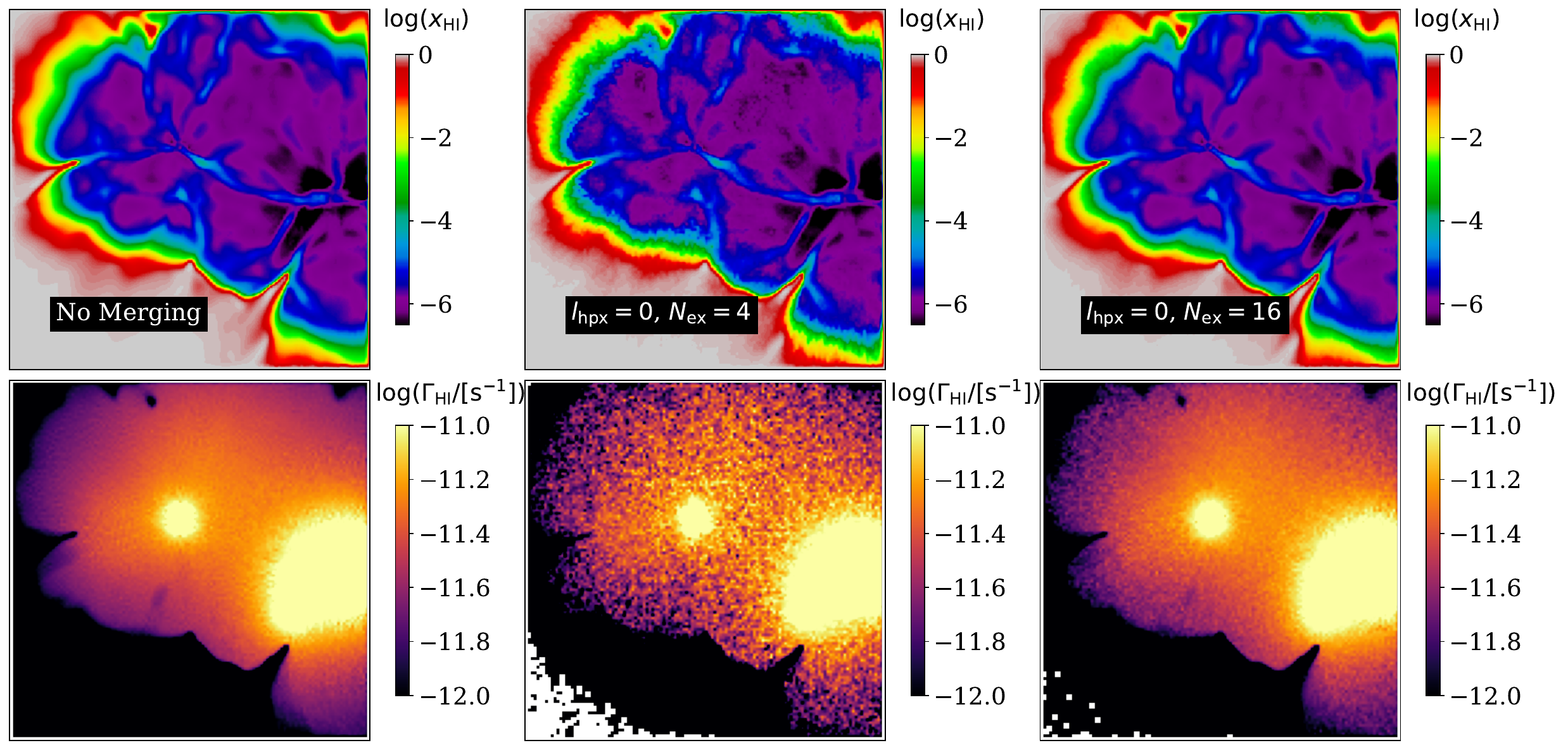}
    \caption{Visualization of the HI fraction (top) and $\Gamma_{\rm HI}$ (bottom) for the merging scenarios shown in Figure~\ref{fig:merging_global_test}.  In the ``no merging'' case, both fields are smooth.  The second case displays significant shot noise in the $\Gamma_{\rm HI}$ field due to the sparsity of rays, which is also visible (to a lesser degree) in the $x_{\rm HI}$ field.  The third case shows intermediate results, with a small amount of noise visible in the $\Gamma_{\rm HI}$ near the edges of ionized regions.  Crucially, the large-scale fluctuations in $\Gamma_{\rm HI}$ and $x_{\rm HI}$ and the structure of the ionized bubble remain similar to the no merging case even with aggressive merging, suggesting that these large-scale features are mostly insensitive to small-scale noise caused by merging.  }
    \label{fig:cosmo_HI_vis_merging}
\end{figure*}

We show slices through the $x_{\rm HI}$ (top) and $\Gamma_{\rm HI}$ (bottom) fields in Figure~\ref{fig:cosmo_HI_vis_merging}, with the No Merging, $N_{\rm ex} = 4$, and $N_{\rm ex} = 16$ models shown from left to right.  The No Merging and $N_{\rm ex} = 16$ cases have similar $x_{\rm HI}$ fields.  A modest amount of noise is seen in the $\Gamma_{\rm HI}$ field for the latter, particularly  near the edges of ionized regions.  In the $N_{\rm ex} = 4$ case, however, the $\Gamma_{\rm HI}$ field is quite noisy, and some of the noise is visible in the $x_{\rm HI}$ field.  This is primarily shot noise caused by the relatively small number of rays (and independent directions) being tracked.  Importantly, the structure of the ionized bubble, and the large-scale fluctuations in $x_{\rm HI}$ and $\Gamma_{\rm HI}$, are mostly unaffected by the shot noise even for $N_{\rm ex} = 4$.  This demonstrates that over many time steps, the effects of this noise average out and have little effect on the growth of ionized regions or the large-scale features of the radiation field.  The insensitivity of large-scale gas properties to this noise is encouraging, since in many cosmological contexts only the large-scale features of the radiation field are important for modeling observables (e.g. the 21 cm signal~\citep{Mesinger2007} and large-scale fluctuations in the Ly$\alpha$ forest~\citep{Bosman2021}).  

\subsection{Test against \textsc{RadHydro}}
\label{subsec:hy_cosmo}

In this section, we test \textsc{FlexRT} against the \textsc{RadHydro} code of Ref.~\cite{Trac2007}, which has a ray tracing algorithm similar to FlexRT (including merging).  We use a box with length $L = 200$ $h^{-1}$Mpc, large-enough to capture realistically the clustering of dark matter halos.  We first run an N-body simulation with $N = 3600^3$ particles to get dark matter halos with masses $\geq 3 \times 10^9$ $h^{-1}M_{\odot}$, which become the sources for our RT simulation.  Next, we ran a hydrodynamics simulation using the same initial density fluctuations and $N = 1024^3$ gas cells, and smoothed the gas density field to $N = 200^3$ for the RT calculation.  We use a single snapshot of the halo field at $z = 7$ for the simulation, and post-process the density field at that redshift with an ionizing emissivity per unit volume\footnote{Halos are assigned UV luminosities by abundance-matching to the UV luminosity function of Ref.~\cite{Finkelstein2019}, and emissivities of individual halos are proportional to their luminosities.  } of $\dot{N}_{\gamma} = 7.96 \times 10^{50}$ ph/s/cMpc$^{3}$.  For simplicity, we set the gas temperature to a constant $T = 10^4$ K.  Our splitting parameters for our fiducial \textsc{FlexRT} run are $l_{\rm source} = 1$ and $N_{\rm ray}^{\min} = 2$, and our merging parameters are $l_{\rm merge} = 1$ and $N_{\rm ex} = 16$.  These choices give a reasonably close match to the default splitting/merging scheme parameters used in \textsc{RadHydro} (see Ref.~\cite{Trac2007} for details).  We have also run a ``High-res'' simulation with \textsc{FlexRT} that has $N_{\rm ex} = 128$.  

\begin{figure}
    \centering
    \includegraphics[scale=0.36]{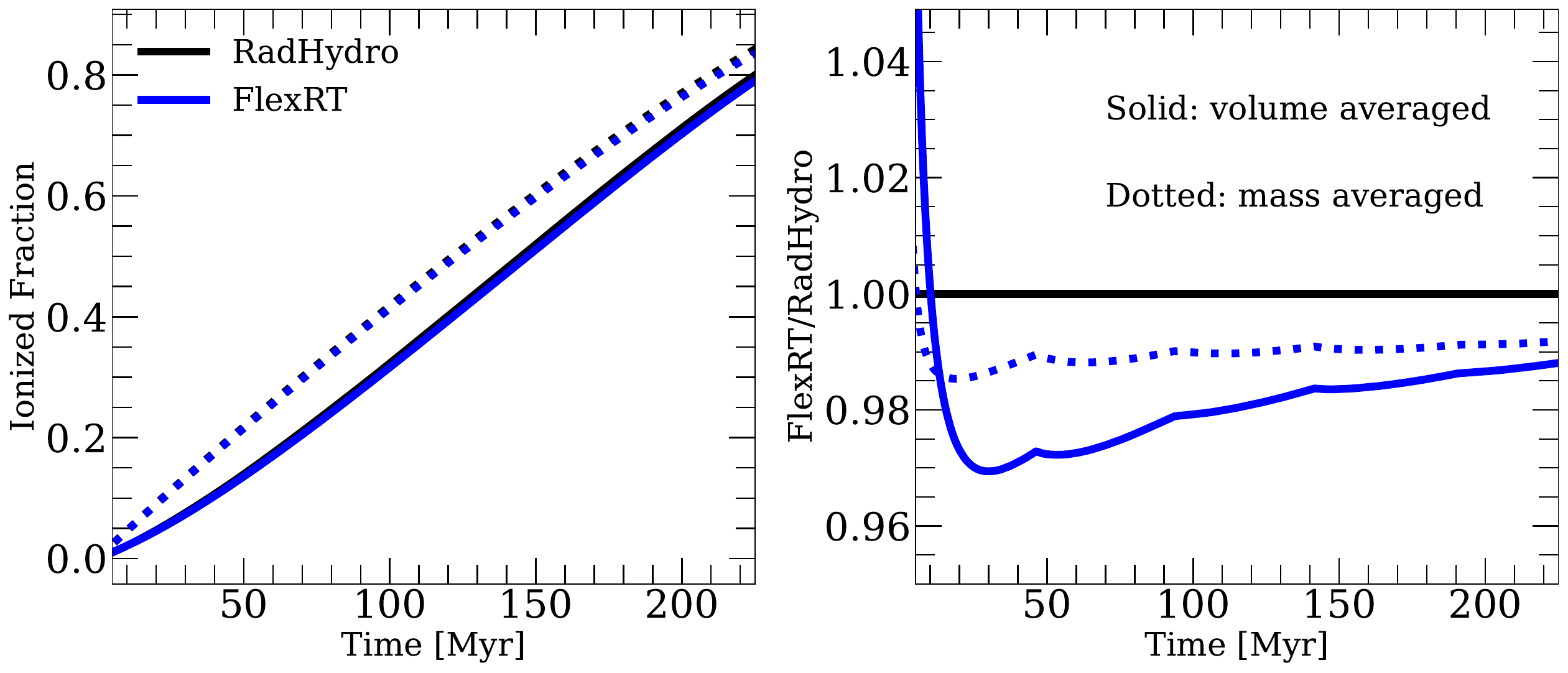}
    \caption{Evolution of the ionized fraction with time in our comparison test of \textsc{FlexRT} (black) and \textsc{RadHydro} (blue).  Solid (dashed) curves denote volume (mass)-weighted averages.  The bottom panel shows the ratio of the \textsc{FlexRT} and \textsc{RadHydro} results.  The ionization history in \textsc{FlexRT} lags $2-3\%$ percent behind that of \textsc{RadHydro}, possibly due to differences in the codes' ionization solvers.  }
    \label{fig:ion_history}
\end{figure}

\begin{figure*}
    \centering
    \includegraphics[scale=0.26]{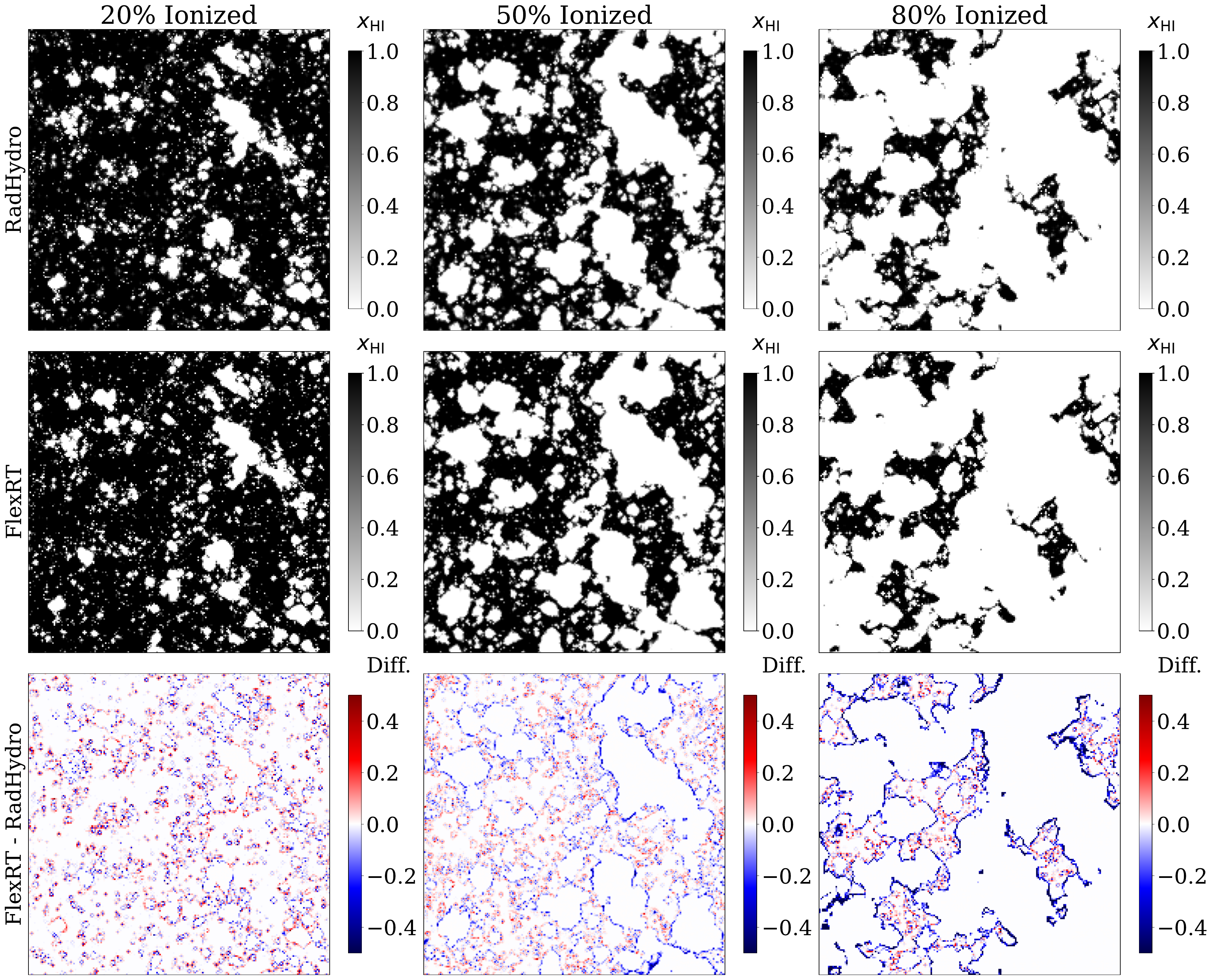}
    \caption{Slices through the neutral fraction field from \textsc{RadHydro} (top) and \textsc{FlexRT} (middle) at $20\%$, $50\%$ and $80\%$ volume ionized fractions (left to right).  Black (white) regions are neutral (ionized).  The bottom row shows the difference between the two.  Blue (red) cells have higher neutral fractions in \textsc{RadHydro} (\textsc{FlexRT}).  At fixed ionized fraction, the large-scale features of the ionization field are in very good agreement.  Inspection of the difference maps reveals slight morphological differences between the codes, especially at $80\%$ ionized - see text for details.  }
    \label{fig:ion_vis}
\end{figure*}

\begin{figure*}
    \centering
    \includegraphics[scale=0.21]{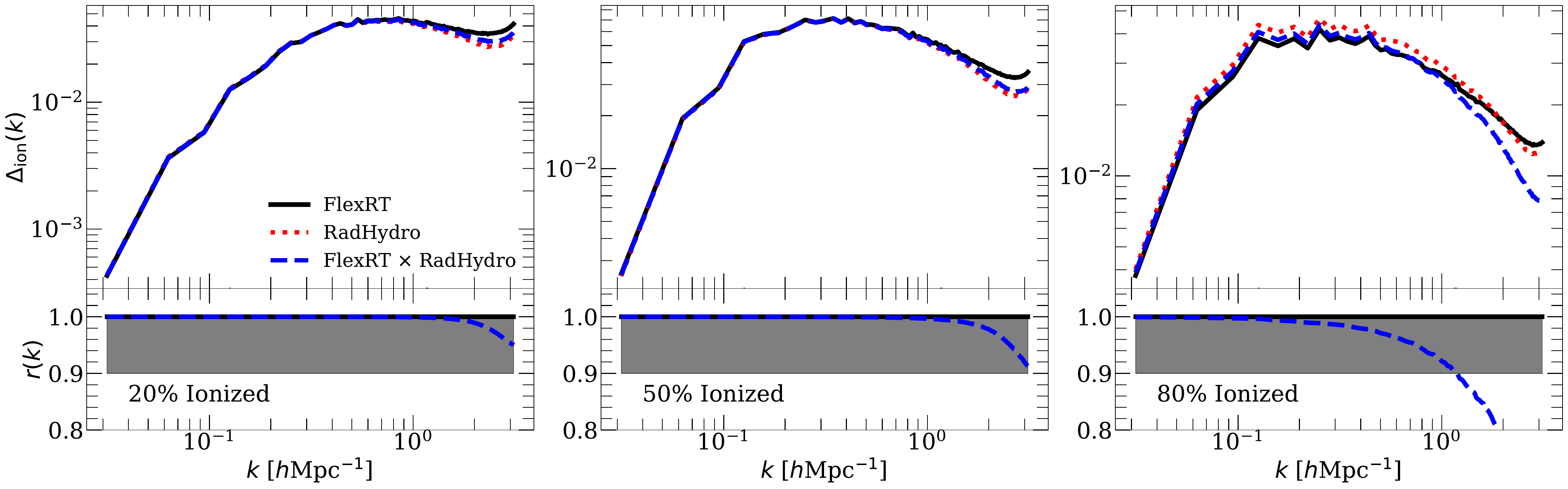}
    \caption{Top: auto power spectrum of the ionization field $\Delta_{\rm ion}(k)$ for \textsc{FlexRT} and \textsc{RadHydro} (solid black and dotted red), and their cross-correlation (dashed blue).  Bottom: their cross-correlation coefficient, $r(k)$ (Eq.~\ref{eq:cross_corr_coeff}).  From left to right, the columns show results at $20\%$, $50\%$, and $80\%$ ionized.  At $20\%$ and $50\%$ ionized, the power spectra are almost indistinguishable, with $r(k)$ differing from unity only close to the Nyquist wavenumber ($k_{\rm nyq} = \pi / \Delta x_{\rm RT} \approx 3$ $h$Mpc$^{-1}$).  At $80\%$, the codes are still in good agreement, but \textsc{RadHydro} has slightly more ionization power at intermediate scales.  This is probably due to differences in how the codes handle ray merging as the ionized fraction nears unity.  }
    \label{fig:ion_power}
\end{figure*}

We show the ionized fraction as a function of time in the top panel of Figure~\ref{fig:ion_history} for \textsc{RadHydro} (black) and \textsc{FlexRT} (blue), and their ratio in the bottom panel.  The solid (dashed) lines denote volume (mass)-weighted averages.  The differences between \textsc{FlexRT} and \textsc{RadHydro} are a few percent or less.  The slight disagreement may arise from the codes' different ionization solvers.  Specifically, \textsc{RadHydro} assumes the optically thin limit of Eq.~\ref{eq:gamma_initial_standard} and does not account for evolution of the neutral fraction during a time step (Eq.~\ref{eq:ion_balance_eqn}-\ref{eq:avg_xi}) when calculating the opacities of cells.  This is an important difference for our test setup because the large RT cells ($1$ $h^{-1}$Mpc) result in large opacities in partially neutral cells ($\tau >> 1$) and large time steps ($\Delta t = 0.6$ Myr). Another, related effect could be that of HealPix structural artifacts, discussed in \S\ref{subsec:raycasting}.  \textsc{RadHydro} does not randomly rotate the HealPix spheres used to cast rays on every time step\footnote{Instead, it assigns each source a single set of random rotation angles, which are used throughout the simulation.  }.  This could lead to differences in the recombination rate around isolated sources early in the simulation before ionized regions overlap.  

Figure~\ref{fig:ion_vis} visualizes the ionization field for \textsc{RadHydro} (top) and \textsc{FlexRT} (middle) at volume-averaged ionized fractions of $20\%$, $50\%$, and $80\%$ (left to right).  White (black) regions denote ionized (neutral) gas.  The bottom panels show the linear differences between the codes, with blue (red) regions being more neutral in \textsc{RadHydro} (\textsc{FlexRT}).  We find excellent agreement in the morphology of ionized regions at all three ionized fractions.  Indeed, the maps are so similar that we must refer to the difference maps to easily detect differences.  At $20\%$ ionized, the largest ionized bubbles are slightly smaller (more neutral) in \textsc{FlexRT}, as indicated by their mostly red boundaries.  Smaller ionized bubbles, on the other hand are more neutral in \textsc{RadHydro} (that is, they have blue boundaries).  The opposite is true in the other two maps.  The prominent blue boundaries of neutral islands at $80\%$ ionized reflect that the two maps are not at exactly the same ionized fraction ($80\%$ vs. $82\%$ for \textsc{RadHydro} and FlexRT, respectively).  These slight, small-scale morphological differences may arise from differences in the codes' ionization solvers, and/or differences in ray splitting/merging.  

\begin{figure*}
    \centering
    \includegraphics[scale=0.225]{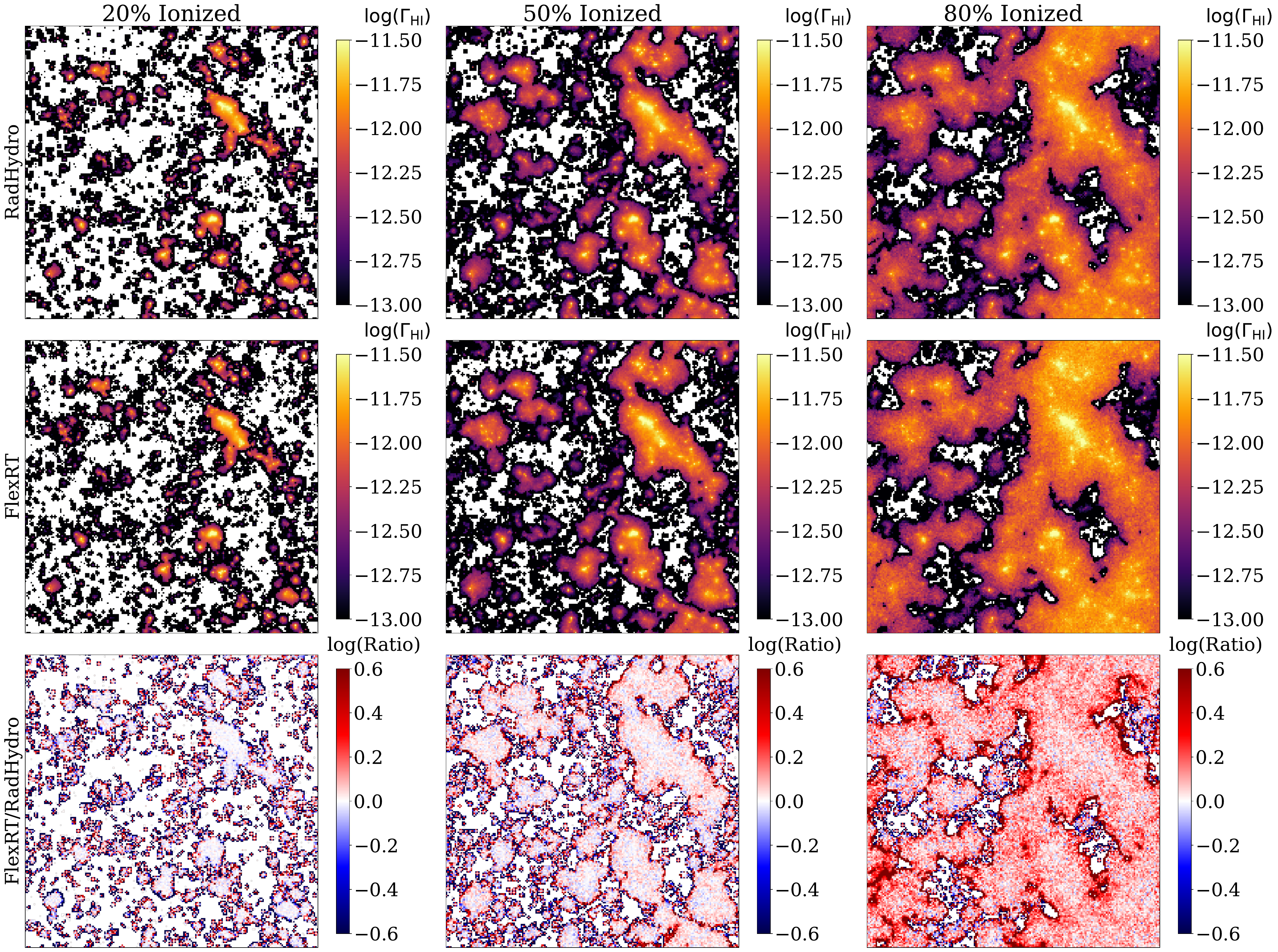}
    \caption{Visualization of the $\Gamma_{\rm HI}$ fields in \textsc{RadHydro} and \textsc{FlexRT}, in the same format as Figure~\ref{fig:ion_vis}.  The bottom panel shows the ratio (logarithmic difference) in $\Gamma_{\rm HI}$.  Note that blue (red) regions have higher $\Gamma_{\rm HI}$ in \textsc{RadHydro} (\textsc{FlexRT}).  At $20\%$ ionized, the $\Gamma_{\rm HI}$ fluctuations almost completely trace the ionization fluctuations.  At $50\%$ ionized, some additional structure within ionized regions becomes evident, tracing the clustering of the sources.  \textsc{RadHydro} and \textsc{FlexRT} are indistinguishable in these cases.  At $80\%$ ionized both $\Gamma_{\rm HI}$ fields start to show small-scale noise, arising from merging (see Figure~\ref{fig:merging_vis} and discussion).  In the difference map, we also see some differences on larger scales, with \textsc{FlexRT} having higher $\Gamma_{\rm HI}$, especially near the edges of ionized regions.  }
    \label{fig:gamma_vis}
\end{figure*}

\begin{figure*}
    \centering
    \includegraphics[scale=0.21]{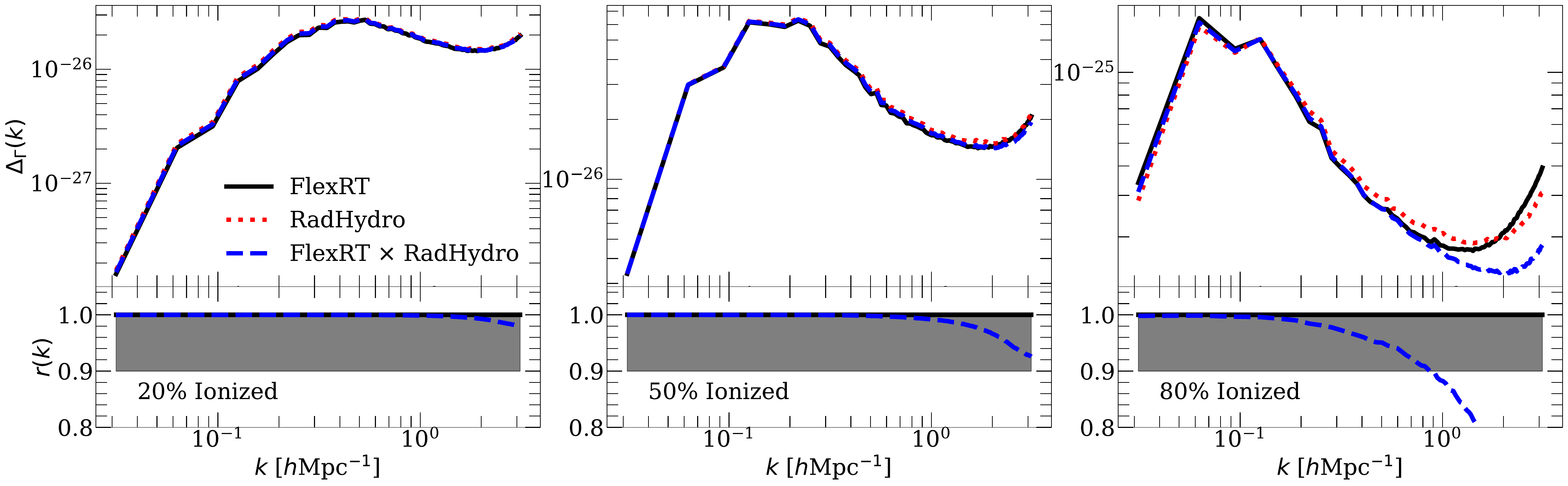}
    \caption{Power spectra of $\Gamma_{\rm HI}$, $\Delta_{\Gamma}$, in the same format as Figure~\ref{fig:ion_power}.  At $20\%$ ionized, $\Delta_{\Gamma}$ has nearly the same shape as $\Delta_{\rm ion}$, and the codes match just as well.  At $50\%$ ionized, $\Delta_{\Gamma}$ peaks at a smaller $k$ than $\Delta_{\rm ion}$ due to fluctuations within large ionized bubbles, and the codes remain in tight agreement.  At $80\%$ ionized, we see a significant up-turn at large $k$ in both codes due to shot noise from merging, which is slightly more significant in \textsc{FlexRT}.  \textsc{FlexRT} also has slightly less power at $0.3 < k/[{\rm hMpc^{-1}}] < 1$. The $r(k)$ is slightly worse than for the ionization field (Figure~\ref{fig:ion_power}).  }
    \label{fig:gamma_power}
\end{figure*}

Figure~\ref{fig:ion_power} quantifies the agreement in the ionization field.  The top row shows the dimensionless power spectrum of the ionization field fluctuations vs. $k$ for the same ionized fractions in Figure~\ref{fig:ion_vis}.  The black solid, red dotted, and blue dashed curves show the auto power for \textsc{FlexRT} and \textsc{RadHydro}, and their cross-power, respectively.  In the bottom panels, we show the correlation coefficient, defined as 
\begin{equation}
    \label{eq:cross_corr_coeff}
    r(k) = \frac{P_{\rm \textsc{FlexRT},\textsc{RadHydro}}(k)}{\sqrt{P_{\rm \textsc{FlexRT}}(k) P_{\rm \textsc{RadHydro}}(k)}}
\end{equation}
where the numerator is the cross-power and the denominator contains the auto terms.  For perfectly correlated (un-correlated, anti-correlated) fields, $r(k) = 1$ ($0$, $-1$).  The shaded region in the bottom panel denotes a $10\%$ deviation from unity.  At $20\%$ and $50\%$ ionized, the \textsc{FlexRT} and \textsc{RadHydro} ionized fields are almost perfectly correlated at all scales, with a $5-10\%$ deviation emerging only near the Nyquist wavenumber ($\approx 3$ $h$Mpc$^{-1}$).  At $80\%$ ionized, $5-10\%$ deviations from $r = 1$ are evident at $k > 0.3$ $h$Mpc$^{-1}$. \textsc{RadHydro} also has slightly more power on nearly all scales.  These differences are likely due to differences in how rays are merged (see subsequent discussion) and the slight mis-match in the ionized fractions of the $80\%$ ionized outputs. 

We next consider the large-scale fluctuations in the $\Gamma_{\rm HI}$ field, which are important to capture correctly for a number of reionization applications (e.g. the Ly$\alpha$ forest,~\citep{Davies2016,Bosman2021}).  Figure~\ref{fig:gamma_vis} visualizes the fluctuations in $\Gamma_{\rm HI}$ in the same format as the ionization fluctuations in Figure~\ref{fig:ion_vis}.  In this case, the bottom row shows the ratio (logarithmic difference) of the fields.  The locations of the ionizing sources, near the centers of ionized bubbles, can be identified by the bright spots in the $\Gamma_{\rm HI}$ maps.  At $20\%$ ionized, the $\Gamma_{\rm HI}$ fluctuations closely trace the ionization fluctuations.  At $50\%$ ionized, some additional structure is visible within ionized regions, which traces the spatial distribution of the ionizing sources.  In both cases, the codes are in good agreement.  
At $80\%$ ionized, some differences are evident.  First, both codes begin to display noise in the $\Gamma_{\rm HI}$ field, which arises from merging (see Figure~\ref{fig:merging_vis} and discussion).  The noise is slightly more noticeable in \textsc{FlexRT} than \textsc{RadHydro}, suggesting that \textsc{FlexRT} is tracking fewer rays\footnote{Recall from Figure~\ref{fig:ray_count_merging_example} that the number of rays in FlexRT does not saturate quite to $N_{\rm ray}^{\max}$, whereas in \textsc{RadHydro} it does.  This may account for the slightly higher noise in FlexRT}. However, we also see differences on larger scales in this comparison.  The ratio map shows that ionized regions generally have higher $\Gamma_{\rm HI}$ in FlexRT, especially at the edges of ionized bubbles.  This could be partially due to the slight mismatch in neutral fractions mentioned earlier.  It also may be that the merging of rays is beginning to affect large-scale $\Gamma_{\rm HI}$ fluctuations.  Recently, Ref.~\cite{Wu2021} argued that $\Gamma_{\rm HI}$ fluctuations on scales smaller than the ionizing photon mean free path could be affected by differences in the angular resolution of the radiation field (see their Fig. 3 and related discussion).  Since FlexRT and \textsc{RadHydro} do not parameterize or treat merging in exactly the same way, large-scale fluctuations in the codes might be affected differently.  

Figure~\ref{fig:gamma_power} compares the power spectrum of $\Gamma_{\rm HI}$, $\Delta_{\Gamma}(k)$, in the same format as Figure~\ref{fig:ion_power}.  At $20\%$ ionized, $\Delta_{\Gamma}(k)$ has almost the same shape as $\Delta_{\rm ion}(k)$, since the early fluctuations in $\Gamma_{\rm HI}$ are dominated by fluctuations in the ionization field.  Unsurprisingly, \textsc{FlexRT} and \textsc{RadHydro} agree very well, as they do in Figure~\ref{fig:ion_power}.  At $50\%$ ionized, the shape of $\Delta_{\Gamma}(k)$ begins to differ from that of $\Delta_{\rm ion}(k)$, peaking at slightly smaller $k$.  This is because the largest ionized bubbles also have the brightest sources, and hence the largest $\Gamma_{\rm HI}$.  This enhances large-scale power in the $\Gamma_{\rm HI}$ field relative to the ionization field.  Again, the level of agreement between \textsc{FlexRT} and \textsc{RadHydro} is similar to that in Figure~\ref{fig:ion_power}, showing that \textsc{FlexRT} is capturing the large-scale $\Gamma_{\rm HI}$ fluctuations correctly.  

\begin{figure}
    \centering
    \includegraphics[scale=0.25]{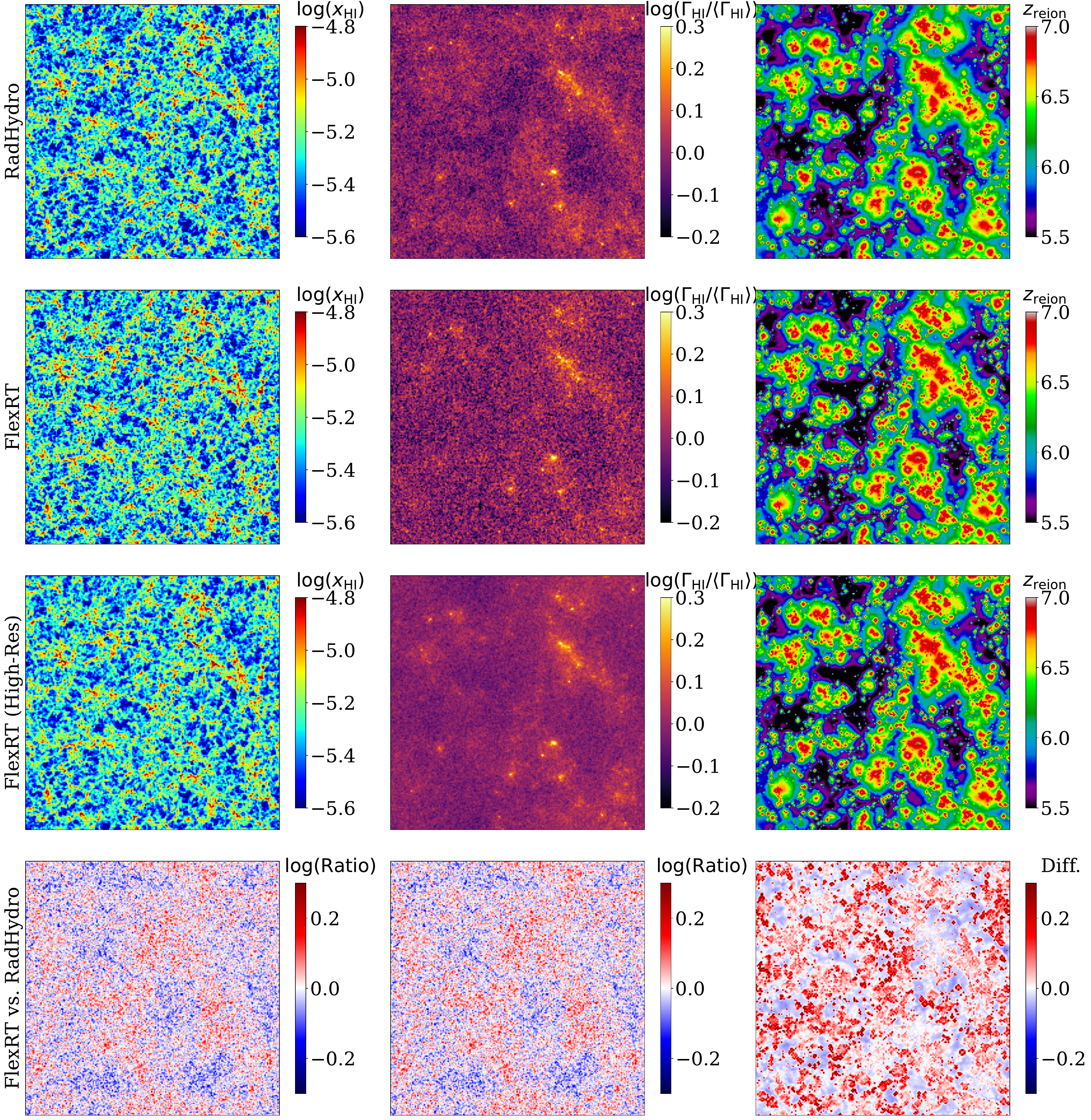}
    \caption{Visualization of gas properties after the completion of reionization ($100\%$ ionized).  From left to right, we show the residual $x_{\rm HI}$, $\Gamma_{\rm HI}$, and the reionization redshift ($z_{\rm reion}$) field.  The third row shows a high-resolution version of the FlexRT run, with $N_{\rm ex} = 128$.  The bottom row shows ratio maps for the $x_{\rm HI}$ and $\Gamma_{\rm HI}$ fields and a difference map for the $z_{\rm reion}$ fields (using the high-resolution FlexRT run).  Fluctuations in $x_{\rm HI}$ are dominated by density, so they agree well between the codes.  The $\Gamma_{\rm HI}$ fields are similar, but with shot noise now conspicuous.  There is more (less) noise in the fiducial (high-res) FlexRT run than in \textsc{RadHydro}.  We also see that FlexRT shows less significant large-scale fluctuations in its $\Gamma_{\rm HI}$ field (for both resolutions).  The $z_{\rm reion}$ fields are in excellent agreement, showing their relative insensitivity to the effects of merging.  In addition to small-scale noise, some large-scale features are notable in the ratio maps of $\Gamma_{\rm HI}$.  The $z_{\rm reion}$ difference map betrays minor differences in ionized morphology and reionization history.  Small bubbles embedded in late-ionizing regions tend to ionize earlier in \textsc{FlexRT}, and the surrounding voids ionize earlier in \textsc{RadHydro}.  }
    \label{fig:100_vis}
\end{figure}

At $80\%$ ionized, the peak in $\Delta_{\Gamma}(k)$ has moved to yet smaller $k$, approaching the box scale.  The power also drops off more rapidly at small scales, since ionized bubbles have begun overlapping and the UVB has started to homogenize.  The up-turn at $k > 1$ $h$Mpc$^{-1}$ is due to shot noise in the radiation field due to ray merging.  This noise is visible in the $\Gamma_{\rm HI}$ maps for both codes, but is slightly more prominent in \textsc{FlexRT}.  From the bottom panel, we see that the correlation between the codes is slightly worse than it was for the ionization maps.  The drop-off in $r(k)$ at large $k$ coincides with the up-turn in power at large $k$, reflecting the ``ray noise'' discussed previously.  \textsc{FlexRT} also has slightly less power at $0.3 < k < 1$ $h$Mpc$^{-1}$, reflecting the differences at large-scales discussed above.  

Figure~\ref{fig:100_vis} shows the residual HI fraction (left), $\Gamma_{\rm HI}$ (middle), and the reionization redshift ($z_{\rm reion}$) field (right) just after reionization has finished ($100\%$ ionized).  The third row shows a higher-resolution version of the FlexRT test, with $N_{\rm ex} = 128$.  The bottom row shows ratio maps for $x_{\rm HI}$ and $\Gamma_{\rm HI}$, and a difference map for $z_{\rm reion}$, for the fiducial resolution \textsc{FlexRT} run and \textsc{RadHydro}.  We show $\Gamma_{\rm HI}$ normalized by its mean value to more easily compare spatial fluctuations.  Fluctuations in the HI fraction on intermediate scales are mainly set by the density field ($x_{\rm HI} \propto \Delta$), which is the same in both codes, so we see good agreement in the left panels.  However, the large-scale features in $\Gamma_{\rm HI}$ differ noticeably between FlexRT and \textsc{RadHydro}, as highlighted by the large-scale features visible in the ratio maps.  We have checked, by running tests in boxes with fewer RT cells and much higher ray tracing resolution, that these large scale differences arise from merging - that is, the codes converge to the same answer in the no-merging limit.

\begin{figure}
    \centering
    \includegraphics[scale=0.30]{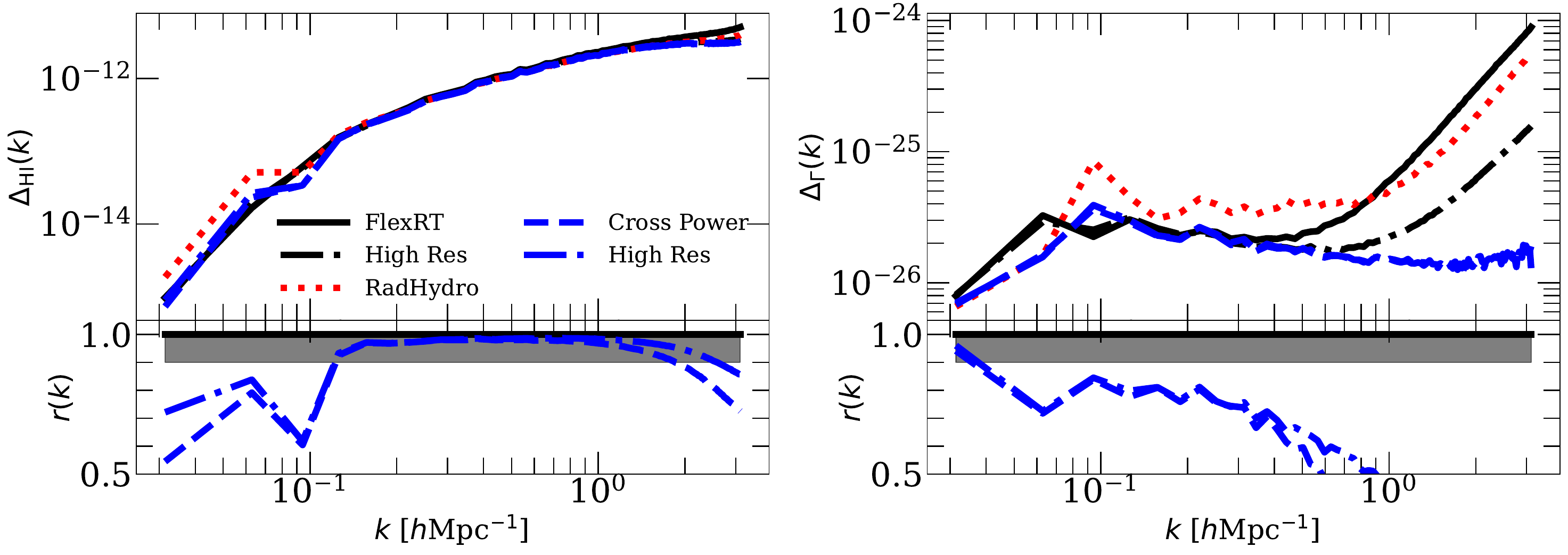}
\caption{Power spectrum and cross-correlation of the residual $x_{\rm HI}$ (left) and $\Gamma_{\rm HI}$ (right) fields after the end of reionization.  The dot-dashed curves show results for the high-resolution FlexRT run.  The bottom panels show $r(k)$.  Here, we see significant differences, especially in the $\Gamma_{\rm HI}$ power.  There is more large-scale power at $k < 0.5 {\rm hMpc^{-1}}$ in \textsc{RadHydro}, which affects fluctuations in the residual $x_{\rm HI}$ at the largest scales.  The higher resolution FlexRT run matches well with the fiducial case at large scales, just with less small-scale noise power.  }
    \label{fig:overlap_power}
\end{figure}

Figure~\ref{fig:overlap_power} shows the power spectrum of the residual HI fraction $\Delta_{\rm HI}(k)$ (left) and $\Delta_{\Gamma}(k)$ (left) at $100\%$ ionized.  The dot-dashed lines show results for the high-resolution FlexRT simulation.  The bottom row shows $r(k)$.  At intermediate scales ($0.1 < k /[h{\rm Mpc}^{-1} < 1]$, where density fluctuations dominate the fluctuations in $x_{\rm HI}$, we find good agreement in $\Delta_{\rm HI}(k)$.  However, $\Delta_{\Gamma}(k)$ is appreciably different on all scales, with $r(k) \lesssim 0.8$.  \textsc{RadHydro} has more power than both of the FlexRT runs at $k < 0.5 {\rm hMpc^{-1}}$.  The up-turn at small $k$ in both codes is due to shot noise, which is more (less) prominent in the fiducial (high-resolution) \textsc{FlexRT} run than in \textsc{RadHydro}.  Although the precise reason for the differences at larger scales is unclear, we are sure that they arise in some way due to merging.  We have confirmed that in tests without any merging, FlexRT and \textsc{RadHydro} give nearly identical $\Delta_{\Gamma}$.   

The fact that different ray merging schemes can result in differing large-scale $\Gamma_{\rm HI}$ fluctuations after reionization ends is of potential concern for applications that are sensitive to these fluctuations.   These include the Ly$\alpha$ forest of high-redshift quasars~\citep{Davies2016,Nasir2020}.  As we showed in Figure~\ref{fig:merging_vis}, this noise can be mitigated by increasing the number of directions tracked (parameterized by $l_{\rm pix}$) and exempting more rays from merging (controlled by $N_{\rm ex}$).  Adjusting these parameters will help to mitigate the ray noise, at the cost of computational efficiency.  Fortunately, we have found in previous work~\citep[][]{Cain2023,Gangolli2024} that the large-scale fluctuations in the forest are reasonably insensitive to the choice of merging parameters, $l_{\rm merge}$ and $N_{\rm ex}$.  Ref.~\cite[][]{Gangolli2024} rigorously quantified this in the context of the IGM opacity/density relation in their Appendix A.  The ability to perform such  tests easily highlights the flexibility of \textsc{FlexRT}.  

\section{Conclusions}
\label{sec:conc}


We have introduced a new ray-tracing radiative transfer code, \textsc{FlexRT}, which is aimed at enabling fast, efficient parameter space studies of reionization.  \textsc{FlexRT} combines adaptive ray tracing with a highly flexible treatment of the intergalactic ionizing opacity. This not only gives the user a great deal
of control over how the IGM is modeled; it also provides a way to reduce the computational cost of a \textsc{FlexRT} simulation by orders of magnitude while still modeling the small-scale physics of the IGM. Alternatively, the user may tune up the angular and spatial resolution of the algorithm to run a more traditional reionization simulation.  

We have validated \textsc{FlexRT} against a number of standard test problems common in the literature.  We have shown that both the ``Standard'' and ``Generalized Opacity'' treatments of the ionizing opacity in \textsc{FlexRT} are in good agreement with existing codes, given an appropriate prescription for the opacity in the latter.  We have run additional tests to validate the accuracy of \textsc{FlexRT}'s adaptive ray tracing scheme.  We find that even when modeling the radiation field with low angular resolution, \textsc{FlexRT} accurately captures the large-scale growth of ionized regions and the ionizing background fluctuations within them during reionization.  Our main conclusions are summarized below: 

\begin{itemize}

    \item We have run the static density field tests described in Paper I of the Cosmological Radiative Transfer Comparison Project~\citep[CRTCP,][]{Iliev2006}.  In Test \#1, which models the growth of a spherically symmetric HII region around a point source in a uniform density, isothermal medium, we found excellent agreement between \textsc{FlexRT} and CRTCP results.  We find few-percent level agreement between standard and Generalized Opacity mode in the I-front position, with the latter $< 1\%$ away from the analytic prediction.  
 
    \item In Test \#2, which includes temperature evolution, we find slightly larger ($\sim 10\%$-level) differences in the temperature profiles in the HII regions, and larger differences in the pre-heating of neutral gas ahead of the I-front.  However, our results remain broadly consistent with the CRTCP results.  

    \item Test \#3 considers a case where a plane-parallel I-front is shadowed by a dense, partially self-shielding clump.  We find good agreement between both modes of \textsc{FlexRT} and CRTCP codes in the $x_{\rm HI}$ profile in the clump behind the I-front.  Generalized Opacity mode fails to capture the ionization and temperature structure in the neutral part of the clump because it does not account for pre-heating, while Standard mode agrees qualitatively with most of the CRTCP codes.  

    \item Finally, Test \#4 considers a scenario with multiple sources in a cosmological density field.  In this test, we considered only Standard mode, since results from the IFT code (to which Generalized Opacity mode is most comparable) were unavailable.  We find broad agreement between Standard mode and the CRTCP codes, but with some differences.  The main differences are the use of the finite speed of light in \textsc{FlexRT}, and the pre-heating of neutral gas by hard ionizing photons.  We also compared \textsc{FlexRT} to RAMSES-RT, which uses the finite speed of light and agrees much better as a result.  

    \item Using a cosmological reionization simulation, we demonstrated the regime in which the moving-screen I-front approximation in Generalized Opacity mode approaches the results of Standard mode.  We found that when the RT cell size, $\Delta x_{\rm RT}$, is much larger than typical I-front widths, ionization maps from the two modes are almost indistinguishable.   Standard mode should be used for problems in which the I-fronts are resolved
and their detailed structure is important. 

    \item We validated the ray merging scheme in \textsc{FlexRT} by repeating CRTCP Test \#4 with several sets of merging parameters.  We found that lowering the angular resolution of the ray tracing significantly increases shot noise in the radiation field.  However, the average ionization fraction and $\Gamma_{\rm HI}$, and their large-scale fluctuations, are affected at the $1-2\%$ level or less.   

    \item Finally, we tested \textsc{FlexRT} against the \textsc{RadHydro} code of Ref.~\cite{Trac2007}, which uses a similar ray tracing algorithm, in a cosmological reionization simulation.  We find good agreement in the morphology of ionized regions and the large-scale fluctuations in $\Gamma_{\rm HI}$ throughout the bulk of reionization.  After the end of reionization, we find modest differences in the in the large-scale $\Gamma_{\rm HI}$ fluctuations, with $r(k) < 0.7$ at $k > 0.3$ $h$Mpc$^{-1}$.  The differences arise due to effects of ray merging, which can be mitigated by adjusting the parameters that control the angular resolution of the radiation field.  

\end{itemize}

The true utility of the Generalized Opacity formalism presented here is to allow for accurate treatments of the IGM opacity when relevant physics (e.g. small-scale clumping and self-shielding) is un-resolved.  In the second paper of this series, we will present and validate an improved version of the sub-grid IGM opacity model described in~\cite{Cain2021,Cain2022b,Cain2023}.

\acknowledgments

The authors thank Hy Trac for helpful comments on this manuscript and for providing the \textsc{\textsc{RadHydro}} code, which inspired much of the ray tracing machinery in \textsc{FlexRT}. The authors thank Joakim Rosdahl for providing the \textsc{RAMSES-RT} results for CRTCP Test \#4.  C.C. acknowledges support from the Beus Center for Cosmic Foundations while this work was being completed.  A.D. and C.C. were supported by NASA 19-ATP19-0191 and NSF AST-2045600. 

\bibliography{references.bib}

\providecommand{\href}[2]{#2}\begingroup\raggedright\begin{thebibliography}{10}

\bibitem{Atek2023}
H.~{Atek}, I.~{Labb{\'e}}, L.J.~{Furtak}, I.~{Chemerynska}, S.~{Fujimoto},
  D.J.~{Setton} et~al., \emph{{Most of the photons that reionized the Universe
  came from dwarf galaxies}},
  \href{https://doi.org/10.1038/s41586-024-07043-6}{\emph{\nat} {\bfseries 626}
  (2024) 975} [\href{https://arxiv.org/abs/2308.08540}{{\ttfamily
  2308.08540}}].

\bibitem{Cameron2023}
A.J.~{Cameron}, H.~{Katz}, C.~{Witten}, A.~{Saxena}, N.~{Laporte} and
  A.J.~{Bunker}, \emph{{Nebular dominated galaxies: insights into the stellar
  initial mass function at high redshift}},
  \href{https://doi.org/10.1093/mnras/stae1547}{\emph{\mnras} (2024) }
  [\href{https://arxiv.org/abs/2311.02051}{{\ttfamily 2311.02051}}].

\bibitem{Saxena2023}
A.~{Saxena}, A.J.~{Bunker}, G.C.~{Jones}, D.P.~{Stark}, A.J.~{Cameron},
  J.~{Witstok} et~al., \emph{{JADES: The production and escape of ionizing
  photons from faint Lyman-alpha emitters in the epoch of reionization}},
  \href{https://doi.org/10.1051/0004-6361/202347132}{\emph{\aap} {\bfseries
  684} (2024) A84} [\href{https://arxiv.org/abs/2306.04536}{{\ttfamily
  2306.04536}}].

\bibitem{Endsley2022b}
R.~{Endsley} and D.P.~{Stark}, \emph{{Strong Lyman-{\ensuremath{\alpha}}
  emission in an overdense region at z = 6.8: a very large (R 3 physical Mpc)
  ionized bubble in COSMOS?}},
  \href{https://doi.org/10.1093/mnras/stac524}{\emph{\mnras} {\bfseries 511}
  (2022) 6042} [\href{https://arxiv.org/abs/2112.14779}{{\ttfamily
  2112.14779}}].

\bibitem{Whitler2024}
L.~{Whitler}, D.P.~{Stark}, R.~{Endsley}, Z.~{Chen}, C.~{Mason}, M.W.~{Topping}
  et~al., \emph{{Insight from JWST/Near Infrared Camera into galaxy
  overdensities around bright Lyman-alpha emitters during reionization:
  implications for ionized bubbles at z 9}},
  \href{https://doi.org/10.1093/mnras/stae516}{\emph{\mnras} {\bfseries 529}
  (2024) 855} [\href{https://arxiv.org/abs/2305.16670}{{\ttfamily
  2305.16670}}].

\bibitem{Coulton2024}
W.R.~{Coulton}, T.~{Schutt}, A.S.~{Maniyar}, E.~{Schaan}, R.~{An}, Z.~{Atkins}
  et~al., \emph{{The Atacama Cosmology Telescope: Detection of Patchy Screening
  of the Cosmic Microwave Background}},
  \href{https://doi.org/10.48550/arXiv.2401.13033}{\emph{arXiv e-prints} (2024)
  arXiv:2401.13033} [\href{https://arxiv.org/abs/2401.13033}{{\ttfamily
  2401.13033}}].

\bibitem{Raghunathan2024}
S.~{Raghunathan}, P.A.R.~{Ade}, A.J.~{Anderson}, B.~{Ansarinejad},
  M.~{Archipley}, J.E.~{Austermann} et~al., \emph{{First Constraints on the
  Epoch of Reionization Using the non-Gaussianity of the Kinematic
  Sunyaev-Zel\{'\}dovich Effect from the South Pole Telescope and
  Herschel-SPIRE Observations}},
  \href{https://doi.org/10.48550/arXiv.2403.02337}{\emph{arXiv e-prints} (2024)
  arXiv:2403.02337} [\href{https://arxiv.org/abs/2403.02337}{{\ttfamily
  2403.02337}}].

\bibitem{2019JCAP...02..056A}
P.~{Ade}, J.~{Aguirre}, Z.~{Ahmed}, S.~{Aiola}, A.~{Ali}, D.~{Alonso} et~al.,
  \emph{{The Simons Observatory: science goals and forecasts}},
  \href{https://doi.org/10.1088/1475-7516/2019/02/056}{\emph{\jcap} {\bfseries
  2019} (2019) 056} [\href{https://arxiv.org/abs/1808.07445}{{\ttfamily
  1808.07445}}].

\bibitem{Jain2023}
D.~{Jain}, T.R.~{Choudhury}, S.~{Raghunathan} and S.~{Mukherjee},
  \emph{{Probing the physics of reionization using kinematic Sunyaev-Zeldovich
  power spectrum from current and upcoming cosmic microwave background
  surveys}}, \href{https://doi.org/10.1093/mnras/stae748}{\emph{\mnras}
  {\bfseries 530} (2024) 35}
  [\href{https://arxiv.org/abs/2311.00315}{{\ttfamily 2311.00315}}].

\bibitem{2017PASP..129d5001D}
D.R.~{DeBoer}, A.R.~{Parsons}, J.E.~{Aguirre}, P.~{Alexander}, Z.S.~{Ali},
  A.P.~{Beardsley} et~al., \emph{{Hydrogen Epoch of Reionization Array
  (HERA)}},
  \href{https://doi.org/10.1088/1538-3873/129/974/045001}{\emph{\pasp}
  {\bfseries 129} (2017) 045001}
  [\href{https://arxiv.org/abs/1606.07473}{{\ttfamily 1606.07473}}].

\bibitem{Koopmans2015}
L.~{Koopmans}, J.~{Pritchard}, G.~{Mellema}, J.~{Aguirre}, K.~{Ahn},
  R.~{Barkana} et~al., \emph{{The Cosmic Dawn and Epoch of Reionisation with
  SKA}},  in \emph{Advancing Astrophysics with the Square Kilometre Array
  (AASKA14)}, p.~1, Apr., 2015, \href{https://doi.org/10.22323/1.215.0001}{DOI}
  [\href{https://arxiv.org/abs/1505.07568}{{\ttfamily 1505.07568}}].

\bibitem{HERA2021b}
Z.~{Abdurashidova}, J.E.~{Aguirre}, P.~{Alexander}, Z.S.~{Ali}, Y.~{Balfour},
  R.~{Barkana} et~al., \emph{{HERA Phase I Limits on the Cosmic 21 cm Signal:
  Constraints on Astrophysics and Cosmology during the Epoch of Reionization}},
  \href{https://doi.org/10.3847/1538-4357/ac2ffc}{\emph{\apj} {\bfseries 924}
  (2022) 51} [\href{https://arxiv.org/abs/2108.07282}{{\ttfamily 2108.07282}}].

\bibitem{HERA2022}
{HERA Collaboration}, Z.~{Abdurashidova}, T.~{Adams}, J.E.~{Aguirre},
  P.~{Alexander}, Z.S.~{Ali} et~al., \emph{{Improved Constraints on the 21 cm
  EoR Power Spectrum and the X-Ray Heating of the IGM with HERA Phase I
  Observations}}, \href{https://doi.org/10.3847/1538-4357/acaf50}{\emph{\apj}
  {\bfseries 945} (2023) 124}
  [\href{https://arxiv.org/abs/2210.04912}{{\ttfamily 2210.04912}}].

\bibitem{Kulkarni2019}
G.~{Kulkarni}, L.C.~{Keating}, M.G.~{Haehnelt}, S.E.I.~{Bosman}, E.~{Puchwein},
  J.~{Chardin} et~al., \emph{{Large Ly {\ensuremath{\alpha}} opacity
  fluctuations and low CMB {\ensuremath{\tau}} in models of late reionization
  with large islands of neutral hydrogen extending to z < 5.5}},
  \href{https://doi.org/10.1093/mnrasl/slz025}{\emph{\mnras} {\bfseries 485}
  (2019) L24} [\href{https://arxiv.org/abs/1809.06374}{{\ttfamily
  1809.06374}}].

\bibitem{Keating2020}
L.C.~{Keating}, G.~{Kulkarni}, M.G.~{Haehnelt}, J.~{Chardin} and D.~{Aubert},
  \emph{{Constraining the second half of reionization with the Ly
  {\ensuremath{\beta}} forest}},
  \href{https://doi.org/10.1093/mnras/staa1909}{\emph{\mnras} {\bfseries 497}
  (2020) 906} [\href{https://arxiv.org/abs/1912.05582}{{\ttfamily
  1912.05582}}].

\bibitem{Nasir2020}
F.~Nasir and A.~D’Aloisio, \emph{Observing the tail of reionization: neutral
  islands in the z = 5.5 lyman-$\alpha$ forest},
  \href{https://doi.org/10.1093/mnras/staa894}{\emph{Monthly Notices of the
  Royal Astronomical Society} {\bfseries 494} (2020) 3080–3094}.

\bibitem{Bosman2021}
S.E.I.~{Bosman}, F.B.~{Davies}, G.D.~{Becker}, L.C.~{Keating}, R.L.~{Davies},
  Y.~{Zhu} et~al., \emph{{Hydrogen reionization ends by z = 5.3:
  Lyman-{\ensuremath{\alpha}} optical depth measured by the XQR-30 sample}},
  \href{https://doi.org/10.1093/mnras/stac1046}{\emph{\mnras} {\bfseries 514}
  (2022) 55} [\href{https://arxiv.org/abs/2108.03699}{{\ttfamily 2108.03699}}].

\bibitem{Zhu2021}
Y.~{Zhu}, G.D.~{Becker}, S.E.I.~{Bosman}, L.C.~{Keating}, H.M.~{Christenson},
  E.~{Ba{\~n}ados} et~al., \emph{{Chasing the Tail of Cosmic Reionization with
  Dark Gap Statistics in the Ly{\ensuremath{\alpha}} Forest over 5 < z < 6}},
  \href{https://doi.org/10.3847/1538-4357/ac26c2}{\emph{\apj} {\bfseries 923}
  (2021) 223} [\href{https://arxiv.org/abs/2109.06295}{{\ttfamily
  2109.06295}}].

\bibitem{Zhu2022}
Y.~{Zhu}, G.D.~{Becker}, S.E.I.~{Bosman}, L.C.~{Keating}, V.~{D'Odorico},
  R.L.~{Davies} et~al., \emph{{Long Dark Gaps in the Ly{\ensuremath{\beta}}
  Forest at z < 6: Evidence of Ultra-late Reionization from XQR-30 Spectra}},
  \href{https://doi.org/10.3847/1538-4357/ac6e60}{\emph{\apj} {\bfseries 932}
  (2022) 76} [\href{https://arxiv.org/abs/2205.04569}{{\ttfamily 2205.04569}}].

\bibitem{Becker2021}
G.D.~{Becker}, A.~{D'Aloisio}, H.M.~{Christenson}, Y.~{Zhu}, G.~{Worseck} and
  J.S.~{Bolton}, \emph{{The mean free path of ionizing photons at 5 < z < 6:
  evidence for rapid evolution near reionization}},
  \href{https://doi.org/10.1093/mnras/stab2696}{\emph{\mnras} {\bfseries 508}
  (2021) 1853} [\href{https://arxiv.org/abs/2103.16610}{{\ttfamily
  2103.16610}}].

\bibitem{Cain2021}
C.~{Cain}, A.~{D'Aloisio}, N.~{Gangolli} and G.D.~{Becker}, \emph{{A Short Mean
  Free Path at z = 6 Favors Late and Rapid Reionization by Faint Galaxies}},
  \href{https://doi.org/10.3847/2041-8213/ac1ace}{\emph{\apjl} {\bfseries 917}
  (2021) L37} [\href{https://arxiv.org/abs/2105.10511}{{\ttfamily
  2105.10511}}].

\bibitem{Zhu2023}
Y.~{Zhu}, G.D.~{Becker}, H.M.~{Christenson}, A.~{D'Aloisio}, S.E.I.~{Bosman},
  T.~{Bakx} et~al., \emph{{Probing Ultralate Reionization: Direct Measurements
  of the Mean Free Path over 5 < z < 6}},
  \href{https://doi.org/10.3847/1538-4357/aceef4}{\emph{\apj} {\bfseries 955}
  (2023) 115} [\href{https://arxiv.org/abs/2308.04614}{{\ttfamily
  2308.04614}}].

\bibitem{Gaikwad2023}
P.~{Gaikwad}, M.G.~{Haehnelt}, F.B.~{Davies}, S.E.I.~{Bosman}, M.~{Molaro},
  G.~{Kulkarni} et~al., \emph{{Measuring the photoionization rate, neutral
  fraction, and mean free path of H I ionizing photons at 4.9
  {\ensuremath{\leq}} z {\ensuremath{\leq}} 6.0 from a large sample of XShooter
  and ESI spectra}},
  \href{https://doi.org/10.1093/mnras/stad2566}{\emph{\mnras} {\bfseries 525}
  (2023) 4093} [\href{https://arxiv.org/abs/2304.02038}{{\ttfamily
  2304.02038}}].

\bibitem{Davies2023}
F.B.~{Davies}, S.E.I.~{Bosman}, P.~{Gaikwad}, F.~{Nasir}, J.F.~{Hennawi},
  G.D.~{Becker} et~al., \emph{{Constraints on the Evolution of the Ionizing
  Background and Ionizing Photon Mean Free Path at the End of Reionization}},
  \href{https://doi.org/10.3847/1538-4357/ad1d5d}{\emph{\apj} {\bfseries 965}
  (2024) 134} [\href{https://arxiv.org/abs/2312.08464}{{\ttfamily
  2312.08464}}].

\bibitem{Roth2023}
J.T.~{Roth}, A.~{D'Aloisio}, C.~{Cain}, B.~{Wilson}, Y.~{Zhu} and
  G.D.~{Becker}, \emph{{The effect of reionization on direct measurements of
  the mean free path}},
  \href{https://doi.org/10.1093/mnras/stae1194}{\emph{\mnras} {\bfseries 530}
  (2024) 5209} [\href{https://arxiv.org/abs/2311.06348}{{\ttfamily
  2311.06348}}].

\bibitem{Davies2018}
F.B.~Davies, J.F.~Hennawi, E.~Ba{\~{n}}ados, Z.~Luki{\'{c}}, R.~Decarli, X.~Fan
  et~al., \emph{Quantitative constraints on the reionization history from the
  {IGM} damping wing signature in two quasars at z $>$ 7},
  \href{https://doi.org/10.3847/1538-4357/aad6dc}{\emph{The Astrophysical
  Journal} {\bfseries 864} (2018) 142}.

\bibitem{Durovcikova2024}
D.~{{\v{D}}urov{\v{c}}{\'\i}kov{\'a}}, A.-C.~{Eilers}, H.~{Chen},
  S.~{Satyavolu}, G.~{Kulkarni}, R.A.~{Simcoe} et~al., \emph{{Chronicling the
  Reionization History at 6 {\ensuremath{\lesssim}} z {\ensuremath{\lesssim}} 7
  with Emergent Quasar Damping Wings}},
  \href{https://doi.org/10.3847/1538-4357/ad4888}{\emph{\apj} {\bfseries 969}
  (2024) 162} [\href{https://arxiv.org/abs/2401.10328}{{\ttfamily
  2401.10328}}].

\bibitem{2009arXiv0912.0201L}
{LSST Science Collaboration}, P.A.~{Abell}, J.~{Allison}, S.F.~{Anderson},
  J.R.~{Andrew}, J.R.P.~{Angel} et~al., \emph{{LSST Science Book, Version
  2.0}}, \href{https://doi.org/10.48550/arXiv.0912.0201}{\emph{arXiv e-prints}
  (2009) arXiv:0912.0201} [\href{https://arxiv.org/abs/0912.0201}{{\ttfamily
  0912.0201}}].

\bibitem{2016arXiv161100036D}
{DESI Collaboration}, A.~{Aghamousa}, J.~{Aguilar}, S.~{Ahlen}, S.~{Alam},
  L.E.~{Allen} et~al., \emph{{The DESI Experiment Part I: Science,Targeting,
  and Survey Design}},
  \href{https://doi.org/10.48550/arXiv.1611.00036}{\emph{arXiv e-prints} (2016)
  arXiv:1611.00036} [\href{https://arxiv.org/abs/1611.00036}{{\ttfamily
  1611.00036}}].

\bibitem{Zhu2024}
Y.~{Zhu}, G.D.~{Becker}, S.E.I.~{Bosman}, C.~{Cain}, L.C.~{Keating}, F.~{Nasir}
  et~al., \emph{{Damping wing-like features in the stacked Ly
  {\ensuremath{\alpha}} forest: Potential neutral hydrogen islands at z < 6}},
  \href{https://doi.org/10.1093/mnrasl/slae061}{\emph{\mnras} {\bfseries 533}
  (2024) L49} [\href{https://arxiv.org/abs/2405.12275}{{\ttfamily
  2405.12275}}].

\bibitem{Spina2024}
B.~{Spina}, S.E.I.~{Bosman}, F.B.~{Davies}, P.~{Gaikwad} and Y.~{Zhu},
  \emph{{Damping wings in the Lyman-{\ensuremath{\alpha}} forest: A
  model-independent measurement of the neutral fraction at 5.4 < z < 6.1}},
  \href{https://doi.org/10.1051/0004-6361/202450798}{\emph{\aap} {\bfseries
  688} (2024) L26} [\href{https://arxiv.org/abs/2405.12273}{{\ttfamily
  2405.12273}}].

\bibitem{Haiman2000}
Z.~{Haiman}, T.~{Abel} and P.~{Madau}, \emph{{Photon Consumption in Minihalos
  during Cosmological Reionization}},
  \href{https://doi.org/10.1086/320232}{\emph{\apj} {\bfseries 551} (2001) 599}
  [\href{https://arxiv.org/abs/astro-ph/0009125}{{\ttfamily
  astro-ph/0009125}}].

\bibitem{Duffy2014}
A.R.~{Duffy}, J.S.B.~{Wyithe}, S.J.~{Mutch} and G.B.~{Poole}, \emph{{Low-mass
  galaxy formation and the ionizing photon budget during reionization}},
  \href{https://doi.org/10.1093/mnras/stu1328}{\emph{\mnras} {\bfseries 443}
  (2014) 3435} [\href{https://arxiv.org/abs/1405.7459}{{\ttfamily 1405.7459}}].

\bibitem{Davies2021b}
F.B.~{Davies}, S.E.I.~{Bosman}, S.R.~{Furlanetto}, G.D.~{Becker} and
  A.~{D'Aloisio}, \emph{{The Predicament of Absorption-dominated Reionization:
  Increased Demands on Ionizing Sources}},
  \href{https://doi.org/10.3847/2041-8213/ac1ffb}{\emph{\apjl} {\bfseries 918}
  (2021) L35} [\href{https://arxiv.org/abs/2105.10518}{{\ttfamily
  2105.10518}}].

\bibitem{Cain2022b}
C.~{Cain}, A.~{D'Aloisio}, N.~{Gangolli} and M.~{McQuinn}, \emph{{The
  morphology of reionization in a dynamically clumpy universe}},
  \href{https://doi.org/10.1093/mnras/stad1057}{\emph{\mnras} {\bfseries 522}
  (2023) 2047} [\href{https://arxiv.org/abs/2207.11266}{{\ttfamily
  2207.11266}}].

\bibitem{Pawlik2015}
A.H.~{Pawlik}, J.~{Schaye} and C.~{Dalla Vecchia}, \emph{{Spatially adaptive
  radiation-hydrodynamical simulations of galaxy formation during cosmological
  reionization}}, \href{https://doi.org/10.1093/mnras/stv976}{\emph{\mnras}
  {\bfseries 451} (2015) 1586}
  [\href{https://arxiv.org/abs/1501.01980}{{\ttfamily 1501.01980}}].

\bibitem{Shukla2016}
H.~{Shukla}, G.~{Mellema}, I.T.~{Iliev} and P.R.~{Shapiro}, \emph{{The effects
  of Lyman-limit systems on the evolution and observability of the epoch of
  reionization}}, \href{https://doi.org/10.1093/mnras/stw249}{\emph{\mnras}
  {\bfseries 458} (2016) 135}
  [\href{https://arxiv.org/abs/1602.01144}{{\ttfamily 1602.01144}}].

\bibitem{Mao2019}
Y.~{Mao}, J.~{Koda}, P.R.~{Shapiro}, I.T.~{Iliev}, G.~{Mellema}, H.~{Park}
  et~al., \emph{{The impact of inhomogeneous subgrid clumping on cosmic
  reionization}}, \href{https://doi.org/10.1093/mnras/stz2986}{\emph{\mnras}
  {\bfseries 491} (2020) 1600}
  [\href{https://arxiv.org/abs/1906.02476}{{\ttfamily 1906.02476}}].

\bibitem{Qin2021}
Y.~Qin, A.~Mesinger, S.E.I.~Bosman and M.~Viel, \emph{{Reionization and galaxy
  inference from the high-redshift Ly$\alpha$ forest}},
  \href{https://doi.org/10.1093/mnras/stab1833}{\emph{Monthly Notices of the
  Royal Astronomical Society} {\bfseries 506} (2021) 2390}
  [\href{https://arxiv.org/abs/https://academic.oup.com/mnras/article-pdf/506/2/2390/39136191/stab1833.pdf}{{\ttfamily
  https://academic.oup.com/mnras/article-pdf/506/2/2390/39136191/stab1833.pdf}}].

\bibitem{Maity2023}
B.~{Maity}, A.~{Paranjape} and T.R.~{Choudhury}, \emph{{A fast method of
  reionization parameter space exploration using GPR trained SCRIPT}},
  \href{https://doi.org/10.1093/mnras/stad2984}{\emph{\mnras} {\bfseries 526}
  (2023) 3920} [\href{https://arxiv.org/abs/2305.04839}{{\ttfamily
  2305.04839}}].

\bibitem{Furlanetto2004}
S.R.~{Furlanetto}, M.~{Zaldarriaga} and L.~{Hernquist}, \emph{{The Growth of H
  II Regions During Reionization}},
  \href{https://doi.org/10.1086/423025}{\emph{The Astrophysical Journal}
  {\bfseries 613} (2004) 1}
  [\href{https://arxiv.org/abs/astro-ph/0403697}{{\ttfamily
  astro-ph/0403697}}].

\bibitem{Mesinger2007}
A.~{Mesinger} and S.~{Furlanetto}, \emph{{Efficient Simulations of Early
  Structure Formation and Reionization}},
  \href{https://doi.org/10.1086/521806}{\emph{\apj} {\bfseries 669} (2007) 663}
  [\href{https://arxiv.org/abs/0704.0946}{{\ttfamily 0704.0946}}].

\bibitem{Grieg2015}
B.~{Greig} and A.~{Mesinger}, \emph{{21CMMC: an MCMC analysis tool enabling
  astrophysical parameter studies of the cosmic 21 cm signal}},
  \href{https://doi.org/10.1093/mnras/stv571}{\emph{\mnras} {\bfseries 449}
  (2015) 4246} [\href{https://arxiv.org/abs/1501.06576}{{\ttfamily
  1501.06576}}].

\bibitem{Choudhury2018}
T.R.~{Choudhury} and A.~{Paranjape}, \emph{{Photon number conservation and the
  large-scale 21 cm power spectrum in seminumerical models of reionization}},
  \href{https://doi.org/10.1093/mnras/sty2551}{\emph{\mnras} {\bfseries 481}
  (2018) 3821} [\href{https://arxiv.org/abs/1807.00836}{{\ttfamily
  1807.00836}}].

\bibitem{Trac2021}
H.~{Trac}, N.~{Chen}, I.~{Holst}, M.A.~{Alvarez} and R.~{Cen}, \emph{{AMBER: A
  Semi-numerical Abundance Matching Box for the Epoch of Reionization}},
  \href{https://doi.org/10.3847/1538-4357/ac5116}{\emph{\apj} {\bfseries 927}
  (2022) 186} [\href{https://arxiv.org/abs/2109.10375}{{\ttfamily
  2109.10375}}].

\bibitem{Mellema2006}
G.~{Mellema}, I.T.~{Iliev}, M.A.~{Alvarez} and P.R.~{Shapiro}, \emph{{C
  $^{2}$-ray: A new method for photon-conserving transport of ionizing
  radiation}}, \href{https://doi.org/10.1016/j.newast.2005.09.004}{\emph{\na}
  {\bfseries 11} (2006) 374}
  [\href{https://arxiv.org/abs/astro-ph/0508416}{{\ttfamily
  astro-ph/0508416}}].

\bibitem{Trac2007}
H.~Trac and R.~Cen, \emph{Radiative transfer simulations of cosmic
  reionization. i. methodology and initial results},
  \href{https://doi.org/10.1086/522566}{\emph{The Astrophysical Journal}
  {\bfseries 671} (2007) 1}.

\bibitem{Rosdahl2013}
J.~{Rosdahl}, J.~{Blaizot}, D.~{Aubert}, T.~{Stranex} and R.~{Teyssier},
  \emph{{RAMSES-RT: radiation hydrodynamics in the cosmological context}},
  \href{https://doi.org/10.1093/mnras/stt1722}{\emph{\mnras} {\bfseries 436}
  (2013) 2188} [\href{https://arxiv.org/abs/1304.7126}{{\ttfamily 1304.7126}}].

\bibitem{Gnedin2014}
N.Y.~{Gnedin}, \emph{{Cosmic Reionization on Computers. I. Design and
  Calibration of Simulations}},
  \href{https://doi.org/10.1088/0004-637X/793/1/29}{\emph{\apj} {\bfseries 793}
  (2014) 29} [\href{https://arxiv.org/abs/1403.4245}{{\ttfamily 1403.4245}}].

\bibitem{Ocvirk2016}
P.~{Ocvirk}, N.~{Gillet}, P.R.~{Shapiro}, D.~{Aubert}, I.T.~{Iliev},
  R.~{Teyssier} et~al., \emph{{Cosmic Dawn (CoDa): the First
  Radiation-Hydrodynamics Simulation of Reionization and Galaxy Formation in
  the Local Universe}},
  \href{https://doi.org/10.1093/mnras/stw2036}{\emph{\mnras} {\bfseries 463}
  (2016) 1462} [\href{https://arxiv.org/abs/1511.00011}{{\ttfamily
  1511.00011}}].

\bibitem{Rosdahl2018}
J.~{Rosdahl}, H.~{Katz}, J.~{Blaizot}, T.~{Kimm}, L.~{Michel-Dansac},
  T.~{Garel} et~al., \emph{{The SPHINX cosmological simulations of the first
  billion years: the impact of binary stars on reionization}},
  \href{https://doi.org/10.1093/mnras/sty1655}{\emph{\mnras} {\bfseries 479}
  (2018) 994} [\href{https://arxiv.org/abs/1801.07259}{{\ttfamily
  1801.07259}}].

\bibitem{Kannan2022}
R.~{Kannan}, E.~{Garaldi}, A.~{Smith}, R.~{Pakmor}, V.~{Springel},
  M.~{Vogelsberger} et~al., \emph{{Introducing the THESAN project:
  radiation-magnetohydrodynamic simulations of the epoch of reionization}},
  \href{https://doi.org/10.1093/mnras/stab3710}{\emph{\mnras} {\bfseries 511}
  (2022) 4005} [\href{https://arxiv.org/abs/2110.00584}{{\ttfamily
  2110.00584}}].

\bibitem{Zahn2011}
O.~{Zahn}, A.~{Mesinger}, M.~{McQuinn}, H.~{Trac}, R.~{Cen} and
  L.E.~{Hernquist}, \emph{{Comparison of reionization models: radiative
  transfer simulations and approximate, seminumeric models}},
  \href{https://doi.org/10.1111/j.1365-2966.2011.18439.x}{\emph{\mnras}
  {\bfseries 414} (2011) 727}
  [\href{https://arxiv.org/abs/1003.3455}{{\ttfamily 1003.3455}}].

\bibitem{Qin2022}
Y.~{Qin}, J.S.B.~{Wyithe}, P.A.~{Oesch}, G.D.~{Illingworth}, E.~{Leonova},
  S.J.~{Mutch} et~al., \emph{{Dark-ages reionization and galaxy formation
  simulation XX. The Ly {\ensuremath{\alpha}} IGM transmission properties and
  environment of bright galaxies during the epoch of reionization}},
  \href{https://doi.org/10.1093/mnras/stab3733}{\emph{\mnras} {\bfseries 510}
  (2022) 3858} [\href{https://arxiv.org/abs/2108.03675}{{\ttfamily
  2108.03675}}].

\bibitem{Ocvirk2018}
P.~{Ocvirk}, D.~{Aubert}, J.G.~{Sorce}, P.R.~{Shapiro}, N.~{Deparis},
  T.~{Dawoodbhoy} et~al., \emph{{Cosmic Dawn II (CoDa II): a new
  radiation-hydrodynamics simulation of the self-consistent coupling of galaxy
  formation and reionization}},
  \href{https://doi.org/10.1093/mnras/staa1266}{\emph{\mnras} {\bfseries 496}
  (2020) 4087} [\href{https://arxiv.org/abs/1811.11192}{{\ttfamily
  1811.11192}}].

\bibitem{Doussot2019}
A.~{Doussot}, H.~{Trac} and R.~{Cen}, \emph{{SCORCH. II. Radiation-hydrodynamic
  Simulations of Reionization with Varying Radiation Escape Fractions}},
  \href{https://doi.org/10.3847/1538-4357/aaef75}{\emph{\apj} {\bfseries 870}
  (2019) 18} [\href{https://arxiv.org/abs/1712.04464}{{\ttfamily 1712.04464}}].

\bibitem{Vogelsberger2014}
M.~Vogelsberger, S.~Genel, V.~Springel, P.~Torrey, D.~Sijacki, D.~Xu et~al.,
  \emph{Introducing the illustris project: simulating the coevolution of dark
  and visible matter in the universe},
  \href{https://doi.org/10.1093/mnras/stu1536}{\emph{Monthly Notices of the
  Royal Astronomical Society} {\bfseries 444} (2014) 1518}.

\bibitem{Hassan2022}
S.~{Hassan}, R.~{Dav{\'e}}, M.~{McQuinn}, R.S.~{Somerville}, L.C.~{Keating},
  D.~{Angl{\'e}s-Alc{\'a}zar} et~al., \emph{{Reionization with SIMBA: How Much
  Does Astrophysics Matter in Modeling Cosmic Reionization?}},
  \href{https://doi.org/10.3847/1538-4357/ac69e2}{\emph{\apj} {\bfseries 931}
  (2022) 62} [\href{https://arxiv.org/abs/2109.03840}{{\ttfamily 2109.03840}}].

\bibitem{Iliev2006}
I.T.~{Iliev}, B.~{Ciardi}, M.A.~{Alvarez}, A.~{Maselli}, A.~{Ferrara},
  N.Y.~{Gnedin} et~al., \emph{{Cosmological radiative transfer codes comparison
  project - I. The static density field tests}},
  \href{https://doi.org/10.1111/j.1365-2966.2006.10775.x}{\emph{\mnras}
  {\bfseries 371} (2006) 1057}
  [\href{https://arxiv.org/abs/astro-ph/0603199}{{\ttfamily
  astro-ph/0603199}}].

\bibitem{Planck2018}
{Planck Collaboration}, N.~{Aghanim}, Y.~{Akrami}, M.~{Ashdown}, J.~{Aumont},
  C.~{Baccigalupi} et~al., \emph{{Planck 2018 results. VI. Cosmological
  parameters}}, \href{https://doi.org/10.1051/0004-6361/201833910}{\emph{\aap}
  {\bfseries 641} (2020) A6}
  [\href{https://arxiv.org/abs/1807.06209}{{\ttfamily 1807.06209}}].

\bibitem{Levermore1984}
C.~Levermore, \emph{Relating eddington factors to flux limiters},
  \href{https://doi.org/https://doi.org/10.1016/0022-4073(84)90112-2}{\emph{Journal
  of Quantitative Spectroscopy and Radiative Transfer} {\bfseries 31} (1984)
  149}.

\bibitem{Gnedin2001}
N.Y.~{Gnedin} and T.~{Abel}, \emph{{Multi-dimensional cosmological radiative
  transfer with a Variable Eddington Tensor formalism}},
  \href{https://doi.org/10.1016/S1384-1076(01)00068-9}{\emph{\na} {\bfseries 6}
  (2001) 437} [\href{https://arxiv.org/abs/astro-ph/0106278}{{\ttfamily
  astro-ph/0106278}}].

\bibitem{Iliev2009}
I.T.~{Iliev}, D.~{Whalen}, G.~{Mellema}, K.~{Ahn}, S.~{Baek}, N.Y.~{Gnedin}
  et~al., \emph{{Cosmological radiative transfer comparison project - II. The
  radiation-hydrodynamic tests}},
  \href{https://doi.org/10.1111/j.1365-2966.2009.15558.x}{\emph{\mnras}
  {\bfseries 400} (2009) 1283}
  [\href{https://arxiv.org/abs/0905.2920}{{\ttfamily 0905.2920}}].

\bibitem{Kannan2019}
R.~{Kannan}, M.~{Vogelsberger}, F.~{Marinacci}, R.~{McKinnon}, R.~{Pakmor} and
  V.~{Springel}, \emph{{AREPO-RT: radiation hydrodynamics on a moving mesh}},
  \href{https://doi.org/10.1093/mnras/stz287}{\emph{\mnras} {\bfseries 485}
  (2019) 117} [\href{https://arxiv.org/abs/1804.01987}{{\ttfamily
  1804.01987}}].

\bibitem{Wu2021}
X.~{Wu}, M.~{McQuinn} and D.~{Eisenstein}, \emph{{On the accuracy of common
  moment-based radiative transfer methods for simulating reionization}},
  \href{https://doi.org/10.1088/1475-7516/2021/02/042}{\emph{\jcap} {\bfseries
  2021} (2021) 042} [\href{https://arxiv.org/abs/2009.07278}{{\ttfamily
  2009.07278}}].

\bibitem{Hirling2023}
P.~{Hirling}, M.~{Bianco}, S.K.~{Giri}, I.T.~{Iliev}, G.~{Mellema} and
  J.P.~{Kneib}, \emph{{pyC2Ray: A flexible and GPU-accelerated radiative
  transfer framework for simulating the cosmic epoch of reionization}},
  \href{https://doi.org/10.1016/j.ascom.2024.100861}{\emph{Astronomy and
  Computing} {\bfseries 48} (2024) 100861}
  [\href{https://arxiv.org/abs/2311.01492}{{\ttfamily 2311.01492}}].

\bibitem{Abel2002}
T.~{Abel} and B.D.~{Wandelt}, \emph{{Adaptive ray tracing for radiative
  transfer around point sources}},
  \href{https://doi.org/10.1046/j.1365-8711.2002.05206.x}{\emph{\mnras}
  {\bfseries 330} (2002) L53}
  [\href{https://arxiv.org/abs/astro-ph/0111033}{{\ttfamily
  astro-ph/0111033}}].

\bibitem{McQuinn2007}
M.~{McQuinn}, A.~{Lidz}, O.~{Zahn}, S.~{Dutta}, L.~{Hernquist} and
  M.~{Zaldarriaga}, \emph{{The morphology of HII regions during reionization}},
  \href{https://doi.org/10.1111/j.1365-2966.2007.11489.x}{\emph{\mnras}
  {\bfseries 377} (2007) 1043}
  [\href{https://arxiv.org/abs/astro-ph/0610094}{{\ttfamily
  astro-ph/0610094}}].

\bibitem{2019MNRAS.483.1582H}
B.~{Hartley} and M.~{Ricotti}, \emph{{ARC: adaptive ray-tracing with CUDA, a
  new ray tracing code for parallel GPUs}},
  \href{https://doi.org/10.1093/mnras/sty2753}{\emph{\mnras} {\bfseries 483}
  (2019) 1582} [\href{https://arxiv.org/abs/1807.07094}{{\ttfamily
  1807.07094}}].

\bibitem{Zheng2010}
Z.~{Zheng}, R.~{Cen}, H.~{Trac} and J.~{Miralda-Escud{\'e}}, \emph{{Radiative
  Transfer Modeling of Ly{\ensuremath{\alpha}} Emitters. I. Statistics of
  Spectra and Luminosity}},
  \href{https://doi.org/10.1088/0004-637X/716/1/574}{\emph{\apj} {\bfseries
  716} (2010) 574} [\href{https://arxiv.org/abs/0910.2712}{{\ttfamily
  0910.2712}}].

\bibitem{2018ApJ...863L...6V}
E.~{Visbal} and M.~{McQuinn}, \emph{{The Impact of Neutral Intergalactic Gas on
  Ly{\ensuremath{\alpha}} Intensity Mapping during Reionization}},
  \href{https://doi.org/10.3847/2041-8213/aad5e6}{\emph{\apjl} {\bfseries 863}
  (2018) L6} [\href{https://arxiv.org/abs/1807.03370}{{\ttfamily 1807.03370}}].

\bibitem{Ciardi2001}
B.~{Ciardi}, A.~{Ferrara}, S.~{Marri} and G.~{Raimondo}, \emph{{Cosmological
  reionization around the first stars: Monte Carlo radiative transfer}},
  \href{https://doi.org/10.1046/j.1365-8711.2001.04316.x}{\emph{\mnras}
  {\bfseries 324} (2001) 381}
  [\href{https://arxiv.org/abs/astro-ph/0005181}{{\ttfamily
  astro-ph/0005181}}].

\bibitem{Maselli2003}
A.~{Maselli}, A.~{Ferrara} and B.~{Ciardi}, \emph{{CRASH: a radiative transfer
  scheme}},
  \href{https://doi.org/10.1046/j.1365-8711.2003.06979.x}{\emph{\mnras}
  {\bfseries 345} (2003) 379}
  [\href{https://arxiv.org/abs/astro-ph/0307117}{{\ttfamily
  astro-ph/0307117}}].

\bibitem{Cain2023}
C.~{Cain}, A.~{D'Aloisio}, G.~{Lopez}, N.~{Gangolli} and J.T.~{Roth}, \emph{{On
  the rise and fall of galactic ionizing output at the end of reionization}},
  \href{https://doi.org/10.1093/mnras/stae1223}{\emph{\mnras} {\bfseries 531}
  (2024) 1951} [\href{https://arxiv.org/abs/2311.13638}{{\ttfamily
  2311.13638}}].

\bibitem{Gorski1999}
K.M.~{Gorski}, B.D.~{Wandelt}, F.K.~{Hansen}, E.~{Hivon} and A.J.~{Banday},
  \emph{{The HEALPix Primer}}, {\emph{arXiv e-prints} (1999) astro}
  [\href{https://arxiv.org/abs/astro-ph/9905275}{{\ttfamily
  astro-ph/9905275}}].

\bibitem{Sokasian2001}
A.~{Sokasian}, T.~{Abel} and L.E.~{Hernquist}, \emph{{Simulating reionization
  in numerical cosmology}},
  \href{https://doi.org/10.1016/S1384-1076(01)00065-3}{\emph{\na} {\bfseries 6}
  (2001) 359} [\href{https://arxiv.org/abs/astro-ph/0105181}{{\ttfamily
  astro-ph/0105181}}].

\bibitem{Wilson2024a}
B.~{Wilson}, A.~{D'Aloisio}, G.D.~{Becker}, C.~{Cain} and E.~{Visbal},
  \emph{{Quantifying Lyman-$\alpha$ emissions from reionization fronts}},
  \href{https://doi.org/10.48550/arXiv.2406.14622}{\emph{arXiv e-prints} (2024)
  arXiv:2406.14622} [\href{https://arxiv.org/abs/2406.14622}{{\ttfamily
  2406.14622}}].

\bibitem{Hui1997}
L.~{Hui} and N.Y.~{Gnedin}, \emph{{Equation of state of the photoionized
  intergalactic medium}},
  \href{https://doi.org/10.1093/mnras/292.1.27}{\emph{\mnras} {\bfseries 292}
  (1997) 27} [\href{https://arxiv.org/abs/astro-ph/9612232}{{\ttfamily
  astro-ph/9612232}}].

\bibitem{Emberson2013}
J.D.~Emberson, R.M.~Thomas and M.A.~Alvarez, \emph{{THE} {OPACITY} {OF} {THE}
  {INTERGALACTIC} {MEDIUM} {DURING} {REIONIZATION}: {RESOLVING} {SMALL}-{SCALE}
  {STRUCTURE}}, \href{https://doi.org/10.1088/0004-637x/763/2/146}{\emph{The
  Astrophysical Journal} {\bfseries 763} (2013) 146}.

\bibitem{Park2016}
H.~{Park}, P.R.~{Shapiro}, J.-h.~{Choi}, N.~{Yoshida}, S.~{Hirano} and
  K.~{Ahn}, \emph{{The Hydrodynamic Feedback of Cosmic Reionization on
  Small-scale Structures and Its Impact on Photon Consumption During the Epoch
  of Reionization}},
  \href{https://doi.org/10.3847/0004-637X/831/1/86}{\emph{ApJ} {\bfseries 831}
  (2016) 86} [\href{https://arxiv.org/abs/1602.06472}{{\ttfamily 1602.06472}}].

\bibitem{DAloisio2020}
A.~D'Aloisio, M.~McQuinn, H.~Trac, C.~Cain and A.~Mesinger, \emph{Hydrodynamic
  response of the intergalactic medium to reionization},
  \href{https://doi.org/10.3847/1538-4357/ab9f2f}{\emph{The Astrophysical
  Journal} {\bfseries 898} (2020) 149}.

\bibitem{Chan2023}
T.K.~{Chan}, A.~{Ben{\'\i}tez-Llambay}, T.~{Theuns}, C.~{Frenk} and R.~{Bower},
  \emph{{The impact and response of mini-haloes and the interhalo medium on
  cosmic reionization}},
  \href{https://doi.org/10.1093/mnras/stae114}{\emph{\mnras} {\bfseries 528}
  (2024) 1296} [\href{https://arxiv.org/abs/2305.04959}{{\ttfamily
  2305.04959}}].

\bibitem{DAloisio2019}
A.~{D'Aloisio}, M.~{McQuinn}, O.~{Maupin}, F.B.~{Davies}, H.~{Trac},
  S.~{Fuller} et~al., \emph{{Heating of the Intergalactic Medium by Hydrogen
  Reionization}}, \href{https://doi.org/10.3847/1538-4357/ab0d83}{\emph{\apj}
  {\bfseries 874} (2019) 154}
  [\href{https://arxiv.org/abs/1807.09282}{{\ttfamily 1807.09282}}].

\bibitem{Zeng2021}
C.~{Zeng} and C.M.~{Hirata}, \emph{{Nonequilibrium Temperature Evolution of
  Ionization Fronts during the Epoch of Reionization}},
  \href{https://doi.org/10.3847/1538-4357/abca38}{\emph{\apj} {\bfseries 906}
  (2021) 124} [\href{https://arxiv.org/abs/2007.02940}{{\ttfamily
  2007.02940}}].

\bibitem{Susa2000}
H.~{Susa} and T.~{Kitayama}, \emph{{Collapse of low-mass clouds in the presence
  of a UV radiation field}},
  \href{https://doi.org/10.1046/j.1365-8711.2000.03616.x}{\emph{\mnras}
  {\bfseries 317} (2000) 175}
  [\href{https://arxiv.org/abs/astro-ph/0004303}{{\ttfamily
  astro-ph/0004303}}].

\bibitem{Nakamoto2001}
T.~{Nakamoto}, M.~{Umemura} and H.~{Susa}, \emph{{The effects of radiative
  transfer on the reionization of an inhomogeneous universe}},
  \href{https://doi.org/10.1046/j.1365-8711.2001.04008.x}{\emph{\mnras}
  {\bfseries 321} (2001) 593}.

\bibitem{Razoumov2005}
A.O.~{Razoumov} and C.Y.~{Cardall}, \emph{{Fully threaded transport engine: new
  method for multi-scale radiative transfer}},
  \href{https://doi.org/10.1111/j.1365-2966.2005.09409.x}{\emph{\mnras}
  {\bfseries 362} (2005) 1413}
  [\href{https://arxiv.org/abs/astro-ph/0505172}{{\ttfamily
  astro-ph/0505172}}].

\bibitem{Ritzerveld2003}
J.~{Ritzerveld}, V.~{Icke} and E.-J.~{Rijkhorst}, \emph{{Triangulating
  Radiation: Radiative Transfer on Unstructured Grids}},
  \href{https://doi.org/10.48550/arXiv.astro-ph/0312301}{\emph{arXiv e-prints}
  (2003) astro} [\href{https://arxiv.org/abs/astro-ph/0312301}{{\ttfamily
  astro-ph/0312301}}].

\bibitem{Whalen2006}
D.~{Whalen} and M.L.~{Norman}, \emph{{A Multistep Algorithm for the Radiation
  Hydrodynamical Transport of Cosmological Ionization Fronts and Ionized
  Flows}}, \href{https://doi.org/10.1086/499072}{\emph{\apjs} {\bfseries 162}
  (2006) 281} [\href{https://arxiv.org/abs/astro-ph/0508214}{{\ttfamily
  astro-ph/0508214}}].

\bibitem{Rijkhorst2006}
E.J.~{Rijkhorst}, T.~{Plewa}, A.~{Dubey} and G.~{Mellema}, \emph{{Hybrid
  characteristics: 3D radiative transfer for parallel adaptive mesh refinement
  hydrodynamics}},
  \href{https://doi.org/10.1051/0004-6361:20053401}{\emph{\aap} {\bfseries 452}
  (2006) 907} [\href{https://arxiv.org/abs/astro-ph/0505213}{{\ttfamily
  astro-ph/0505213}}].

\bibitem{Alvarez2006}
M.A.~{Alvarez}, V.~{Bromm} and P.R.~{Shapiro}, \emph{{The H II Region of the
  First Star}}, \href{https://doi.org/10.1086/499578}{\emph{\apj} {\bfseries
  639} (2006) 621} [\href{https://arxiv.org/abs/astro-ph/0507684}{{\ttfamily
  astro-ph/0507684}}].

\bibitem{Finkelstein2019}
S.L.~{Finkelstein}, A.~{D'Aloisio}, J.-P.~{Paardekooper}, J.~{Ryan}, Russell,
  P.~{Behroozi}, K.~{Finlator} et~al., \emph{{Conditions for Reionizing the
  Universe with a Low Galaxy Ionizing Photon Escape Fraction}},
  \href{https://doi.org/10.3847/1538-4357/ab1ea8}{\emph{\apj} {\bfseries 879}
  (2019) 36} [\href{https://arxiv.org/abs/1902.02792}{{\ttfamily 1902.02792}}].

\bibitem{Davies2016}
F.B.~{Davies} and S.R.~{Furlanetto}, \emph{{Large fluctuations in the
  hydrogen-ionizing background and mean free path following the epoch of
  reionization}}, \href{https://doi.org/10.1093/mnras/stw931}{\emph{\mnras}
  {\bfseries 460} (2016) 1328}
  [\href{https://arxiv.org/abs/1509.07131}{{\ttfamily 1509.07131}}].

\bibitem{Gangolli2024}
N.~{Gangolli}, A.~{D'Aloisio}, C.~{Cain}, G.D.~{Becker} and H.~{Christenson},
  \emph{{On the correlation between Ly$\alpha$ forest opacity and galaxy
  density in late reionization models}},
  \href{https://doi.org/10.48550/arXiv.2408.08358}{\emph{arXiv e-prints} (2024)
  arXiv:2408.08358} [\href{https://arxiv.org/abs/2408.08358}{{\ttfamily
  2408.08358}}].

\end{thebibliography}\endgroup
\bibliographystyle{JHEP}

\end{document}